\documentclass[preprint2]{aastex}
\pdfoutput=1

\usepackage{amsmath, hyperref, multirow}
\hypersetup{colorlinks=true,linkcolor=black,citecolor=black,filecolor=black,urlcolor=black}

\newlength{\tablerowsep}
\newcommand{\phe}{\phm{=}}

\newcommand{\RA}[3]{#1^\mathrm{h}#2^\mathrm{m}#3^\mathrm{s}}
\newcommand{\DEC}[3]{#1\degree#2'#3''}

\newcommand{\unit}[1]{\expandafter\newcommand\csname#1\endcsname{\mathrm{#1}}}
\unit{cm}
\unit{km}
\unit{kpc}
\unit{Mpc}
\unit{s}
\unit{days}
\newcommand{\gram}{\mathrm{g}}
\newcommand{\A}{\text{\AA}}
\newcommand{\kms}{\km~\s^{-1}}
\newcommand{\magn}{\mathrm{mag}}
\newcommand{\degree}{^\circ}

\newcommand{\Halpha}{\ensuremath{\mathrm{H}\alpha}}
\newcommand{\Hbeta}{\ensuremath{\mathrm{H}\beta}}
\newcommand{\Hgamma}{\ensuremath{\mathrm{H}\gamma}}
\newcommand{\Mpeak}{M_\text{peak}}
\newcommand{\Dm}{\Delta m_{15}}
\newcommand{\gi}{\ensuremath{g{-}i}}
\newcommand{\LL}{\lambda\lambda}
\newcommand{\isotope}[2]{\ensuremath{{}^{#2}\mathrm{#1}}}

\shorttitle{Slow-Speed Supernovae from PTF}
\shortauthors{C.~J.\ White, M.~M.\ Kasliwal, et al.}

\begin{document}

\title{Slow-Speed Supernovae from the Palomar Transient Factory: \\ Two Channels}
\author{Christopher~J.~White,\altaffilmark{1} Mansi~M.~Kasliwal,\altaffilmark{2} Peter~E.~Nugent,\altaffilmark{3,4} Avishay~Gal-Yam,\altaffilmark{5} D.~Andrew~Howell,\altaffilmark{6,7} Mark~Sullivan,\altaffilmark{8} Ariel~Goobar,\altaffilmark{9} Anthony~L.~Piro,\altaffilmark{10} Joshua~S.~Bloom,\altaffilmark{4} Shrinivas~R.~Kulkarni,\altaffilmark{11} Russ~R.~Laher,\altaffilmark{12} Frank~Masci,\altaffilmark{12} Eran~O.~Ofek,\altaffilmark{5} Jason~Surace,\altaffilmark{12} Sagi~Ben-Ami,\altaffilmark{5} Yi~Cao,\altaffilmark{11} S.~Bradley~Cenko,\altaffilmark{13,14} Isobel~M.~Hook,\altaffilmark{15,16} Jakob~J{\"o}nsson,\altaffilmark{17} Thomas~Matheson,\altaffilmark{18} Assaf~Sternberg,\altaffilmark{19,20} Robert~M.~Quimby,\altaffilmark{21} Ofer~Yaron\altaffilmark{5}}
\altaffiltext{1}{Department of Astrophysical Sciences, Princeton University, 4 Ivy Lane, Princeton, NJ 08544, USA}
\altaffiltext{2}{The Observatories, Carnegie Institution for Science, 813 Santa Barbara Street, Pasadena, CA 91101, USA}
\altaffiltext{3}{Computational Cosmology Center, Lawrence Berkeley National Laboratory, 1 Cyclotron Road, Berkeley, CA 94720, USA}
\altaffiltext{4}{Department of Astronomy, University of California, Berkeley, CA 94720-3411, USA}
\altaffiltext{5}{Benoziyo Center for Astrophysics, The Weizmann Institute of Science, Rehovot 76100, Israel}
\altaffiltext{6}{Department of Physics, University of California, Santa Barbara, Broida Hall, Mail Code 9530, Santa Barbara, CA 93106-9530, USA}
\altaffiltext{7}{Las Cumbres Observatory Global Telescope Network, Inc., Santa Barbara, CA 93117, USA}
\altaffiltext{8}{School of Physics and Astronomy, University of Southampton, Southampton, SO17 1BJ, UK}
\altaffiltext{9}{The Oskar Klein Centre, Department of Physics, AlbaNova, Stockholm University, SE-106 91 Stockholm, Sweden}
\altaffiltext{10}{Theoretical Astrophysics, California Institute of Technology, 1200 E California Blvd, M/C 350-17, Pasadena, CA 91125, USA}
\altaffiltext{11}{Cahill Center for Astrophysics, California Institute of Technology, Pasadena, CA 91125, USA}
\altaffiltext{12}{Spitzer Science Center, California Institute of Technology, M/S 314-6, Pasadena, CA 91125, USA}
\altaffiltext{13}{Astrophysics Science Division, NASA Goddard Space Flight Center, Mail Code 661, Greenbelt, MD 20771, USA}
\altaffiltext{14}{Joint Space Science Institute, University of Maryland, College Park, MD 20742, USA}
\altaffiltext{15}{Department of Physics (Astrophysics), University of Oxford, Keble Road, Oxford OX1 3RH, UK}
\altaffiltext{16}{INAF -- Osservatorio Astronomico di Roma, Via Frascati 33, 00040, Monte Porzio Catone (RM), Italy}
\altaffiltext{17}{Savantic AB, Rosenlundsgatan 50, 118 63 Stockholm, Sweden}
\altaffiltext{18}{National Optical Astronomy Observatory, Tucson, AZ 85719-4933, USA}
\altaffiltext{19}{Excellence Cluster Universe, Technische Universit{\"a}t M{\"u}nchen, Boltzmannstr.\ 2, D-85748, Garching, Germany}
\altaffiltext{20}{Max Planck Institute for Astrophysics, Karl Schwarzschild St.\ 1, D-85748 Garching, Germany}
\altaffiltext{21}{Kavli IPMU, The University of Tokyo, Kashiwanoha 5-1-5, Kashiwa 277-8583, Japan}

\begin{abstract}
  Since the discovery of the unusual prototype SN~2002cx, the eponymous class of low-velocity, hydrogen-poor supernovae has grown to include at most another two dozen members identified from several heterogeneous surveys, in some cases ambiguously. Here we present the results of a systematic study of $1077$ hydrogen-poor supernovae discovered by the Palomar Transient Factory, leading to nine new members of this peculiar class. Moreover we find there are two distinct subclasses based on their spectroscopic, photometric, and host galaxy properties: The ``SN~2002cx-like'' supernovae tend to be in later-type or more irregular hosts, have more varied and generally dimmer luminosities, have longer rise times, and lack a \ion{Ti}{2} trough when compared to the ``SN~2002es-like'' supernovae. None of our objects show helium, and we counter a previous claim of two such events. We also find that these transients comprise $5.6_{-3.7}^{+17}\%$ ($90\%$ confidence) of all SNe~Ia, lower compared to earlier estimates. Combining our objects with the literature sample, we propose that these subclasses have two distinct physical origins.
\end{abstract}


\section{Introduction}
\label{sec:introduction}

Thermonuclear supernovae (SNe~Ia) have long served as standardizable candles for precision cosmology \citep{Riess1998, Schmidt1998, Perlmutter1999}. With the advent of synoptic imaging, the numbers of rare and peculiar supernovae have grown. In particular, many have been found that do not obey the standardization procedure outlined in \citet{Phillips1993}, which relates peak luminosity to light curve shape. Here we focus on a class of hydrogen-poor supernovae having the defining feature of unusually low ejecta velocities. This class is identified by similarities to the prototypical member, SN~2002cx \citep{Li2003}, and hence we refer to these objects as 02cx-like.

Perhaps the best studied among 02cx-likes is SN~2005hk \citep[e.g.][]{Chornock2006, Phillips2007, Sahu2008}. Roughly two dozen other candidate members have been announced \citep[e.g.][]{Jha2006, Foley2013}, though about half of these have sparse photometric and spectroscopic coverage. With increased numbers trends have begun to emerge, especially a preference for late-type host galaxies and lower peak luminosities relative to typical SNe~Ia. The lowest peak luminosity and lowest velocity belong to the debated member SN~2008ha \citep{Foley2009, Foley2010, Valenti2009}. \Citet{McClelland2010} claim that SN~2008ha is indeed an 02cx-like based on their discovery of SN~2007qd, whose properties are intermediate between SNe~2002cx and 2008ha. Furthermore, the distinct forest of \ion{Co}{2} lines observed in the near infrared by \citet{Stritzinger2014} reinforces the spectroscopic similarity among SN~2008ha, SN~2010ae, and SN~2005hk.

Recently the low-ejecta velocity (slow-speed) and rapidly fading SN~2002es has been reported \citep{Ganeshalingam2012}. This object has ejecta velocities and a peak luminosity similar to SN~2002cx, and it replicates several of SN~2002cx's characteristic spectral features. However, it is notably different from SN~2002cx in its light curve shape. Furthermore, SN~2002es clearly shows a titanium trough in its spectra---a feature that is not found in SN~2002cx or other members of the developing class. As we discuss in this paper, SN~2002es can be taken as the prototype for an alternate channel leading to low-velocity, hydrogen-poor supernovae.

Here we present slow-speed supernovae discovered by the Palomar Transient Factory \citep[PTF;][]{Rau2009, Law2009}. Searching through the database of $1077$ hydrogen-poor supernovae, we find six new 02cx-like and three new 02es-like transients with velocities less than about $7000~\kms$. We first detail how our final sample of 02cx- and 02es-like objects was chosen in \S\ref{sec:selection}. Section~\ref{sec:analysis} contains the analysis of these objects' properties, divided into host properties (\S\ref{sec:analysis:hosts}), photometry (\S\ref{sec:analysis:light_curves}), spectroscopy (\S\ref{sec:analysis:spectra}), the interplay between photometry and spectroscopy (\S\ref{sec:analysis:combining}), and occurrence rates (\S\ref{sec:analysis:rates}). We follow up with a discussion of the implications of our analysis in \S\ref{sec:discussion} and present our conclusions in \S\ref{sec:conclusion}.

\section{Sample Selection}
\label{sec:selection}

\subsection{The PTF Sample}
\label{sec:selection:ptf}

In order to identify slow-speed supernovae of interest, we thoroughly re-analyzed the spectra in the PTF database. We selected our sample of 02cx- and 02es-like spectra using template matching, after which we further examined the results using ejecta velocities.

First we selected a set of template spectra. We obtained publicly available spectra from the Weizmann Interactive Supernova Data Repository \citep[WISeREP, see][]{Yaron2012}: thirteen of SN~2002cx \citep{Li2003, Jha2006}, $37$ of SN~2005hk \citep{Chornock2006, Phillips2007, Blondin2012}, and three of SN~2008ha \citep{Foley2009}. The latter are used as representatives of the extremely low velocity region of the supernova parameter space. Omitting them would risk missing any other such objects, since very low velocities tend to correspond to more lines being resolved, and such spectra may not match even SN~2002cx very well. Furthermore, we included eleven spectra of SN~2002es (M.~Ganeshalingam, priv.\ comm.). The phases for our templates ranged from $-8~\days$ to $+314~\days$.

For the PTF objects we considered all $1904$ hydrogen-poor spectra of $1077$ distinct supernovae observed between March 2009 and May 2012. Using the Superfit package \citep{Howell2005}, the spectra were compared to each of $296$ supernova spectra in its standard library without any assumptions on phase. Superfit was allowed to choose the redshift that gave the best match in each comparison within $\pm0.02$ of the PTF best estimate for the actual redshift, sampling in increments of $0.002$. Allowing the redshift to vary avoids the situation wherein two otherwise similar spectra are deemed a poor match solely because of differing ejecta velocities. Superfit was also allowed to vary the assumed extinction $A(V)$ between $-3~\magn$ and $3~\magn$, with $R(V)$ fixed at $3.1$. The other Superfit settings employed were a $20~\A$ rebinning of PTF spectra and five iterations of $3\sigma$ clipping on a pixel-by-pixel basis to avoid fitting noise. These parameters were found to result in a low false negative rate when tested with known 02cx-likes.

\begin{figure*}
  \centering
  \includegraphics[width=0.8\textwidth]{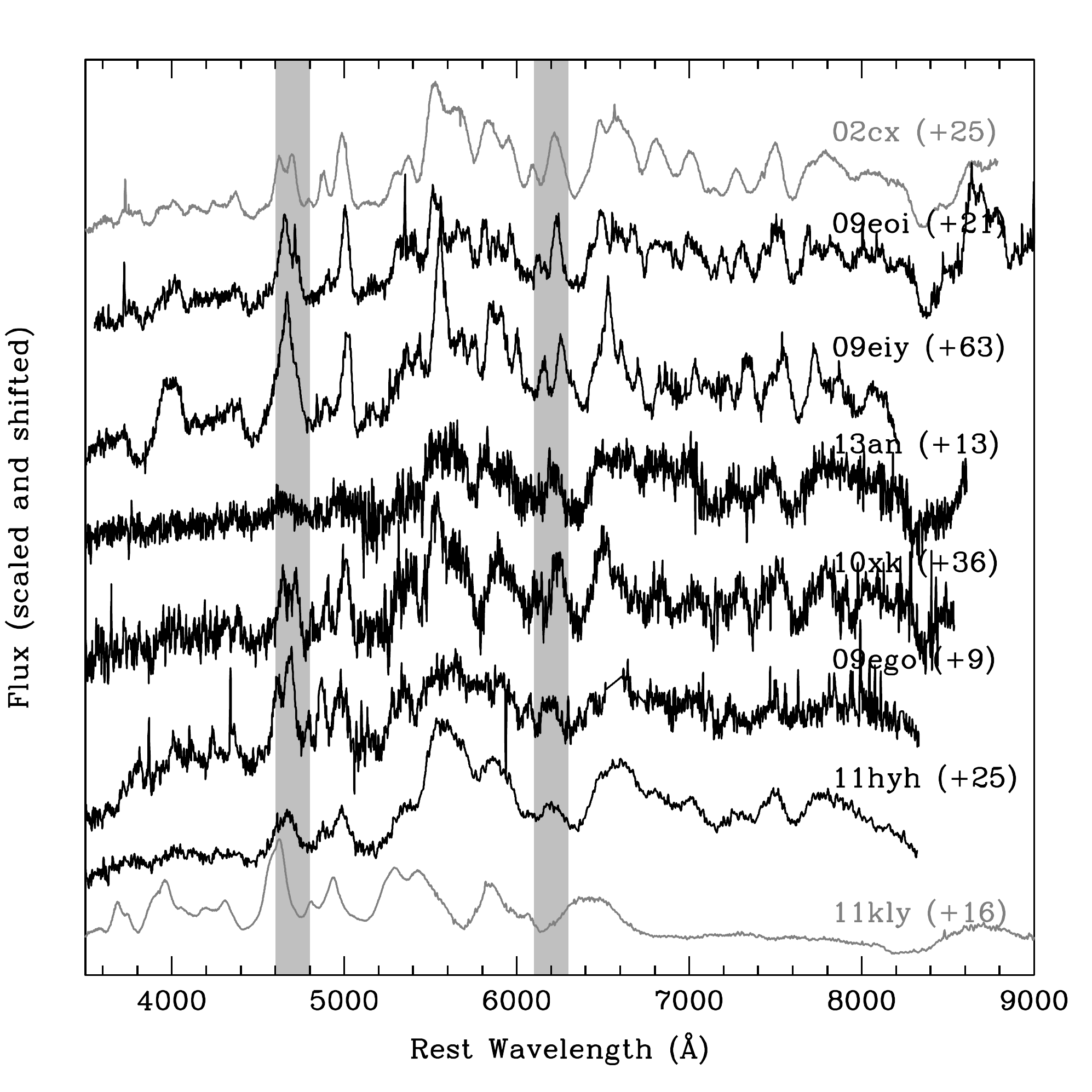}
  \caption{Montage of confirmed 02cx-like spectra. SN~2002cx is shown on top, with a typical SN~Ia shown at the bottom. The gray bands highlight regions of particular interest. Details of observations can be found in Table~\ref{tab:observations_of_spectra}. \label{fig:02cx_spectra}}
\end{figure*}

\begin{figure*}
  \centering
  \includegraphics[width=0.8\textwidth]{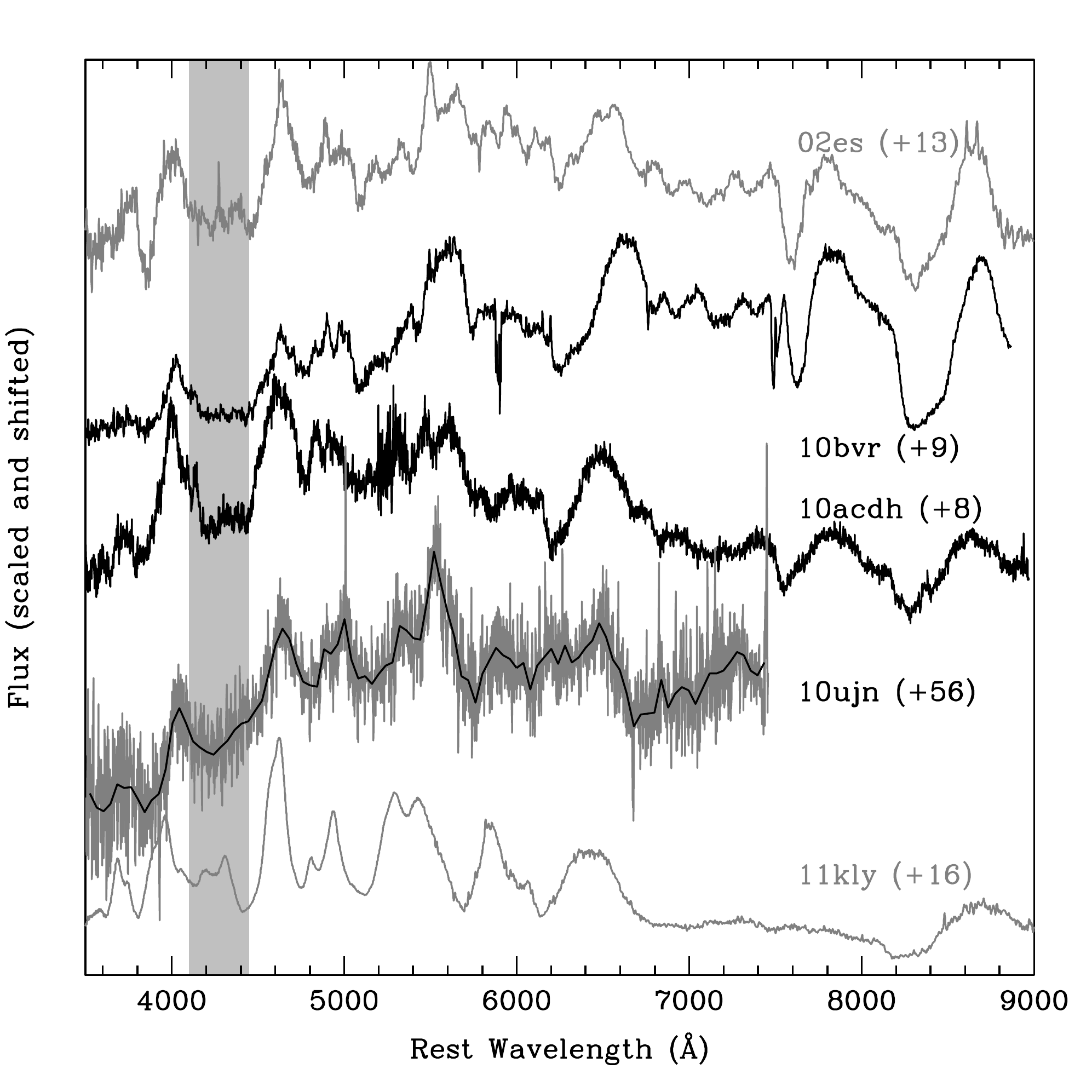}
  \caption{Montage of 02es-like spectra. SN~2002es is shown on top, with a typical SN~Ia shown at the bottom. The gray band highlights the titanium trough. The PTF~10ujn spectrum has been binned for clarity. Details of observations can be found in Table~\ref{tab:observations_of_spectra}. \label{fig:02es_spectra}}
\end{figure*}

For each PTF supernova, Superfit's output is a rank-ordered list of matches to templates, using a figure of merit for goodness of fit. Any PTF supernova with a low-velocity template match among the top fifteen matches was selected for further visual inspection. In performing the visual analysis, four features were considered: the number of peaks between $6000~\A$ and $8000~\A$; a particular peak at a rest wavelength of around $6200~\A$; the resolution of the feature near $4700~\A$ into two peaks; and the presence of a \ion{Ti}{2} trough. See Figures~\ref{fig:02cx_spectra} and \ref{fig:02es_spectra} for montages of spectra illustrating these features. A summary of the appearance of these features in the final sample can be found in Table~\ref{tab:spectrum_properties}.

\placefigure{fig:02cx_spectra}
\placefigure{fig:02es_spectra}
\placetable{tab:spectrum_properties}

\begin{figure*}
  \centering
  \includegraphics[width=0.8\textwidth]{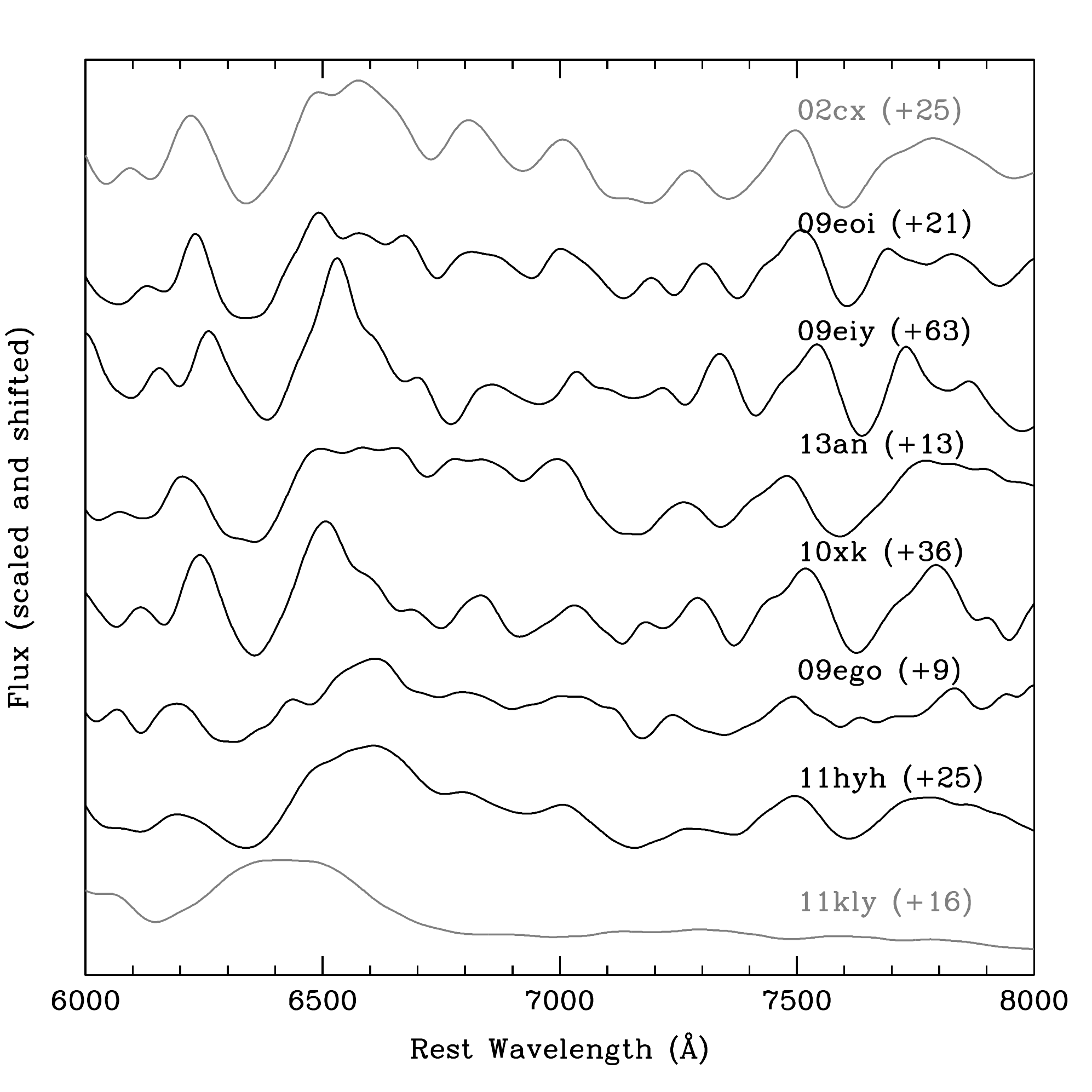}
  \caption{Montage of confirmed 02cx-like spectra, focusing on $6000\text{--}8000~\A$ and smoothed with a Gaussian filter of standard deviation $20~\A$. SN~2002cx is shown on top, with a typical SN~Ia shown at the bottom. Details of observations can be found in Table~\ref{tab:observations_of_spectra}. \label{fig:02cx_zoomed_spectra}}
\end{figure*}

\begin{figure*}
  \centering
  \includegraphics[width=0.8\textwidth]{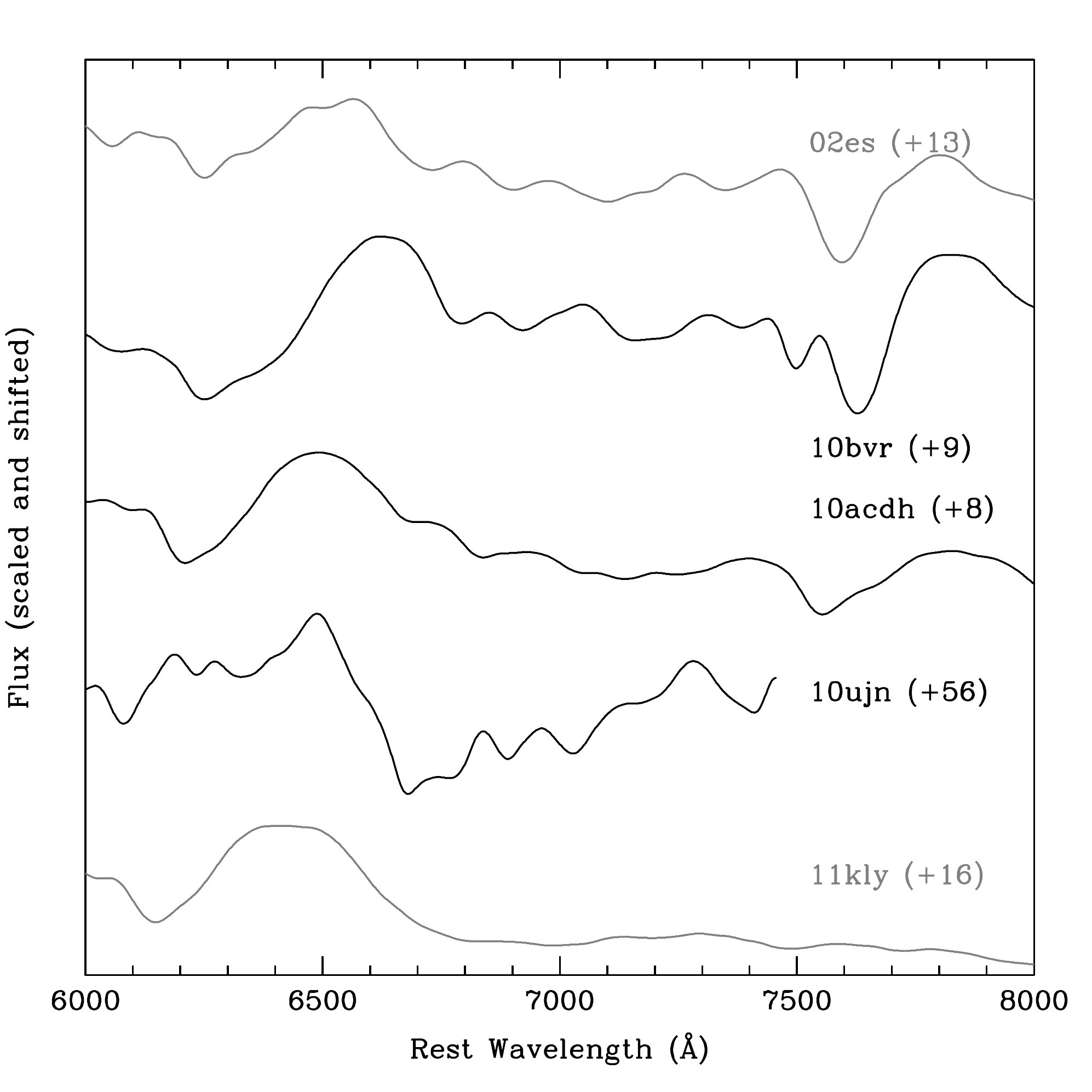}
  \caption{Same as Figure~\ref{fig:02cx_zoomed_spectra}, but showing the matches to SN~2002es. \label{fig:02es_zoomed_spectra}}
\end{figure*}

The reason we employ peak counting is that high velocities will tend to smooth spectra and eliminate features. Thus large numbers of peaks serve as a proxy for identifying low velocities. For this inspection we first pass the spectra through a $20~\A$ Gaussian smoothing filter. This eliminates the high-frequency noise and allows us to unambiguously count features independent of original resolution or signal-to-noise ratio. The smoothed spectra are shown in Figures~\ref{fig:02cx_zoomed_spectra} (02cx-likes) and \ref{fig:02es_zoomed_spectra} (02es-likes).

The nature of the $6200~\A$ feature is not a settled topic in the literature. It may very well be due to adjacent \ion{Fe}{2} absorption, as noted by \citet{Li2003} and \citet{Sahu2008} and modeled by \citet{Branch2004} and \citet{Foley2009}. However, it has also been suggested that [\ion{Co}{3}] \citep{Li2003} or \ion{O}{1} \citep{McClelland2010} could play a role.

\placefigure{fig:02cx_zoomed_spectra}

\placefigure{fig:02es_zoomed_spectra}

As a result of visual inspection, there were several Superfit matches we decided were false positives. The PTF targets 11cfm and 11pzq had spectra initially found to match SN~2002cx by Superfit. However, upon further inspection we concluded their spectra were too noisy to justify any certain placement. Furthermore, PTF~11cfm had an ejecta velocity of around $10{,}000~\kms$. Another preliminary match included PTF~10vzj, but it was rejected upon closer inspection. The last Superfit match that was rejected upon closer inspection is PTF~10xfh (the Type~Ic nature of which is discussed in Cao et al., in prep.).

The names and positions of the final sample are given in Table~\ref{tab:positions}. There are nine matches in all, three of which match SN~2002es while the other six match only SN~2002cx. Overlays of spectra from these objects with matching templates can be found in Appendix~\ref{sec:individual_spectra}. For completeness, each of the nine rejected objects is also given an overlay, in Appendix~\ref{sec:rejected_spectra}, showing the best match we could achieve to a template.

Lest it be thought that the 02cx-likes and 02es-likes are entirely disparate, we note that all three of the SN~2002es matches found were also found to be reasonably good SN~2002cx matches. Indeed, even SN~2002es itself has spectra generally similar to SN~2002cx, resulting in its classification as 02cx-like by \citet{Ganeshalingam2012}. At this stage in our analysis these two classes are distinguished only by the visual presence or absence of the \ion{Ti}{2} trough, with verification of the dichotomy left to the techniques discussed in \S\ref{sec:analysis}.

\placetable{tab:positions}

\subsection{The Slow-Speed Sample in Context}
\label{sec:selection:velocity}

We place our spectroscopically selected sample in context by constructing a histogram of velocities for all PTF Type~Ia supernovae as follows.

First, cuts are made in phase so as to exclude pre-maximum spectra or spectra more than fourteen days post-maximum, as well as any spectra for which there is no reliable phase information. The remaining $837$ spectra are deredshifted using the precise spectroscopic redshift of the host galaxy where available and the approximate supernova redshift otherwise.

Next, each spectrum is plotted and the \ion{Si}{2} $\LL~6348,6373$ absorption feature visually identified. The feature is immediately recognizable in $395$ spectra; the other spectra are not considered further in this analysis. The minimum of the feature is compared to the weighted mean of the two rest wavelengths, $6355~\A$, and the velocity calculated from this shift is taken to be the ejecta velocity.

\begin{figure*}
  \centering
  \includegraphics[width=0.8\textwidth]{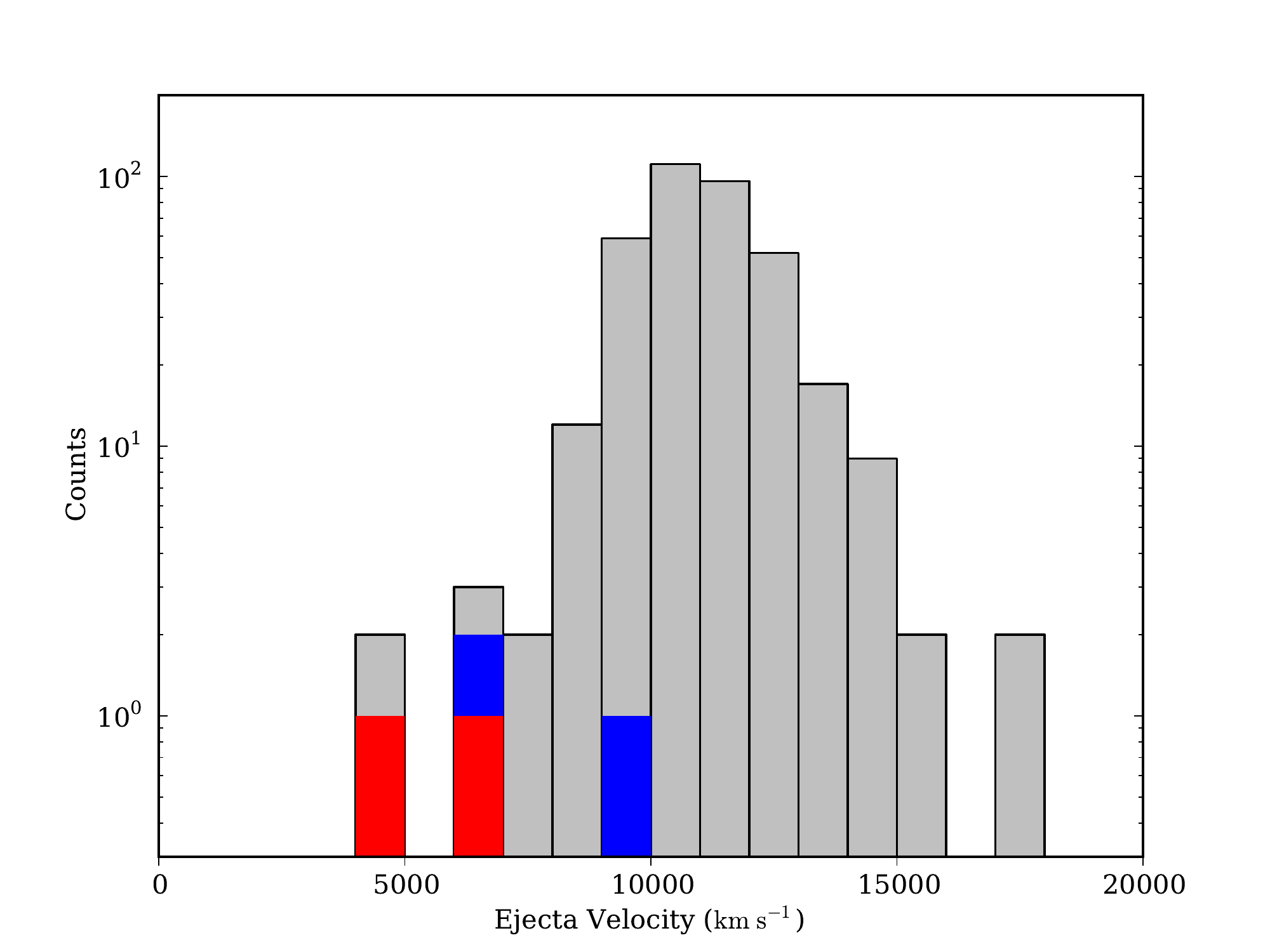}
  \caption{Distribution of ejecta velocities in PTF data. Shown are SNe with phases ranging from maximum brightness to $+14~\days$. The blue region denotes the 02cx-like objects, with the red section counting the 02es-likes. Velocities are averaged in the case of multiple spectra for the same object. Note that many of our 02cx-like and 02es-like spectra are excluded here in adherence to our strict phase cut. Some objects, such as PTF~10pko ($4100~\kms$) and PTF~11sd ($6800~\kms$), have low expansion velocities but are not 02cx-like or 02es-like, as discussed in \S\ref{sec:selection:velocity}. \label{fig:velocity_histogram}}
\end{figure*}

Figure~\ref{fig:velocity_histogram} shows the distribution of velocities obtained this way for the PTF objects. The histogram also shades the sample ultimately selected for being 02es-like (red) or 02cx-like but not 02es-like (blue), following the procedure of \S\ref{sec:selection:ptf}. The velocities for these selected objects are obtained in a more precise, robust way, as described in Appendix~\ref{sec:robust_velocities}, comparing the 02cx-likes to the SN~2002cx $+12$ spectrum, and the 02es-likes to the SN~2002es $+6$ spectrum. As can be seen, our sample does indeed lie in the low-velocity part of the distribution. Thus there is agreement between the spectrum-matching selection described in the previous section and a simple analysis based solely on an inferred velocity from a single feature.

\placefigure{fig:velocity_histogram}

There were five PTF objects whose \ion{Si}{2} velocities were found to be particularly low, but which were nonetheless excluded from our final sample on the basis of being overall poor matches to the templates (overlays shown in Appendix~\ref{sec:rejected_spectra}). The spectrum of PTF~09aly has flat-bottomed profiles, making the velocity obtained too untrustworthy to include in the distribution. Another two objects with low ejecta velocities are PTF~10pko and PTF~10xfv, though both suffer from large uncertainties in host redshift, meaning they could very well be normal SNe~Ia scattered into the low-velocity bins by chance. PTF~10pko, moreover, does not show a very clear phase in the spectrum obtained; Superfit finds reasonable matches to templates spanning two weeks in phase. Both PTF~11sd and 09dav have low \ion{Si}{2} velocities, but neither shows any spectral similarities to SN~2002cx and company other than just low velocity. Separately we note that PTF~09dav is the lowest-velocity member of a rare class of calcium-rich gap transients \citep{Sullivan2011, Kasliwal2012}. We choose not to include them because we have no \emph{a priori} reason to believe \emph{all} low-velocity objects, no matter how dissimilar their spectra, should be inseparably grouped together.

A detailed analysis of the \ion{Si}{2}, \ion{Ca}{2}, and \ion{C}{2} features in the SN Ia sample from PTF is presented in Maguire et al., in prep.

\subsection{The Literature Sample}
\label{sec:selection:literature}

Recently, \citealt{Foley2013} compiled an exhaustive literature sample of 25 supernovae from the past two decades that could potentially
be 02cx-likes.  We find that the two helium-rich supernovae in this sample, SNe~2004cs and 2007J, are likely not 02cx-like but rather core-collapse Type~IIb. As discussed in Appendix~\ref{sec:helium_interlopers}, SN~2004cs shows \Halpha{} absorption and emission broader than galaxy lines. Furthermore, the light curve of SN~2004cs is consistent with Type~IIb \citep{Arcavi2012}. Similarly, SN~2007J shows \Halpha{} in absorption and is a good spectral match to the prototypical SN~IIb 1993J. For more details, we refer the reader to the appendix. Since the two helium-rich objects are fully consistent with hydrogen-rich core-collapse supernovae, we do not consider them to have implications for the physical origin of the 02cx-likes. 

In addition to spectral similarity to SN~2002cx, \citealt{Foley2013} required the sample to be photometrically ``underluminous,'' which may include a selection bias against the more luminous, yet kinematically and spectroscopically similar, members of this class. Hence, we combed through the PTF database with a photometry-agnostic sieve. 

Given the publicly available data, we are only able to consider ten of the remaining 23 supernovae in this literature sample for comparisons in this paper: SNe~1991bj, 2002cx, 2003gq, 2004gw, 2005hk, 2006hn, 2007qd, 2008ge, 2008ha, and 2009ku. Not all objects have well-sampled photometry and spectroscopy.

\section{Analysis}
\label{sec:analysis}

\subsection{Host Galaxies}
\label{sec:analysis:hosts}

\begin{figure*}
  \rotate
  \centering
  \includegraphics[width=\textwidth]{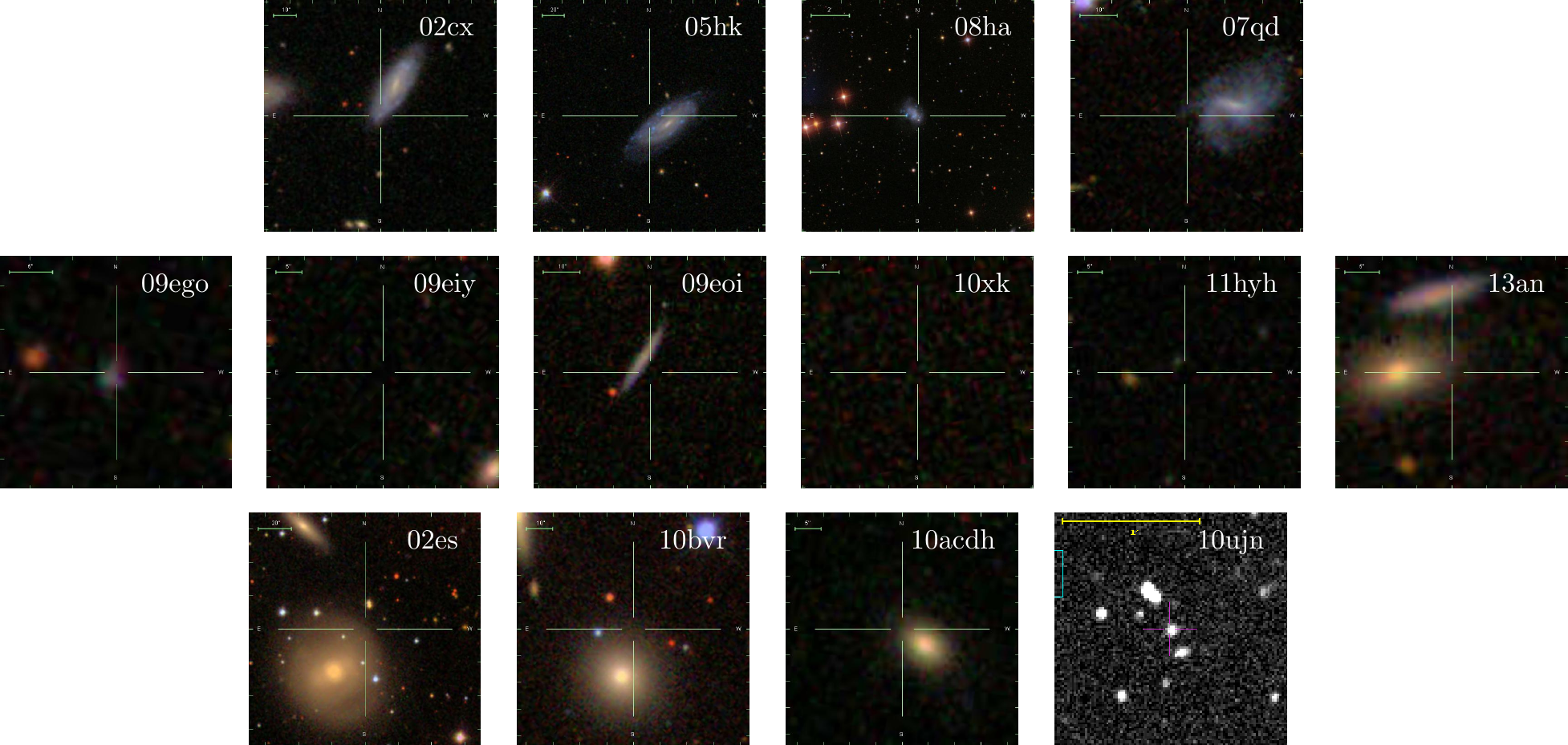}
  \caption{Images of hosts for 02cx-like and 02es-like objects. The top row shows four objects previously published in the literature. The second row shows the six 02cx-like objects in our sample. The bottom row shows the host for 02es itself (leftmost), followed by hosts for our three 02es-like objects. All images are from SDSS except for the host of PTF~10ujn, where the PTF reference image is used. All SDSS images are scaled so as to be $44~\kpc$ on a side at the host's distance, and the last image is $206~\kpc$ on each side. \label{fig:hosts}}
\end{figure*}

The environment in which a class of supernovae is found can place strong constraints on theories about their progenitors. We present a collage of host galaxies in Figure~\ref{fig:hosts} and summarize host galaxy properties in Table~\ref{tab:hosts}.

\placefigure{fig:hosts}

\placetable{tab:hosts}

Comparing the PTF sample of 02cx-likes (row~2, Figure~\ref{fig:hosts}) to the literature sample of 02cx-likes (row~1, Figure~\ref{fig:hosts}), we find that the host galaxies are relatively less luminous and more dwarf-like. At the same time the host galaxies of the 02es-likes (row~3, Figure~\ref{fig:hosts}) are relatively more luminous and redder than those of the 02cx-likes.

Quantitatively, the $R$-band luminosities for 02cx-like hosts in the literature range from $-17~\magn$ to $-21~\magn$, while our sample ranges from being fainter than $-14~\magn$ up to $-21~\magn$. On the other hand, the hosts of 02es-likes span a range of $-19~\magn$ to $-22~\magn$. Moreover, the hosts of 02es-likes are redder than \gi{} of $1.2~\magn$, while the 02cx-likes' hosts are on the blue side of this boundary.

There are possible outliers to these overall trends. iPTF~13an has the reddest host of the 02cx-likes; all others have \gi{} values blueward of $1.0~\magn$. We note, however, that there are at least two other galaxies close by and at a similar redshift, and it is possible that this supernova in fact belongs to a slightly bluer galaxy. However, SN~2008ge was also reported in a galaxy without significant star formation \citep{Foley2013}, so it may be that 02cx-likes are not restricted to occur \emph{only} in star-forming regions. The implications of this are discussed in \S\ref{sec:discussion:two_channels}.

\begin{figure*}
  \centering
  \includegraphics[width=\textwidth]{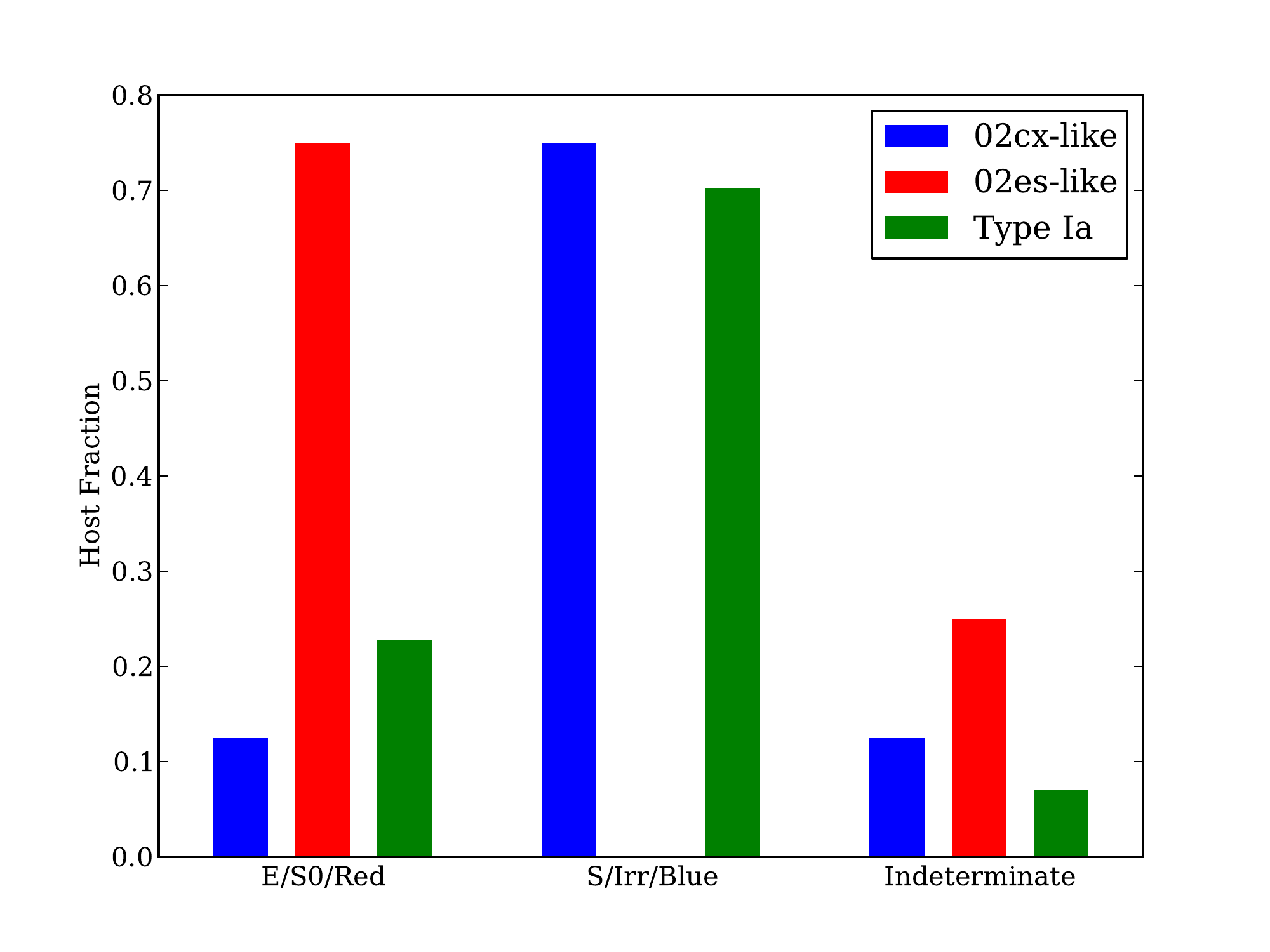}
  \caption{Distribution of host morphological types, with color used as a stand-in where the type is not published. The three distributions are calculated from sixteen (SN~2002cx), four (SN~2002es), and $57$ (Type~Ia) objects, the latter of which are given in \citet{Gallagher2005}. Note that we are limited by small-number statistics with our sample. \label{fig:host_distribution}}
\end{figure*}

Our data indicates a wide range of galaxy properties associated with 02cx-like supernovae, as well as a connection between 02es-like events and luminous galaxies with old stellar populations. To illustrate this graphically, we present a histogram of host types in Figure~\ref{fig:host_distribution}. Because we only have a small sample to work with, we divide host galaxies into just three categories---ellipticals and S0 galaxies, spirals and irregulars, and indeterminate galaxies. Where we cannot find a classification in the literature (i.e.\ where there is no entry in the last column of Table~\ref{tab:hosts}) we fall back to \gi{} colors to sort bluer galaxies into the spiral category and redder galaxies into the elliptical category. In cases where no data is available we place the object in the ``indeterminate'' bin.

\placefigure{fig:host_distribution}

For comparison, we include in the histogram a sample of generic Type~Ia hosts. The data for this population comes from \citet{Gallagher2005}, where a set of $57$ SNe~Ia is studied. The distribution of hosts for 02cx-likes is heavily biased toward star-forming galaxies, whereas 02es-likes tend to be located in early-type hosts. This is reminiscent of the observation that overluminous Type~Ia supernovae tend to be preferentially found in late-type hosts \citep{Hamuy1996} when compared to subluminous SNe~Ia preferentially found in early-type galaxies \citep{Howell2001}.

\subsection{Light Curves}
\label{sec:analysis:light_curves}

\begin{figure*}
  \centering
  \includegraphics[width=0.8\textwidth]{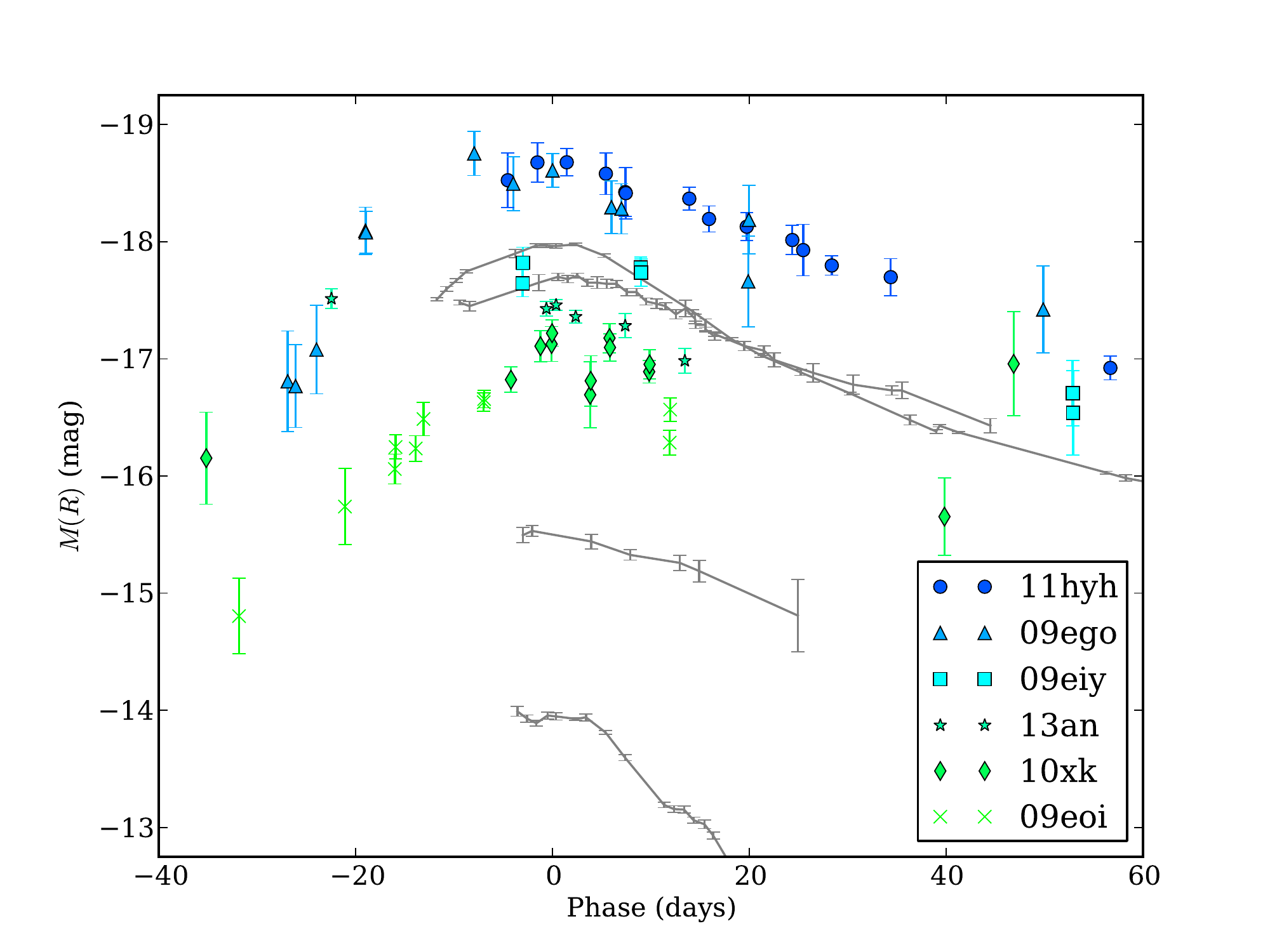}
  \caption{Light curves for the PTF sample. The gray curves show SNe~2005hk, 2002cx, 2007qd, and 2008ha from top to bottom, i.e.\ in order of decreasing peak luminosity. \label{fig:light_curves_confirmed}}
\end{figure*}

\begin{figure*}
  \centering
  \includegraphics[width=0.8\textwidth]{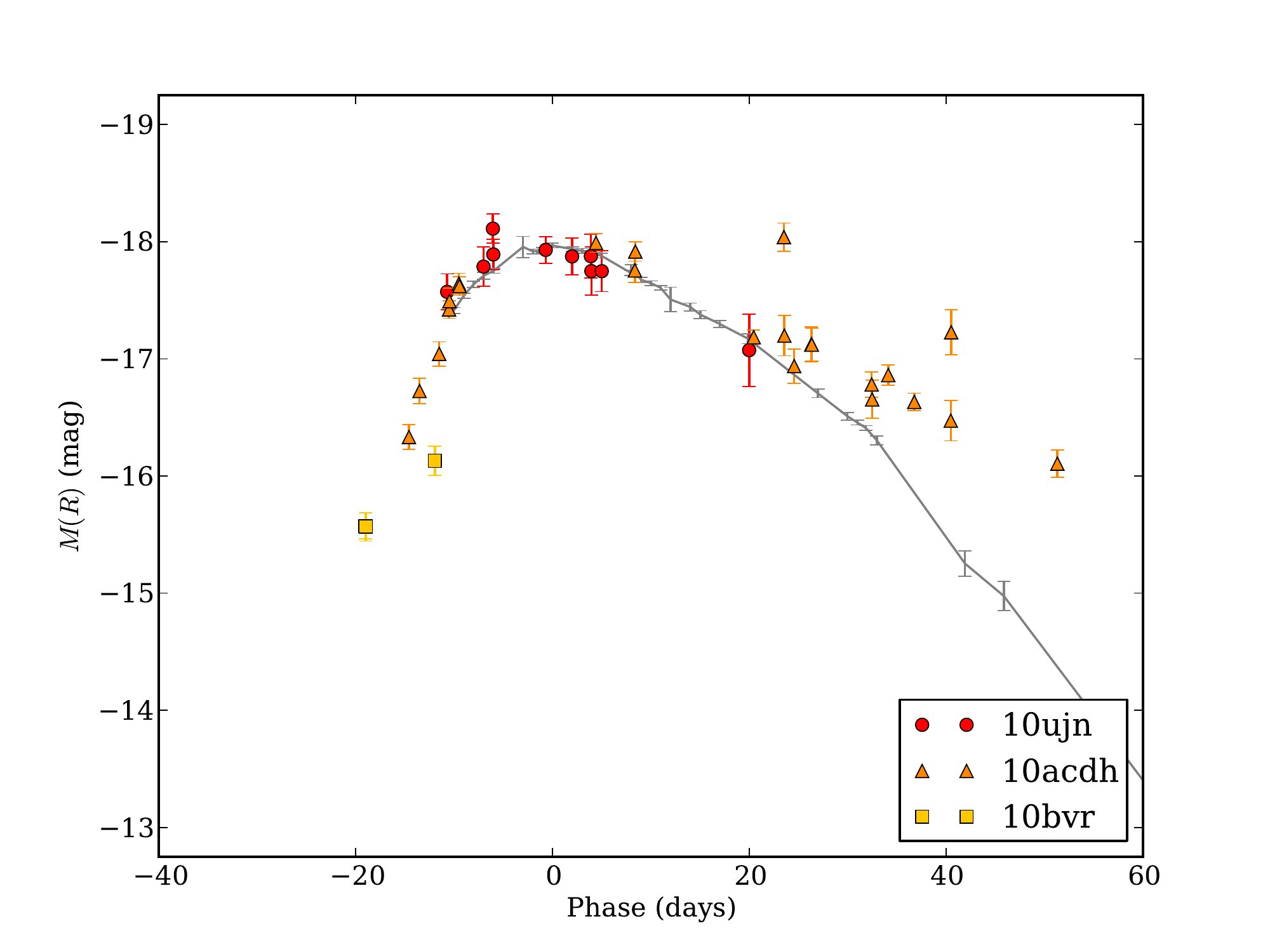}
  \caption{Light curves for the 02es-like objects in the PTF sample. The gray points are for SN~2002es itself. Note how close the peak magnitudes are to one another. In the case of PTF~10bvr, there is insufficient photometry to constrain the peak. \label{fig:light_curves_02es}}
\end{figure*}

Apparent brightnesses for the PTF targets were obtained in $R$-band using the 48-inch Samuel Oschin telescope. Image subtraction using the PTFIDE pipeline \citep{Masci2013} and reference images \citep{Laher2014} was used to eliminate host light. PSF photometry was performed, calibrated relative to SDSS data \citep{Ofek2012}. The light curves are presented in Figures~\ref{fig:light_curves_confirmed} (for the 02cx-likes) and \ref{fig:light_curves_02es} (for the 02es-likes). Photometric properties can be found in Table~\ref{tab:light_curves}.

\placefigure{fig:light_curves_confirmed}

\placefigure{fig:light_curves_02es}

\placetable{tab:light_curves}

For hosts without redshift-independent distances, distance moduli were calculated from host redshifts assuming a flat cosmology with parameters $H_0 = 70.4~\kms\Mpc^{-1}$, $\Omega_\mathrm{m} = 0.273$, and $\Omega_\Lambda = 0.728$, as given in \citet{Jarosik2011}. Applying these to get absolute luminosities, with no corrections made for extinction, one can immediately see that the 02cx-likes' luminosities span a wide range, from $-13~\magn$ to $-19~\magn$. Furthermore, with the inclusion of PTF~09eoi and PTF~10xk, there is no longer such an extreme gap between SN~2008ha and the rest of this class. On the other hand, the well-determined peak luminosities of the 02es-likes are confined to the small range of $-17.9~\magn$ to $-18.1~\magn$, toward the bright end of the 02cx-like range.

Some members of the sample were seen early enough to record the rise in the light curve. For simplicity, we fit the pre-maximum datapoints to a parabola. A visual inspection is used to determine which datapoints to use in the fit, but beyond this there is no prior placed on the location of the peak. Here we define the rise time as the time between maximum brightness and when the fit is at $1.5~\magn$ below maximum, and these are the values given in the fourth column of Table~\ref{tab:light_curves}.\footnotemark{} We observe a trend of 02cx-like rise times being longer than those of the 02es-likes.

\footnotetext{Note that the value of $16~\days$ obtained for 02es agrees with the estimate of $16\pm3~\days$ obtained in \citet{Ganeshalingam2012} using other means.}

For most of our sample we are able to measure decline rates. This is simply taken to be the slope of a linear fit to the post-maximum data, where we exclude observations from very late times where the slope of the curve is likely to have changed. The 02cx-likes are seen to have a spread in decline rates, from $0.2~\magn$ to $1.0~\magn$ over the first fifteen days after maximum. Again the spread in properties for the 02es-likes is smaller, lying between $0.57~\magn$ and $0.70~\magn$ over fifteen days.

\begin{figure*}
  \centering
  \includegraphics[width=0.8\textwidth]{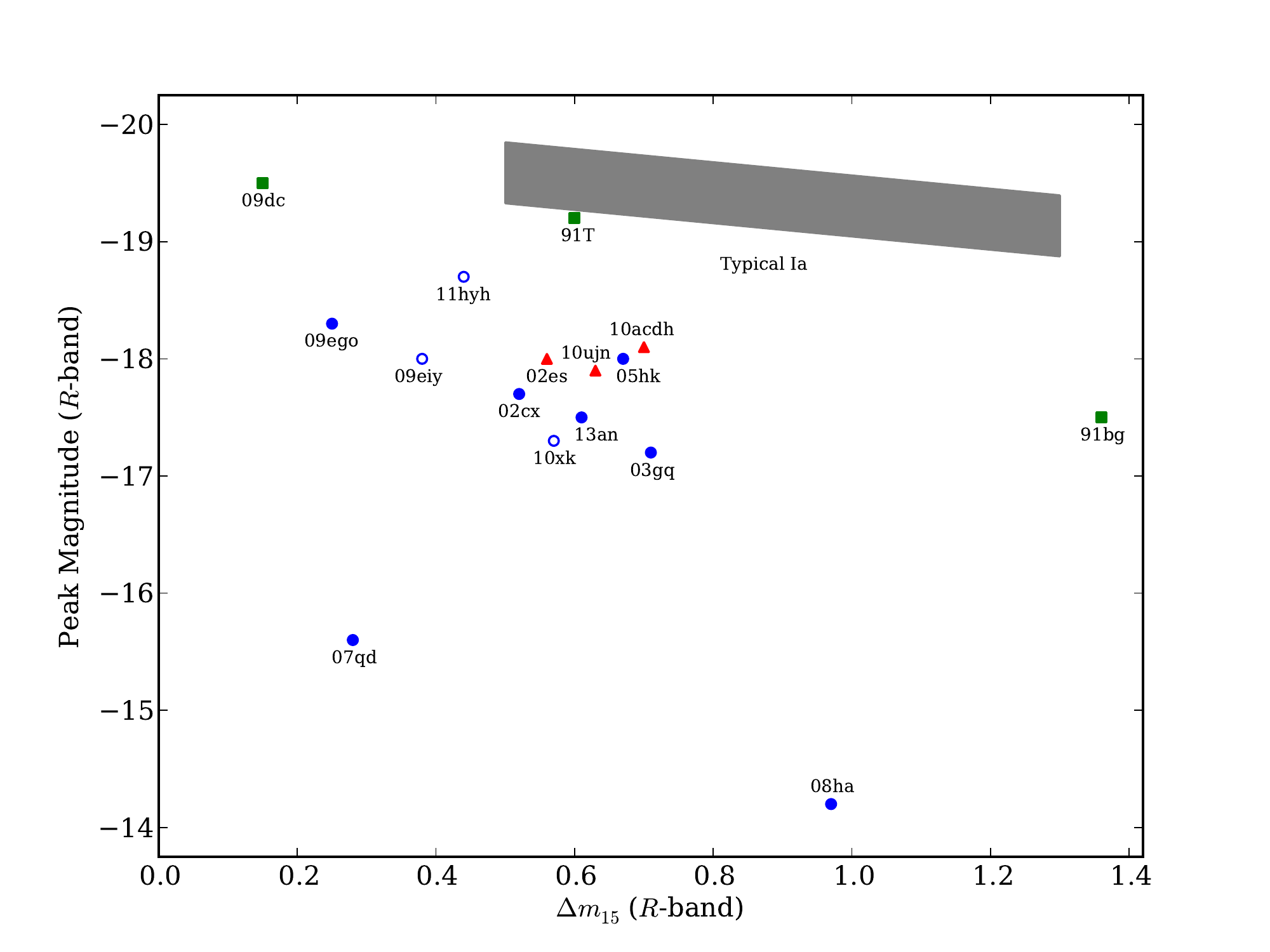}
  \caption{Relation between peak luminosity and decline rate in $R$-band (see Table~\ref{tab:light_curves}). Blue circles indicate 02cx-like data, red triangles are for 02es-like objects, and green squares denote other supernovae discussed in the text. Open symbols denote some doubt regarding the host redshift. The gray band shows the Phillips relation in $R$-band with parameters obtained from Table~3 of \citet{Prieto2006}, extrapolated to $\Dm = 0.5$ and truncated at $\Dm = 1.3$. In particular, this region shades a $2\sigma$ scatter about the best-fit line. \label{fig:decline_luminosity}}
\end{figure*}

With multiple properties measured, we can begin to plot the locations of our sample in various slices of parameter space. One set of parameters of particular interest is that of decline rate $\Delta m_{15}$ and absolute luminosity $M$. These are the very parameters employed in the Phillips relation \citep{Phillips1993} that has proven key to cosmology. This plot is shown in Figure~\ref{fig:decline_luminosity}. For the 02cx-likes, one can see both the offset from typical SNe~Ia and the overall diversity of the class reflected in this diagram. The 02es-like objects are also offset in the same way but do not display as much diversity. There is no simple correlation between these two properties in $R$-band \citep[cf.][]{McClelland2010}.

\placefigure{fig:decline_luminosity}

For aiding comparison, key members of several other proposed sub-classes of supernovae are also plotted. SNe~1991T and 1991bg are prototypical overluminous and underluminous examples, respectively. The former has approximately the same decline rate as our sample, the latter has a similar luminosity, but neither is a particularly good match in this parameter space. We also include a point for SN~2009dc, a so-called super-Chandrasekar Type~Ia whose spectrum at one month past maximum is similar to that of SN~2002cx \citep{Silverman2011}. Its location in Figure~\ref{fig:decline_luminosity} suggests that it \emph{could} be at the more luminous, slower-declining end of a loose distribution.

We pause to note that most of PTF's photometry is done in $R$-band, while decline rates and peak luminosities for normal SNe~Ia are often reported in $V$- and $B$-bands. We do not have reliable bolometric corrections for the full sample to work in other bands. We caution that working in $R$-band may cause well known bright events such as SN~2009dc to appear somewhat less luminous. Given that SN~2005hk, for example, was redder post-$B$-maximum than representative normal and overluminous SNe~Ia \citep[cf.\ Figure~5 in][]{Sahu2008}, it may be that our sample would appear even less luminous next to other Type~Ia events if bolometric luminosities or luminosities in bluer bands were used.

\subsection{Spectra}
\label{sec:analysis:spectra}

We now consider the spectra for these objects. The log of all PTF spectral observations is given in Table~\ref{tab:observations_of_spectra}.

\placetable{tab:observations_of_spectra}

One spectrum for each object is shown in either Figure~\ref{fig:02cx_spectra} (02cx-likes) or Figure~\ref{fig:02es_spectra} (02es-likes). These are arranged in order of increasing absolute magnitude from top to bottom, and the features of interest discussed in \S\ref{sec:selection} are highlighted.

In addition to their lack of hydrogen, none of our sample show any clear signs of helium. This is in agreement with the spectral modeling done in \citet{Branch2004}, where no helium was needed to match the important features. It is also in accord with the lack of helium in the entire Type~Iax sample in \citet{Foley2013} save for SNe~2004cs and 2007J, which we believe do not belong in the same class (see Appendix~\ref{sec:helium_interlopers}).

The most notable difference between the 02cx-likes and 02es-likes is the appearance of the \ion{Ti}{2} trough between about $4100$ and $4400~\A$ only in the latter. This feature is simply not present in any of the 02cx-likes, indicating a difference between the two classes in temperature or possibly composition despite ejecta velocity similarities.

For each PTF spectrum in our sample (as well as some other spectra from the literature), we measure the velocities relative to each of six templates (three SN~2002cx spectra, a SN~2005hk spectrum, a SN~2008ha spectrum, and a SN~2002es spectrum) using the method described in Appendix~\ref{sec:robust_velocities}. The resulting velocity offsets are compiled in Table~\ref{tab:velocities}.

\placetable{tab:velocities}

Four of the templates have absolute velocities reported from spectral modeling \citep{Branch2004, Ganeshalingam2012}, enabling us to convert offsets to absolute velocities in these cases. The absolute velocity so obtained will vary depending on the template used. This reflects the fact that the cross-correlation method employed takes into account the entire spectrum, including more lines than just that of \ion{Si}{2}. These variations are self-consistent in the sense that, e.g., the change from the first column of offsets to the second is always roughly $+1000~\kms$.

From Table~\ref{tab:velocities} one can see that 02es-likes have a range of ejecta velocities (post-maximum) spanning $4000\text{--}7000~\kms$. As has been the case with other properties, the 02cx-likes have a broader range, running from $3000~\kms$ to just under $10{,}000~\kms$.

\begin{figure*}
  \centering
  \includegraphics[width=0.8\textwidth]{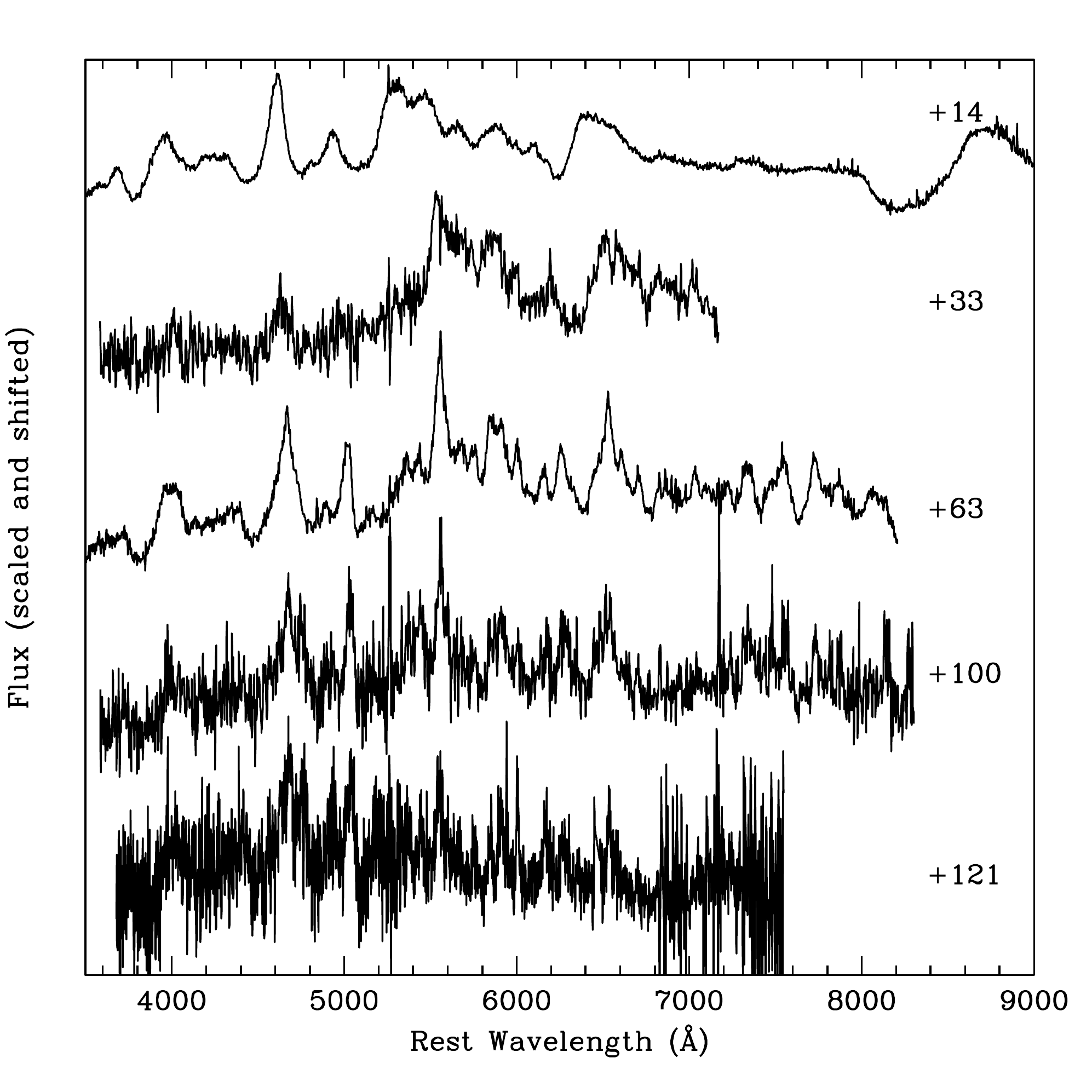}
  \caption{Spectral evolution of PTF~09eiy. Phases labeled are days after peak $R$-band brightness, as defined in Table~\ref{tab:light_curves}. Details of observations can be found in Table~\ref{tab:observations_of_spectra}. \label{fig:09eiy_evolution}}
\end{figure*}

\begin{figure*}
  \centering
  \includegraphics[width=0.8\textwidth]{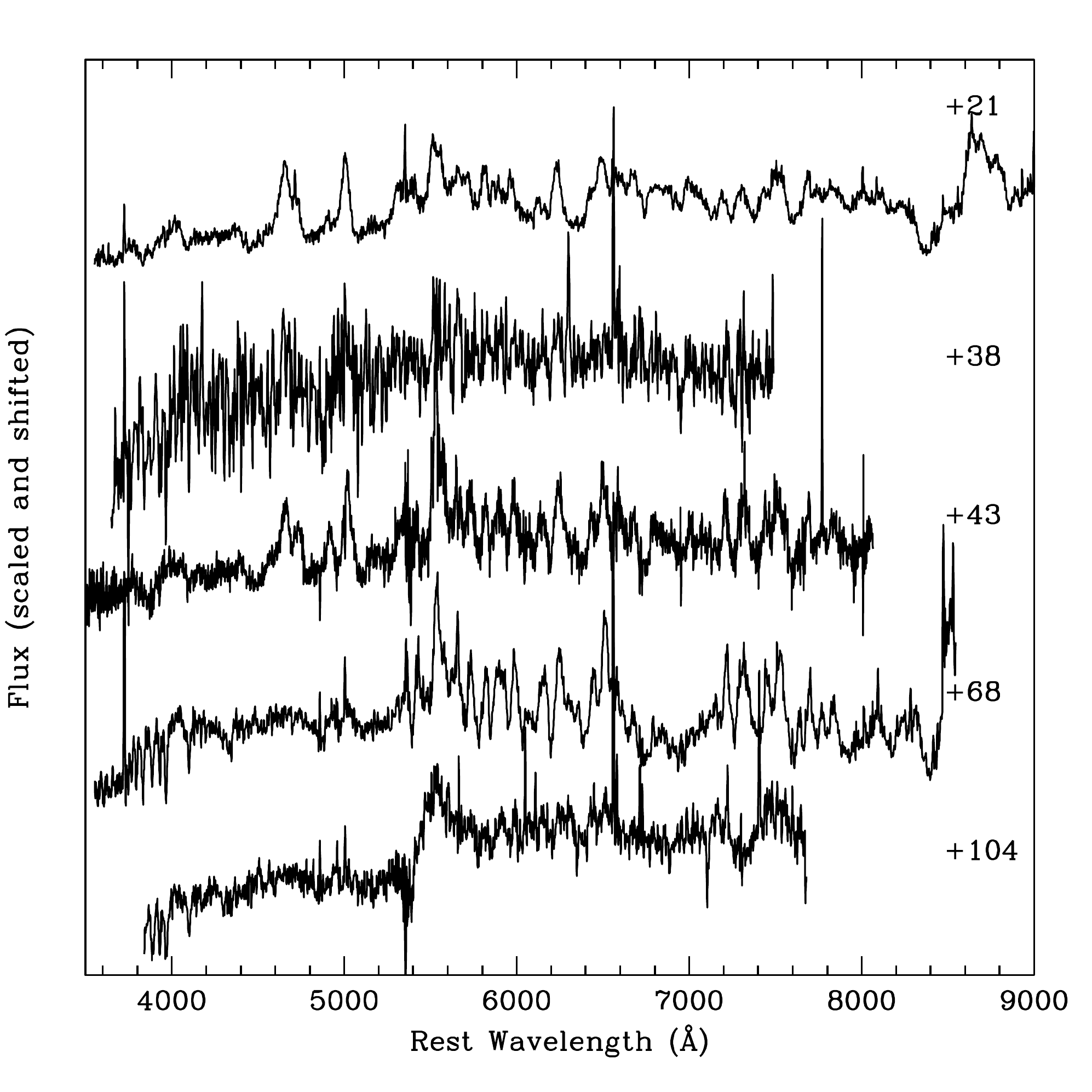}
  \caption{Spectral evolution of PTF~09eoi. Phases labeled are days after peak $R$-band brightness, as defined in Table~\ref{tab:light_curves}. Details of observations can be found in Table~\ref{tab:observations_of_spectra}. \label{fig:09eoi_evolution}}
\end{figure*}

The velocities slow down over the several weeks following maximum light as expected, but the amount of deceleration varies considerably amongst the 02cx-likes. Two of our 02cx-like supernovae, PTF~09eiy and PTF~09eoi, had spectra taken at many different phases. Figures~\ref{fig:09eiy_evolution} and \ref{fig:09eoi_evolution} show their spectral evolution. The former shows rather clear signs of decreasing velocities, as might be expected from the drop in velocity from roughly $9600~\kms$ at $+14~\days$ to $3900~\kms$ at $+121~\days$. The features become narrower and more of them are resolved as time passes. In particular, the characteristic $6200~\A$ feature grows more distinctive with time. PTF~09eoi, on the other hand, starts with a low velocity, even by SN~2002cx standards, and shows relatively little velocity evolution over a period of approximately $80~\days$, going from $5400~\kms$ at $+21~\days$ to $4700~\kms$ at $+104~\days$.

\placefigure{fig:09eiy_evolution}

\placefigure{fig:09eoi_evolution}

\begin{figure*}
  \centering
  \includegraphics[width=0.8\textwidth]{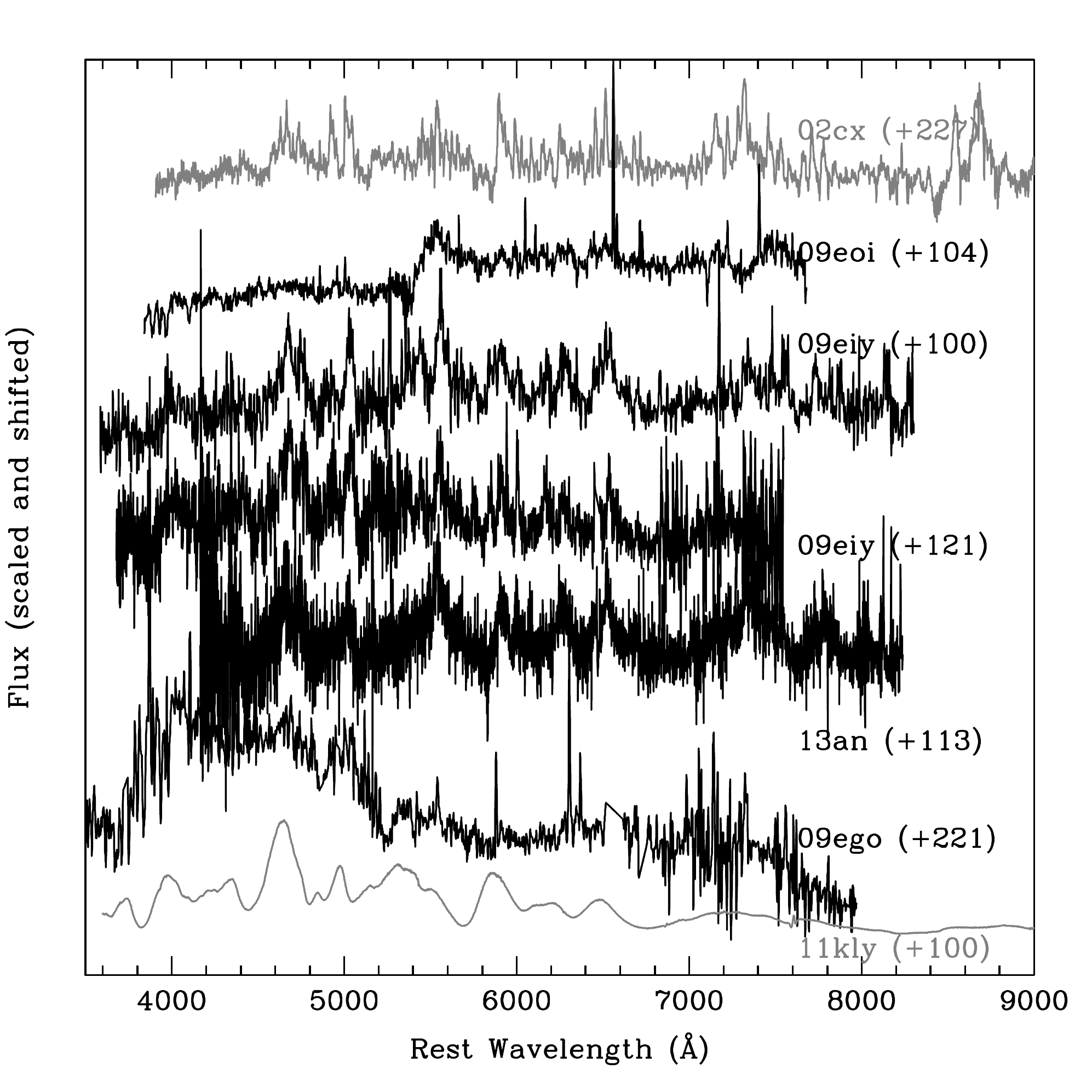}
  \caption{Montage of late-time spectra. SN~2002cx \citep{Jha2006} is shown on top, with a typical SN~Ia shown at the bottom. Details of observations can be found in Table~\ref{tab:observations_of_spectra}. \label{fig:nebular_spectra}}
\end{figure*}

We can also consider the late-phase spectra in their own right. Figure~\ref{fig:nebular_spectra} shows five spectra for phases later than $+100~\days$ obtained for four of the 02cx-like objects in our sample. As was observed by \citet{Jha2006} for the case of SN~2002cx, these objects have many narrow lines indicative of low expansion velocities.

\placefigure{fig:nebular_spectra}

\subsection{Combining Photometry and Spectroscopy}
\label{sec:analysis:combining}

When photometric and spectroscopic data is taken together, Arnett's Law \citep{Arnett1982} can be used to estimate the mass of the ejecta. In particular, the analysis given in \citet{Pinto2000} shows (via their equations 4, 17, and 43) that a plot of magnitude as a function of time is expected to be a parabola near the peak, with
\begin{equation*}
  m - m_\mathrm{peak} \propto -\frac{v_\mathrm{ej}}{M_\mathrm{ej}\kappa} t^2 \text{,}
\end{equation*}
where time is measured from the peak and $\kappa$ is an effective opacity.

Taking normal SN~Ia values to be $M_\mathrm{ej} = 1.4~M_\sun$, $v_\mathrm{ej} = 10^4~\kms$, $\kappa = 0.13~\cm^2~\gram^{-1}$ \citep[as in][]{Pinto2000}, and $t_\mathrm{rise} = 18~\days$ \citep[as in][whose alternate definition of rise time agrees with ours in the case of SN~2002cx]{Ganeshalingam2012}, we can construct curves of constant mass in the $v_\mathrm{ej}{-}t_\mathrm{rise}$ plane for any assumed opacity, as is done in Figure~\ref{fig:velocity_rise}. Also plotted are points for members of our sample for which rise times could be computed. These rise times are defined in \S\ref{sec:analysis:light_curves} and reported in Table~\ref{tab:light_curves}. For the ejecta velocities, we use the earliest post-maximum velocity for the event given in Table~\ref{tab:velocities}, where for these purposes we use the comparisons to the SN~2002cx $+12$ spectrum for the 02cx-likes, while we use the comparisons to the SN~2002es $+6$ spectrum for the 02es-likes.

\placefigure{fig:velocity_rise}

From the figure one can immediately infer that the 02cx-likes and 02es-likes must have different ejecta masses and/or opacities. Moreover, even without calculating absolute masses one can see that the 02cx-likes have a broader range of ejecta masses compared to the 02es-likes. In fact, the three 02es-likes fall extremely close to the same contour of constant ejecta mass for any fixed opacity.

Figure~\ref{fig:velocity_luminosity} shows the relation between \emph{peak} luminosity\footnotemark{} and the time evolution of ejecta velocity. One can see that our sample is slow speed in nature even at early phases. Some of the 02cx-likes, such as SN~2003gq \citep{Foley2013} and PTF~09eiy, show a dramatic change in velocity over the course of several months. Others such as PTF~09ego and PTF~09eoi slow down little over this timeframe. It should be noted here that in some cases but not all, e.g.\ 09eoi but not 09eiy, the velocity relative to a fixed template increases slightly in the latest phases (see Table~\ref{tab:velocities}). As a result, velocities that are only measured infrequently can mask significant evolution.

\footnotetext{Few supernovae have enough spectra to construct a well-sampled, time-dependent curve in velocity-magnitude space.}

\placefigure{fig:velocity_luminosity}

The peak luminosities are seen to be as diverse as the ejecta velocities; most but not all of the sample luminosities lie below those of typical SNe~Ia. Both the 02cx-likes and 02es-likes follow a loose trend of more luminous events having generally faster ejecta.

\subsection{Rates}
\label{sec:analysis:rates}

Our data also lends itself to calculating relative rates of slow-speed supernovae in the Type~Ia population. Excluding iPTF~13an, which was detected in a different data-taking period than the rest of the sample, we have eight objects at redshifts $z < 0.12$ (Table~\ref{tab:hosts}). The PTF database contains $683$ SNe~Ia within that redshift limit. As the somewhat underluminous nature of these classes means we may miss a greater fraction of them than we do for typical Type~Ia supernovae in a volume-limited survey, the rate should be greater than about $8/683 \approx 1\%$. On the other hand, PTF can reliably detect transients at $20.5~\magn$ relative magnitude, and so we should be complete down to an absolute luminosity of $-17~\magn$ ($-14~\magn$) by placing an upper limit to the distance modulus of $\mu = 37.5~\magn$ ($\mu = 34.5~\magn$), corresponding to a redshift of $z = 0.070$ ($z = 0.018$). (Here we are assuming negligible extinction.) We detect six (one) slow-speed supernovae out of $304$ (eighteen) Type~Ia supernovae at $z < 0.070$ ($z < 0.018$). Thus, we can place an upper limit on the rate for supernovae brighter than $-17~\magn$ ($-14~\magn$) of $2\%$ ($6\%$). We therefore estimate the rates of slow-speed events in the Type~Ia population to be $2.0\%$ if their luminosities are brighter than $-17~\magn$ and $5.6\%$ if their luminosities are brighter than $-14~\magn$.

We can take this analysis a step further and give confidence intervals for the rate. Given a rate $r$, the probability of finding $k$ objects of interest in a sample of $N$ is given by the binomial distribution
\begin{equation}
  P_r(k; N) = \binom{N}{k} r^k (1-r)^{N-k} \text{.}
\end{equation}
Assuming a uniform prior on $[0, 1]$, the probability of the rate being less than some value $r_0$ given an observed fraction $k/N$ is
\begin{equation}
  P_{k;N}(r < r_0) = \frac{\int_0^{r_0} P_r(k; N) \, \mathrm{d}r}{\int_0^1 P_r(k; N) \, \mathrm{d}r} \text{.}
\end{equation}
Consider the dimmer cutoff, in which we have $N = 18$, $k = 1$. As $P_{1;18}(r < 31\%) = 99\%$, our rates are significantly lower than the central value of $31_{-13}^{+17}\%$ reported in \citet{Foley2013}. Even using the lower end of their interval, $31\% - 13\% = 18\%$, our data predicts the actual rate to be lower with $P_{1;18}(r < 18\%) = 88\%$ confidence. Note that \citeauthor{Foley2013}\ obtain their rates using correction factors as large as $2$, working under the assumption that 02cx-likes are missed more frequently than normal Type~Ia events even after limiting the survey volume. Given the homogeneous nature of our survey and data analysis, we do not need to employ such corrections. As $P_{1;18}(r < 23\%) = 95\%$ and $P_{1;18}(r < 1.9\%) = 5\%$, we report a $90\%$ confidence interval for the rate of $5.6_{-3.7}^{+17}\%$. This is consistent with the volume-limited value of $5\%$ quoted in \citet{Li2011}.

\section{Discussion}
\label{sec:discussion}

\subsection{A Summary of the Slow-Speed Sample}
\label{sec:discussion:summary}

Combining the PTF slow-speed sample with the published literature sample, there is an overall dichotomy of properties, with two distinct channels environmentally, photometrically, and spectroscopically (see Table~\ref{tab:comparison}). Both channels are low-ejecta velocity explosions of the Type~Ia variety (as determined from the features in their hydrogen-poor spectra) with certain additional spectral similarities not found in all SNe~Ia. However, there are remarkable differences. SN~2002cx and its closest relatives span a wider range of peak luminosities; have a broader range of ejecta masses and/or opacities; and often appear in late-type hosts, including dwarfs. On the other hand, SN~2002es and related transients display a narrower range of higher peak luminosities; have very similar ejecta masses and opacities; and exclusively reside in luminous, early-type hosts with minimal star formation. Furthermore, despite the otherwise similar spectra, the presence or absence of the \ion{Ti}{2} trough is a clear and simple way to discriminate between these two cases.

\placetable{tab:comparison}

The \ion{Ti}{2} feature is the same one distinguishing faint SN~1991bg-like from brighter SN~1991T-like Type~Ia supernovae. It is also remarkable that both 91bg-likes and 02es-likes have older hosts and faster photometric evolution than their counterparts that do not show the \ion{Ti}{2} trough. Furthermore the range in peak luminosities is small for both 91bg-likes and 02es-likes.

Reviewing the evidence, both classes share similarities with canonical Type~Ia explosions. Contrary to \citet{Foley2013} we believe SNe~2004cs and 2007J are not related to either class. The distinguishing feature of both our classes is low ejecta velocity, i.e.\ low kinetic energy per unit ejecta mass. Arnett's Law, if physically applicable and barring an extreme difference in effective opacity, indicates that 02es-likes have lower masses (see Figure~\ref{fig:velocity_rise}), meaning these objects overall have weaker, less energetic explosions. Both classes have somewhat lower peak luminosities relative to typical SNe~Ia, and hence have less radioactive material. They both have a rate that is an order of magnitude lower than Type~Ia supernovae. Finally, the older environments of the 02es-likes suggest a longer delay time compared to the 02cx-likes.

\subsection{Review of Models for SN~2002cx}
\label{sec:discussion:overview}

With these properties in mind, we proceed to discuss the relevant models that have been proposed. When \citet{Li2003} announced the discovery of SN~2002cx, three models were initially put forth as candidates for explaining this event.

The first idea was a variation on the traditional picture of Type~Ia supernovae: A single white dwarf explodes as a deflagration and does not immediately transition to a detonation. These models could explain the underluminous, slow-speed nature of SN~2002cx, but they were not fully accepted due to the chemical compositions they predicted. More recent work by \citet{Phillips2007} and \citet{Jordan2012}, however, has shown that \emph{pure} deflagrations are better candidates for explaining SN~2002cx, with off-center deflagrations leading to appropriate compositions and kinematics. On the other hand, these models cannot explain large ejecta masses, as they leave behind a bound remnant.

The second idea considered by \citeauthor{Li2003}\ is that of two white dwarfs merging---the double-degenerate scenario. Work done by \citet{Nomoto1995} had indicated that the disruption and subsequent accretion of a carbon-oxygen white dwarf onto an oxygen-neon-magnesium white dwarf could result in an underluminous explosion. Attempts were made to explain the low-luminosity supernova SN~1991bg with these models, but they were at the time dismissed in the context of SN~2002cx due to the high expansion velocities they predicted.

The third type of model discussed in \citeauthor{Li2003}\ is that of a low-mass white dwarf being detonated by the ignition of an outer shell of helium. Helium shell detonation has more recently been considered in connection with so-called .Ia models, as discussed in \citet{Bildsten2007}. This explanation has the benefit of producing underluminous, hydrogen-poor explosions. However, .Ia explosions are predicted to have ejecta velocities upward of $10{,}000~\kms$ \citep{Fink2007, Shen2010}. Moreover, they evolve more rapidly than our observed supernovae, and in fact their rise times are too fast even for SN~2002es \citep{Ganeshalingam2012}. A variation in which accretion from a nondegenerate helium star triggers a deflagration is advocated by \citet{Foley2013} for explaining 02cx-likes, including two supernovae we do not consider to be a part of this class (see \S\ref{sec:selection:literature} and Appendix~\ref{sec:helium_interlopers}). We note again that with these objects (which also showed signs of hydrogen) excluded, there is no helium signature in our sample.

We also mention two models not discussed in the original SN~2002cx paper. One recently proposed idea for SN~2002cx events is that of a white dwarf merging with a neutron star or black hole \citep{Fernandez2012}. Here the white dwarf is spread into an accretion disc vulnerable to detonation waves propagating through it. Of particular interest is the low expansion velocity predicted by such a model, due to the gravity of the compact accreting object. Furthermore, the amount of \isotope{Ni}{56} produced is predicted to be less than for a typical Type~Ia, leading to low luminosities. In this preliminary stage, however, the model does not yet make specific spectral predictions, and so we cannot yet determine if the spectral features found in 02cx- and 02es-like objects can be obtained from these.

Finally, the possibility has been raised that SN~2008ha is in fact a core-collapse supernova in disguise \citep{Valenti2009}. They note that the core collapse of a hydrogen-deficient star could reproduce the low-luminosity, low-velocity nature of SN~2008ha. \Citet{Fryer2009, Moriya2010} agree, claiming to have a model wherein a large amount of material falls back onto the remnant. One may be tempted to declare SN~2008ha, whose Type~Ia status has the most doubt, to be entirely unrelated to SN~2002cx and the other members of the family. However, as Figures~\ref{fig:light_curves_confirmed} and \ref{fig:velocity_luminosity} show, the gap between SNe~2002cx and 2008ha is partially bridged by three supernovae: PTF~10xk, PTF~09eoi, and SN~2007qd \citep{McClelland2010}.

\subsection{Two Channels for Two Subclasses}
\label{sec:discussion:two_channels}

Despite the kinematic and certain spectroscopic similarities between 02cx-likes and 02es-likes, there are enough distinctions to make finding a single, satisfactory model difficult. Considering the observed dichotomy between the two subclasses, we suggest that one should look for two distinct theoretical explanations in order to best fit the two sets of observations.

White dwarf deflagration models are currently favored in the literature for explaining 02cx-likes. We find that some deflagration models may explain the relatively weaker (subluminous and low-energy) 02es-like explosions better than they explain the 02cx-likes.

Recently, \citet{Kromer2013} undertook 3D deflagration simulations of white dwarfs in an attempt to explain 02cx-likes. They found low-velocity and low-luminosity explosions with a small amount of \isotope{Ni}{56} (some of which falls back onto a bound remnant). They predict spectra with weak \ion{Si}{2} absorption and prominent iron-group elements at all epochs. All four properties are consistent with both 02cx-likes and 02es-likes. However, they note that their model has difficulty being extended far enough to encompass the extremely low \isotope{Ni}{56} levels in SN~2008ha. Their rise times are also much shorter than those observed in 02cx-likes.

\begin{figure*}
  \centering
  \includegraphics[width=0.8\textwidth]{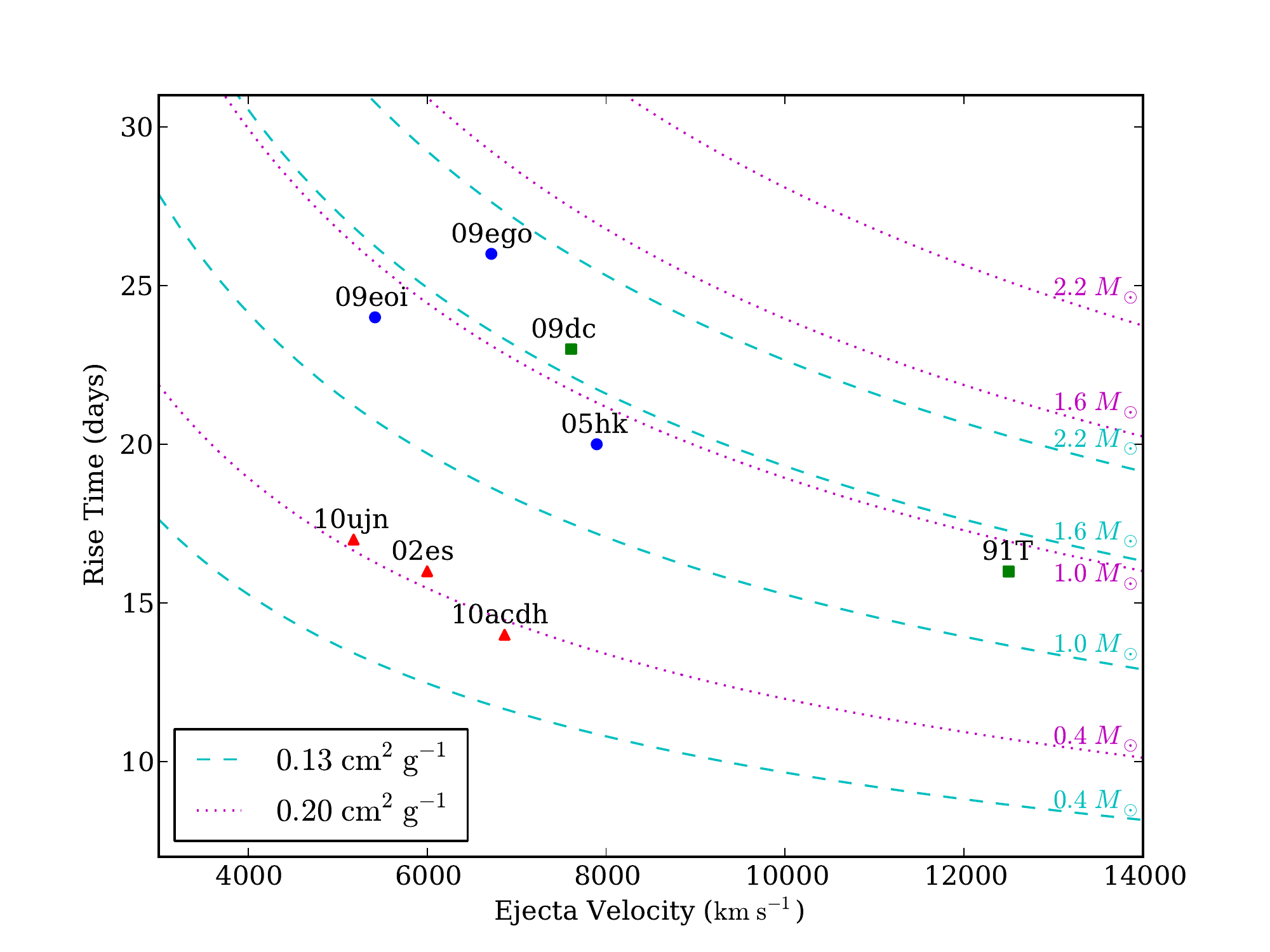}
  \caption{Visualization of Arnett's Law. The symbol scheme is the same as in Figure~\ref{fig:decline_luminosity}. Lines of constant ejecta mass are calculated for two different values of effective opacity. The rise time for SN~2009dc is taken from \citet{Silverman2011}; all others are calculated as described in the text. \label{fig:velocity_rise}}
\end{figure*}

\begin{figure*}
  \centering
  \includegraphics[width=0.8\textwidth]{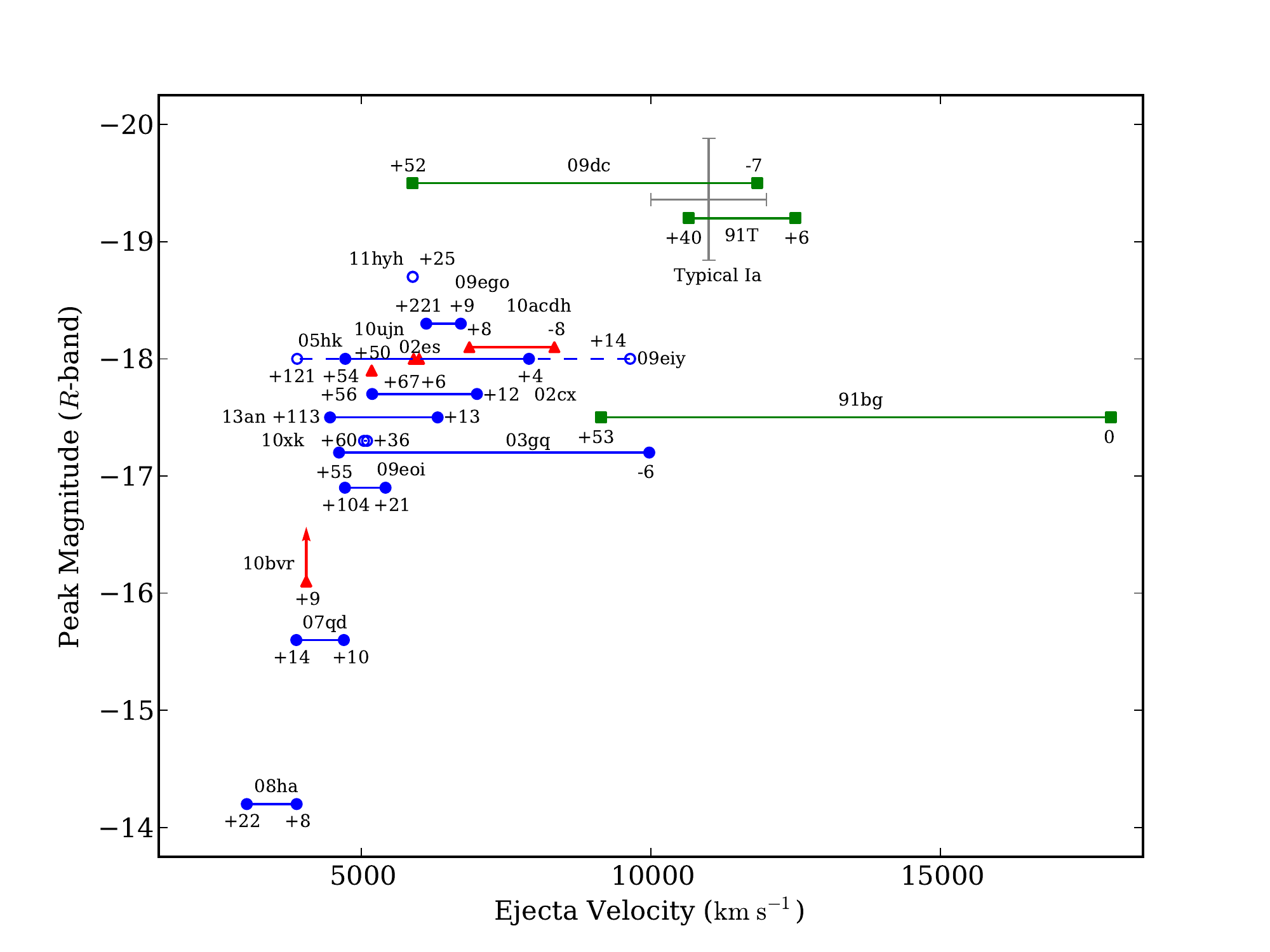}
  \caption{Relation between peak luminosity in $R$ band (see Table~\ref{tab:light_curves}) and ejecta velocity (see Table~\ref{tab:velocities}). The symbol scheme is the same as in Figure~\ref{fig:decline_luminosity}. Typical SNe~Ia have velocities of $10,\!000\text{--}12,\!000~\kms$. The typical peak magnitudes are obtained by projecting the relation in Figure~\ref{fig:decline_luminosity}. \label{fig:velocity_luminosity}}
\end{figure*}

These latter two shortcomings for explaining 02cx-likes are in fact advantages when it comes to explaining 02es-likes, which have a narrower range of luminosities and shorter rise times. The archetypal model used in that study, N5def, results in $0.372~M_\sun$ of ejecta (cf.\ Figure~\ref{fig:velocity_rise}), $0.158~M_\sun$ of which is \isotope{Ni}{56}, and a rise time of $11.2~\days$. Furthermore, while \citet{Kromer2013} do not explicitly report on titanium yields or ejecta temperatures, the synthetic spectrum they calculate for two weeks post-maximum (see their Figure~5) shows signs of the same trough we identify as \ion{Ti}{2} in 02es-like objects.

Further modeling will be needed to secure what is now just speculation based on observations of 02es-likes. One key feature any successful model will have to explain is the low scatter in the photometric properties and the preference for older host galaxies seen in this class.

If both 02es-likes and 02cx-likes are deflagrations, then some parameter, e.g.\ temperature or host environment, must result in distinguishable explosions. Alternatively, 02cx-likes may come from a different channel.

Here we reconsider the double degenerate models for the 02cx-likes. \Citet{Pakmor2010} simulate the merger of equal-mass carbon-oxygen white dwarfs and apply this to SN~1991bg. They find low velocities, large amounts of partially burnt material (up to $1.8~M_\sun$), and luminosities between $-17~\magn$ and $-18~\magn$ in $R$-band. We note that all three features are consistent with both SN~1991bg and the 02cx-likes.

However, the low decline rates found by \citeauthor{Pakmor2010}\ are more consistent with 02cx-likes than SN~1991bg. It is also easier to explain the observed photometric diversity of 02cx-likes in this model---neither white dwarf need be up against the Chandrasekhar limit at the time of merger, so one would expect to see explosions of a range of white dwarf masses. Note that, subject to opacity considerations, our inferred mass for PTF~09ego (see Figure~\ref{fig:velocity_rise}) may require the merger of two relatively massive white dwarfs in this scenario.

As \citeauthor{Pakmor2010}\ point out, unless the initial separation is large and/or there is only one common envelope phase, the merger may happen in a relatively short time, while there is still a high rate of star formation in the galaxy. At the same time, some fraction of the appropriate progenitor binaries may have large enough initial separations that by the time the merger does happen, star formation has turned off in the host.

If however the apparently old environment of SN~2008ge is an exception, then the evidence points to young environments rather than a range. This, combined with the fact that there are known core-collapse events with extremely low ejecta velocities (Appendix~\ref{sec:helium_interlopers}), suggests we cannot rule out core-collapse scenarios for 02cx-likes \citep[e.g.\ fallback;][]{Moriya2010}. Note that PTF~09ego has a large ejecta mass---about $2~M_\sun$ with certain opacity assumptions (see Figure~\ref{fig:velocity_rise})---suggesting the involvement of either two massive white dwarfs or the core collapse of a massive star.

More modeling with 02cx-likes in mind must be done to vet these possibilities. In particular, any successful model must satisfactorily explain the remarkable diversity of this class's properties, as well as the observed host galaxy distribution.

\section{Conclusion}
\label{sec:conclusion}

Using the wealth of data obtained over several years of PTF surveys, we were able to identify nine new low-velocity, hydrogen-poor supernovae. Six of these are spectroscopic matches to the much-discussed SN~2002cx, while the other three bear striking similarity to SN~2002es. Combining our sample with those already reported in the literature (which we number at ten), we see a clear dichotomy of properties emerge: The 02cx-likes come from more varied hosts (that are often star-forming), can be quite underluminous, and have hotter ejecta, while the 02es-likes are found in elliptical hosts, are only slightly underluminous, have cooler ejecta, and also have remarkably fast rise times. It is important to note that our selection and classification is based on generic spectral matching; the persistence of the dichotomy through the photometric and host analyses lends further support to the idea that there are indeed two distinct subclasses. Though constituting only $5.6_{-3.7}^{+17}\%$ of Type~I supernovae (given the luminosity cut described in \S\ref{sec:analysis:rates}), we now have enough examples of these slow-speed transients to classify and understand them.

Prior attempts at explaining these low-velocity phenomena have fallen notably short of capturing all members. We posit that modeling can be done in a consistent way, but not if one is trying to describe physically distinct scenarios with the same models. The 02cx-likes and the 02es-likes are dissimilar enough to warrant two different progenitor scenarios, and neither class should be confused with hydrogen- or helium-rich transients. Finding a model for 02cx-likes remains difficult, but we speculate that the merger of two carbon-oxygen white dwarfs should at least be considered further now that we have a more coherent picture of this class. More convincingly, there are white dwarf deflagration models that fit observations of 02es-likes. These two channels lead to explosions that are similar to but distinguishable from typical Type~Ia supernovae.


We thank M.~H.\ van~Kerkwijk for valuable discussions that improved this manuscript. We also thank I.~Arcavi, C.~Badenes, V.~Bhalerao, K.~Clubb, A.~Cucchiara, A.~De~Cia, O.~Fox, A.~Horesh, D.~Levitan, K.~Mooley, D.~Perley, S.~Tang, S.~Tendulkar, P.~Vreeswijk, and D.~Xu for their contributions to PTF, especially with observations and data reduction.

C.~J.~W.\ began this research as part of the summer student exchange program between Princeton and Carnegie. M.~M.~K.\ acknowledges generous support from the Hubble Fellowship and Carnegie-Princeton Fellowship. A.~L.~P.\ is supported through NSF grants AST-1205732, PHY-1068881, and PHY-1151197, as well as the Sherman Fairchild Foundation. E.~O.~O.\ is incumbent on the Arye Dissentshik career development chair and is grateful for support by grants from the Willner Family Leadership Institute, Ilan Gluzman (Secaucus, NJ), the Israeli Ministry of Science, the Israel Science Foundation, Minerva, Weizmann-UK, and the I-CORE Program of the Planning and Budgeting Committee and the Israel Science Foundation.

This research made use of the Sloan Digital Sky Survey (SDSS), the NASA/IPAC Extragalactic Database (NED), and the SIMBAD database. The PTF spectra for the slow-speed sample will be made available on WISeREP.

\emph{Facilities:} \facility{PO:1.2m}, \facility{PO:1.5m}, \facility{Keck:I (LRIS)}, \facility{Hale (DBSP)}, \facility{ING:Herschel (ISIS)}, \facility{Keck:II (DEIMOS)}, \facility{Magellan:Baade (IMACS)}

\bibliographystyle{apj}
\bibliography{references}

\appendix

\section{Obtaining Robust Velocities via Cross-Correlation}
\label{sec:robust_velocities}

Here we describe our method for measuring precise relative velocities between an input spectrum and a template. This method is robust against weak \ion{Si}{2} lines being difficult to discern amongst many low-velocity features. Each input is compared to the appropriate template using the IRAF routine \verb|fxcor|. The theory behind the routine is given in \citet{Tonry1979}, and we summarize the idea of the procedure and our usage of it here.

First, all spectra are de-redshifted according to their hosts' cataloged redshifts (as listed below in Table~\ref{tab:hosts}). They are then rebinned to a rest frame resolution of $1~\A$, which is in all cases finer than the raw data. Spectra are continuum subtracted with a cubic spline averaged over five points, rejecting data points at the $2\sigma$ level. Five iterations are performed with a one-pixel grow radius.

Next, the spectra are limited to the wavelength range of $4000\text{--}7000~\A$. This was chosen as the maximal range covered by all object and template data. For both the object and the template, this windowed spectrum is Fourier transformed with periodic boundary conditions, apodizing $20\%$ of the window with a cosine bell on either end to prevent discontinuity artifacts stemming from the enforced periodicity. The spectra are then filtered by multiplying these transforms by a ramp function. The ramp rises from zero to unity between wavenumbers $10$ and $15$, filtering out very low frequency variations. It descends back to zero over the wavenumber range $1067\text{--}1600$, excluding any signal with a frequency more than about half the highest achievable spatial frequency given the resolution.

As described in \citet{Tonry1979}, one takes the product of the Fourier transforms and fits a symmetric function---in our case a Gaussian---to the largest peak in this correlation function; the center corresponds to the relative velocity. In practice we identify the peak of interest as the largest one with a reasonable shift---i.e., the ejecta velocity must be positive and is almost certainly less than $20{,}000~\kms$---and fit this peak to a Gaussian as follows. The peak is defined as the contiguous set of points with values at least half that of the largest value in the peak. Each datapoint, corresponding to a single wavenumber, is weighted by a symmetric triangle function that vanishes at the outermost points and reaches its maximum of unity at the center. A least-squares fit is then performed, with the model being a Gaussian with no offset. The shifts of the peaks directly translate into relative ejecta velocities, and these are tabulated in Table~\ref{tab:velocities}. More information could in theory be ascertained from the shape of the peak, but for our purposes here we only use the relative velocity thus obtained.

This method has the advantage of taking all lines into account. Furthermore, the weighting of the lines in contributing to the final velocity scales quadratically in the amplitude of the line. As discussed in \citet{Tonry1979}, this is a desired feature.

\section{The Type~IIb Nature of SNe~2004cs and 2007J}
\label{sec:helium_interlopers}

SN~2004cs was discovered by LOSS \citep{Li2004} and classified as a Type~IIb supernova by \citet{Rajala2005} based on a Keck/LRIS spectrum now available on WISeREP. This spectrum showed strong \ion{He}{1} features, as well as the \Halpha{} line. Recently, \citet{Foley2013} reanalyzed the spectrum, claiming the \Halpha{} feature consists only of host galaxy emission that is furthermore blueshifted compared to nearby lines. They associate this object with the class of 02cx-likes via comparison with SN~2007J. This association is then used to bolster the claim, originally based only on SN~2007J, that some members of this class show helium signatures and that therefore the class as a whole must come from progenitor systems that contain an abundant source of helium (e.g.\ accreting binary systems with helium star donors).

\begin{figure*}
  \centering
  \includegraphics[width=0.8\textwidth]{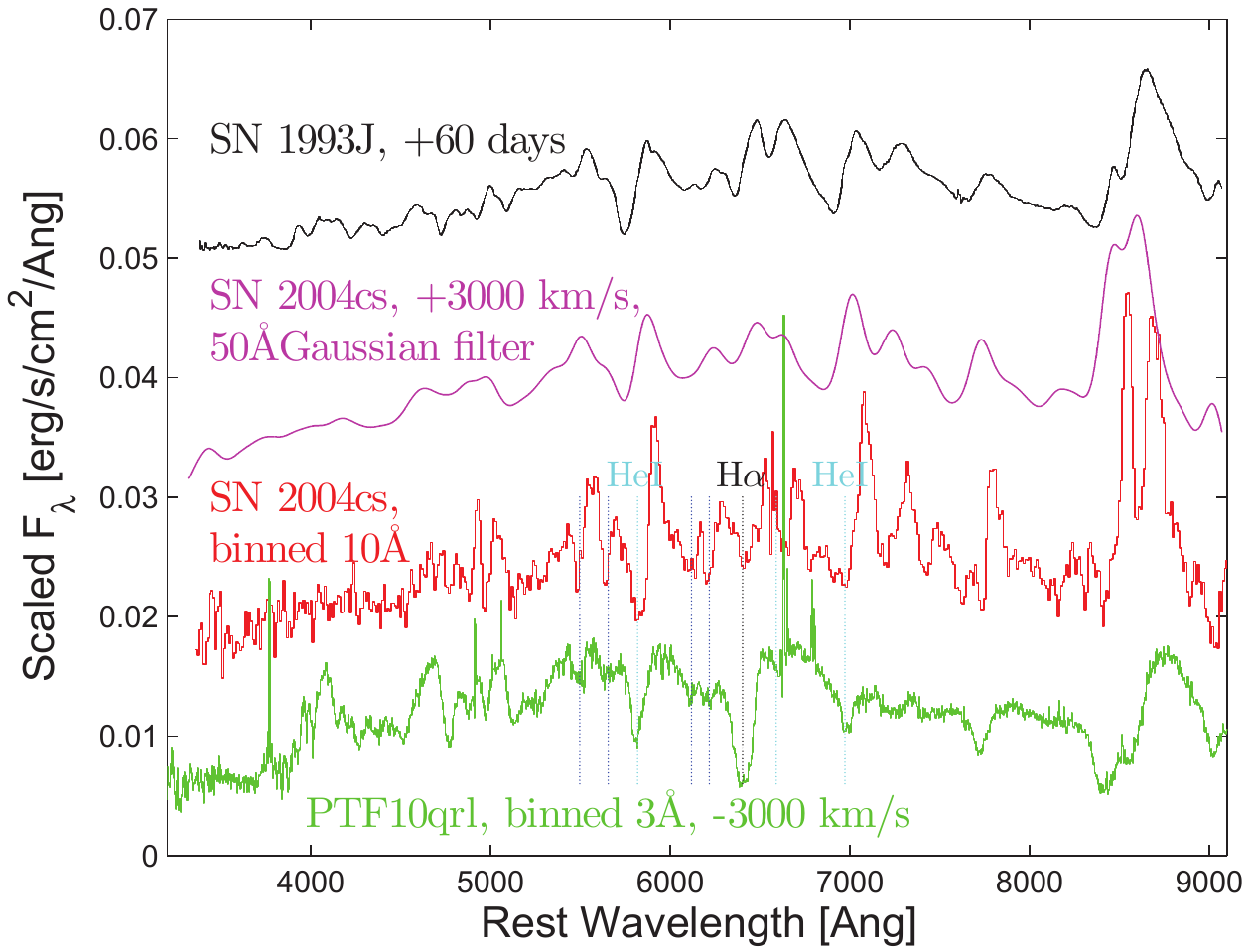}
  \caption{Comparison of spectra of SN~2004cs (red) and PTF~10qrl (green, a low-velocity SN~IIb) showing excellent alignment of features. Absorption lines of hydrogen (\Halpha, black) and \ion{He}{1} ($5876~\A$ and $7065~\A$, cyan) are clearly identified, and additional nearby features (blue) also perfectly align. A broadened and blueshifted version of the SN~2004cs spectrum (magenta) also displays a strong similarity to a spectrum of the prototypical Type~IIb SN~1993J (black). \label{fig:2004cs_spec}}
\end{figure*}

Due to the importance of this issue, we reexamine these two objects here. The spectrum of SN~2004cs is shown in Figure~\ref{fig:2004cs_spec}. In addition to prominent \ion{He}{1} absorption lines, the object displays the \Halpha{} line, both in emission (broader than galaxy lines) and absorption. This line is blueshifted with respect to the host lines and is thus unaffected by possible host galaxy contamination.

\placefigure{fig:2004cs_spec}

In order to clarify line identification, we went through the PTF database and located the two Type~IIb supernovae with the lowest expansion velocities. One of those, PTF~10qrl, has a spectrum similar to that of SN~2004cs, though its expansion velocities are not as low. PTF~10qrl has a very strong and obvious \Halpha{} line, as well as higher order Balmer lines (\Hbeta{} and \Hgamma{}) and \ion{He}{1} lines, so its Type~IIb classification is not in question. This spectrum is also plotted in Figure~\ref{fig:2004cs_spec}, after being redshifted by $3000~\kms$. As can be seen, the spectra are quite similar, with most features aligning very well. The expansion velocity is indeed somewhat higher for hydrogen than for \ion{He}{1}, but this is a known feature of SNe~IIb \citep[e.g.][]{Arcavi2011} reflecting the fact that the thin hydrogen layer lies above the deeper helium-rich layers, and the velocity difference seen in PTF~10qrl is the same as that seen in SN~2004cs.

To further illuminate this issue, we recall the process through which the class of 02cx-likes was associated with SNe~Ia: spectral comparison of artificially broadened and blueshifted low-velocity spectra with those of normal events \citep[e.g.\ in Figure~1 of][]{Foley2013}. We repeat this process for SN~2004cs, convolving the spectrum with a $50~\A$ Gaussian filter and blueshifting the result by $3000~\kms$. Figure~\ref{fig:2004cs_spec} compares the result with a spectrum of the prototypical Type~IIb SN~1993J \citep[][available on WISeREP]{Barbon1995}, and the similarity is striking throughout the entire visible wavelength range, including the clear alignment of the \Halpha{} absorption features.

\begin{figure*}
  \centering
  \includegraphics[width=0.8\textwidth]{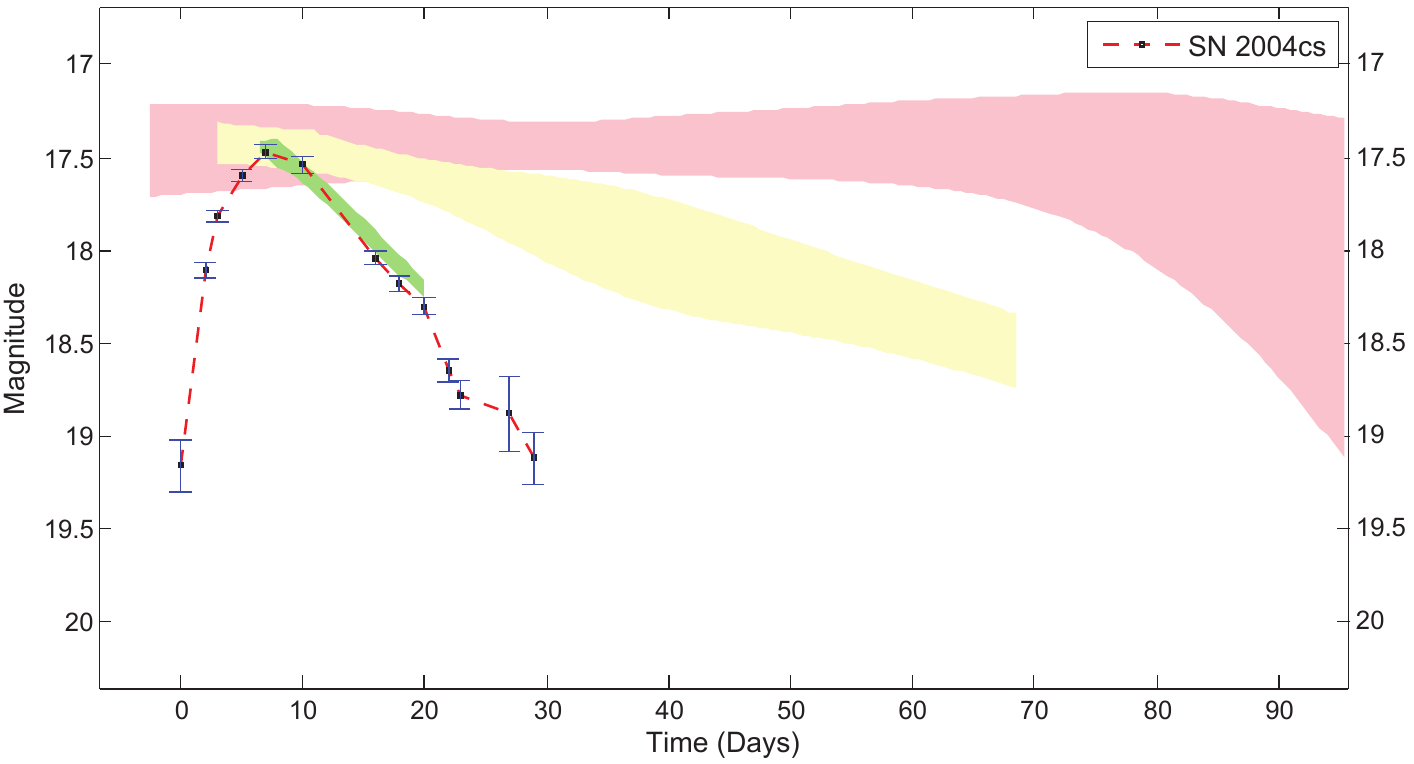}
  \caption{Unfiltered photometry of SN~2004cs \citep{Foley2013} fitting the $R$-band SN~IIb template (green shade) from \citet{Arcavi2012}. The yellow band indicates Type~IIL SNe~II, while the pink represents Type~IIP. \label{fig:2004cs_phot}}
\end{figure*}

Regarding the light curve of SN~2004cs, Figure~16 of \citet{Foley2013} shows this object being something of an outlier in terms of a width-luminosity relation in $V$-band that may exist for the 02cx-likes. We therefore investigate whether the photometry for this object can shed light on its nature. \Citet{Arcavi2012} show that SNe~IIb have a very tight range of decline rates in $R$-band. In Figure~\ref{fig:2004cs_phot} we overlay the light curve of SN~2004cs \citep[][we follow the common practice of assuming unfiltered KAIT datapoints trace $R$-band luminosity]{Foley2013} on the \citet{Arcavi2012} SN~IIb template, finding good agreement.

\placefigure{fig:2004cs_phot}

We conclude that SN~2004cs is indeed a Type~IIb supernova, as it was originally classified by \citet{Rajala2005}. The spectrum provides strong support for this classification, including robust evidence for hydrogen. The photometry provides further support, showing this object to be consistent with SNe~IIb rather than 02cx-like SNe~Ia.

\begin{figure*}
  \centering
  \includegraphics[width=0.8\textwidth]{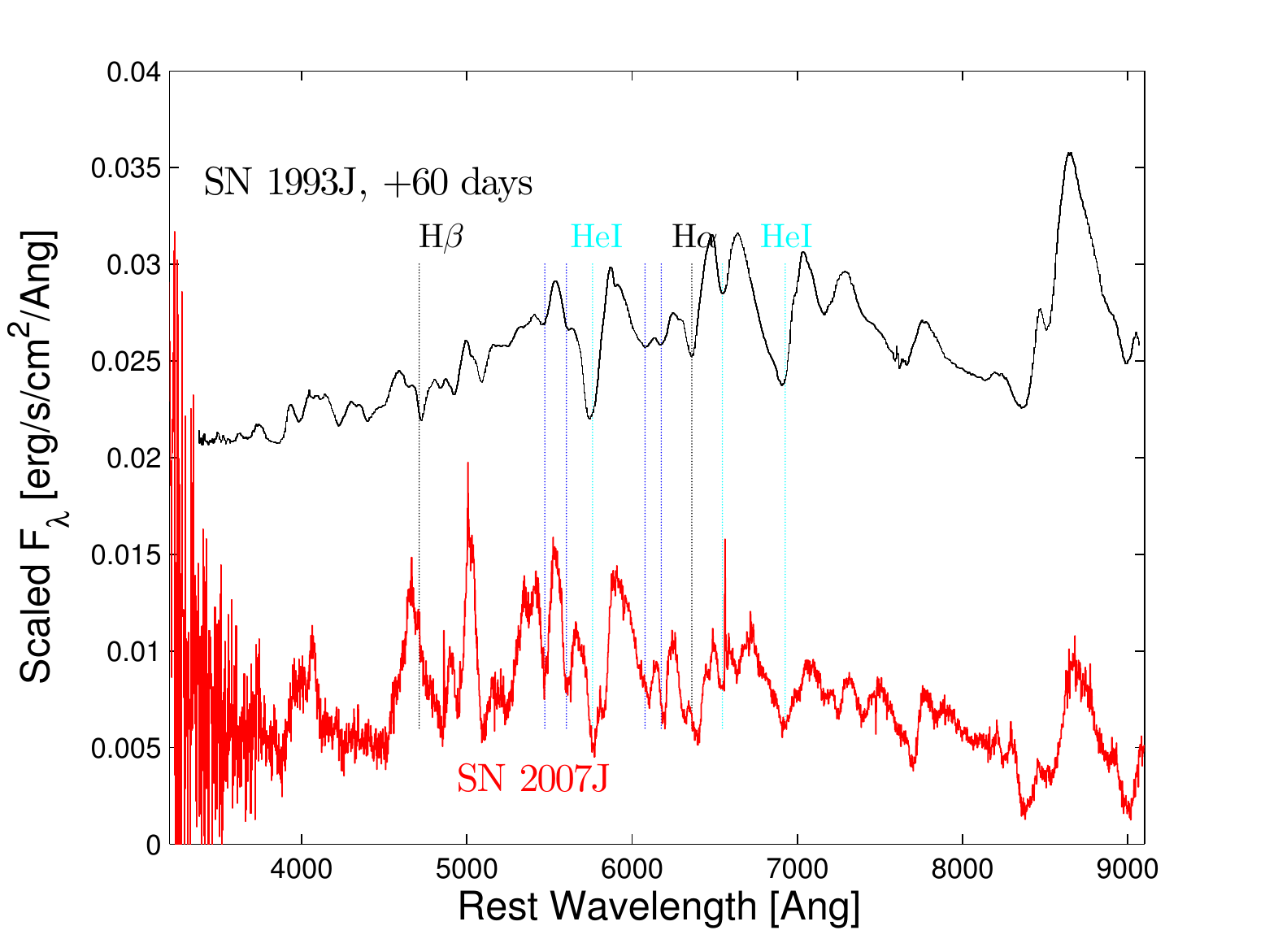}
  \caption{Comparison of the spectrum of SN~2007J (red) with that of SN~1993J (black). The \Halpha{} (black) and \ion{He}{1} ($5876~\A$ and $7065~\A$, cyan) lines are clearly identified, and additional nearby features (blue) align as well. \Hbeta{} is not detected in SN~2007J, perhaps indicating a thinner hydrogen envelope. \label{fig:2007J_spec}}
\end{figure*}

Motivated by our study of SN~2004cs, we now turn to SN~2007J, the similarity to which led \citet{Foley2013} to classify the former as an 02cx-like. In Figure~\ref{fig:2007J_spec} we compare the spectrum of SN~2007J to the Type~IIb SN~1993J. The spectra are indeed quite similar. In particular, an absorption line coincident with \Halpha{} can be seen in the SN~2007J spectrum. The expansion velocity derived from this feature assuming it is \Halpha{} is $9000~\kms$, which is greater than the \ion{He}{1} velocity ($6500~\kms$), as we have already noted is commonly seen in SNe~IIb. One may argue that there is no corresponding \Hbeta{} feature in the SN~2007J spectrum, but this has also been seen in some SNe~IIb \citep[e.g.\ SN~2008ax, see][Figure~5]{Pastorello2008} and probably suggests that the amount of hydrogen is lower in this object than in SN~1993J.

From this analysis we conclude that SN~2007J could also be a Type~IIb supernova, rather than an 02cx-like Type~Ia. We therefore view the claim of a helium connection to the 02cx-class---a connection based only on SN~2004cs and SN~2007J---as questionable. The helium in these two objects should not be taken as a constraint on the progenitor systems of 02cx-like transients.

\placefigure{fig:2007J_spec}

\section{Individual Spectra}
\label{sec:individual_spectra}

Here we present overlays showing the matches between our PTF spectra and SN~2002cx or SN~2002es, as is appropriate. We show a single phase for each PTF object, and the template against which we choose to match it is chosen based on the phase of our spectrum.

\begin{figure*}
  \centering
  \includegraphics[height=0.5\textwidth]{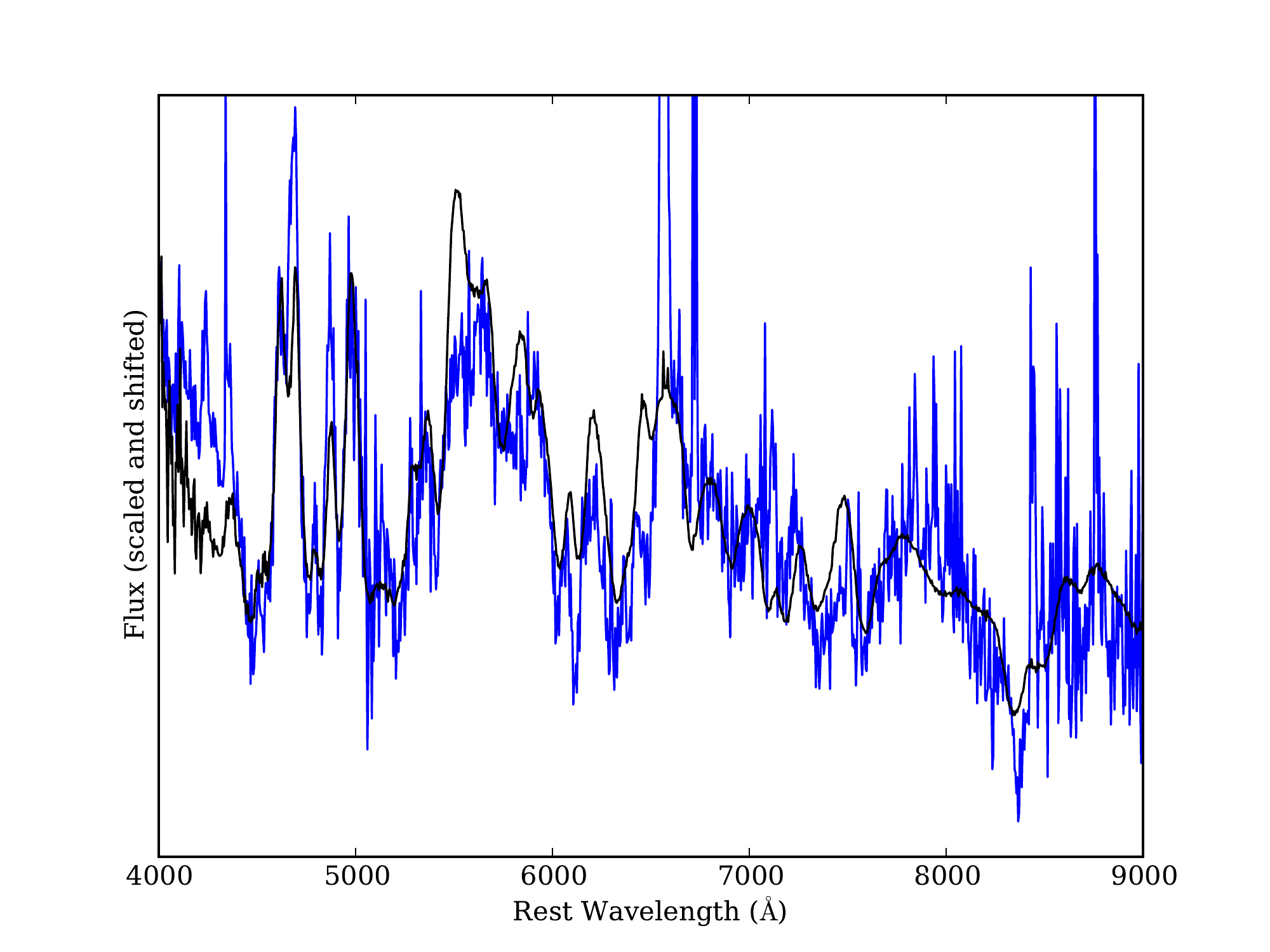}
  \caption{Comparison between PTF~09ego (blue, $+9~\days$) and SN~2002cx (black, $+20~\days$). The former has had galaxy light subtracted via Superfit. \label{fig:09ego_spec}}
\end{figure*}

\begin{figure*}
  \centering
  \includegraphics[height=0.5\textwidth]{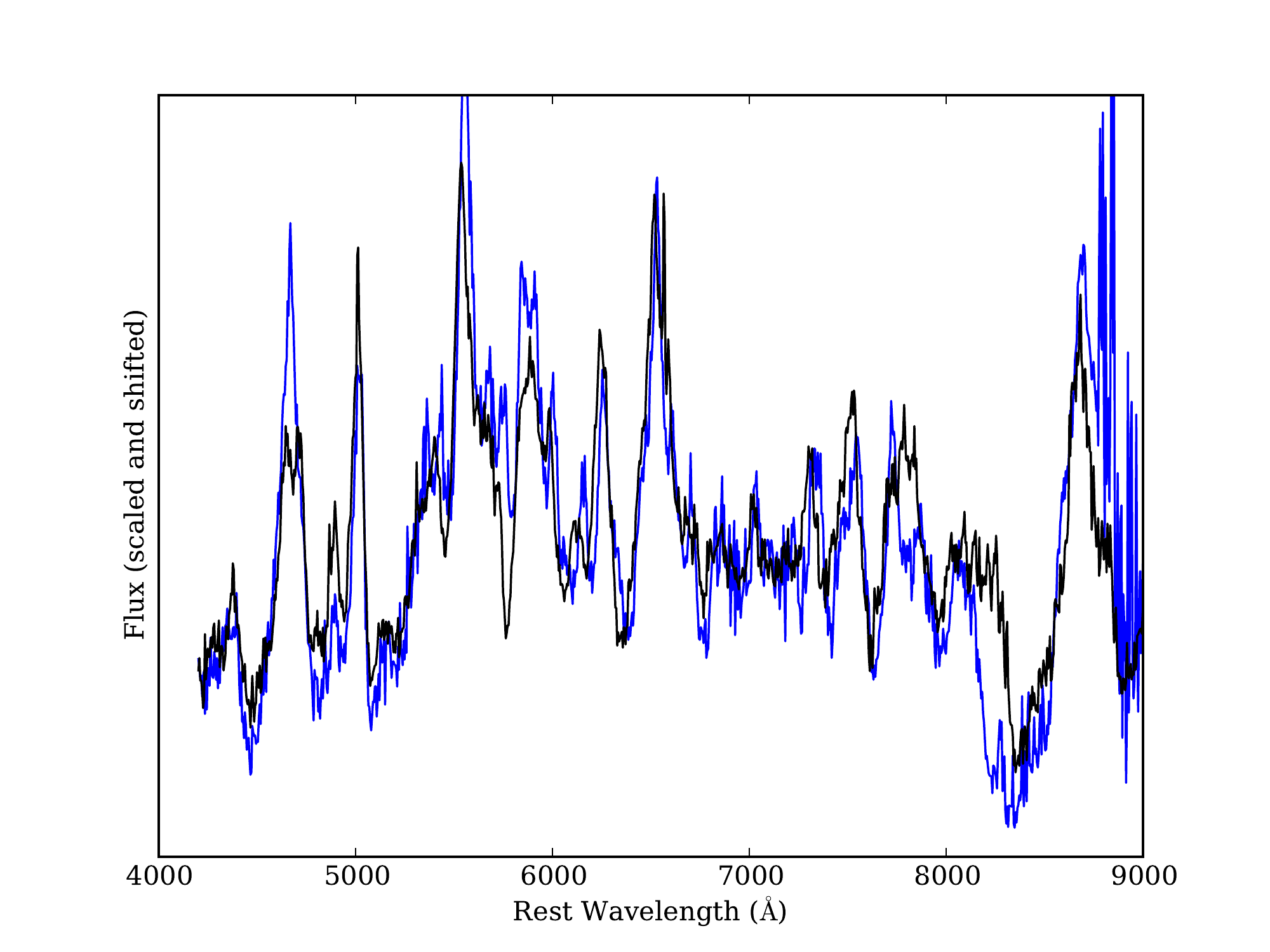}
  \caption{Comparison between PTF~09eiy (blue, $+55~\days$) and SN~2002cx (black, $+56~\days$). The former has had galaxy light subtracted via Superfit, and the fiducial redshift of $0.06$ was used to align spectra. \label{fig:09eiy_spec}}
\end{figure*}

\begin{figure*}
  \centering
  \includegraphics[height=0.5\textwidth]{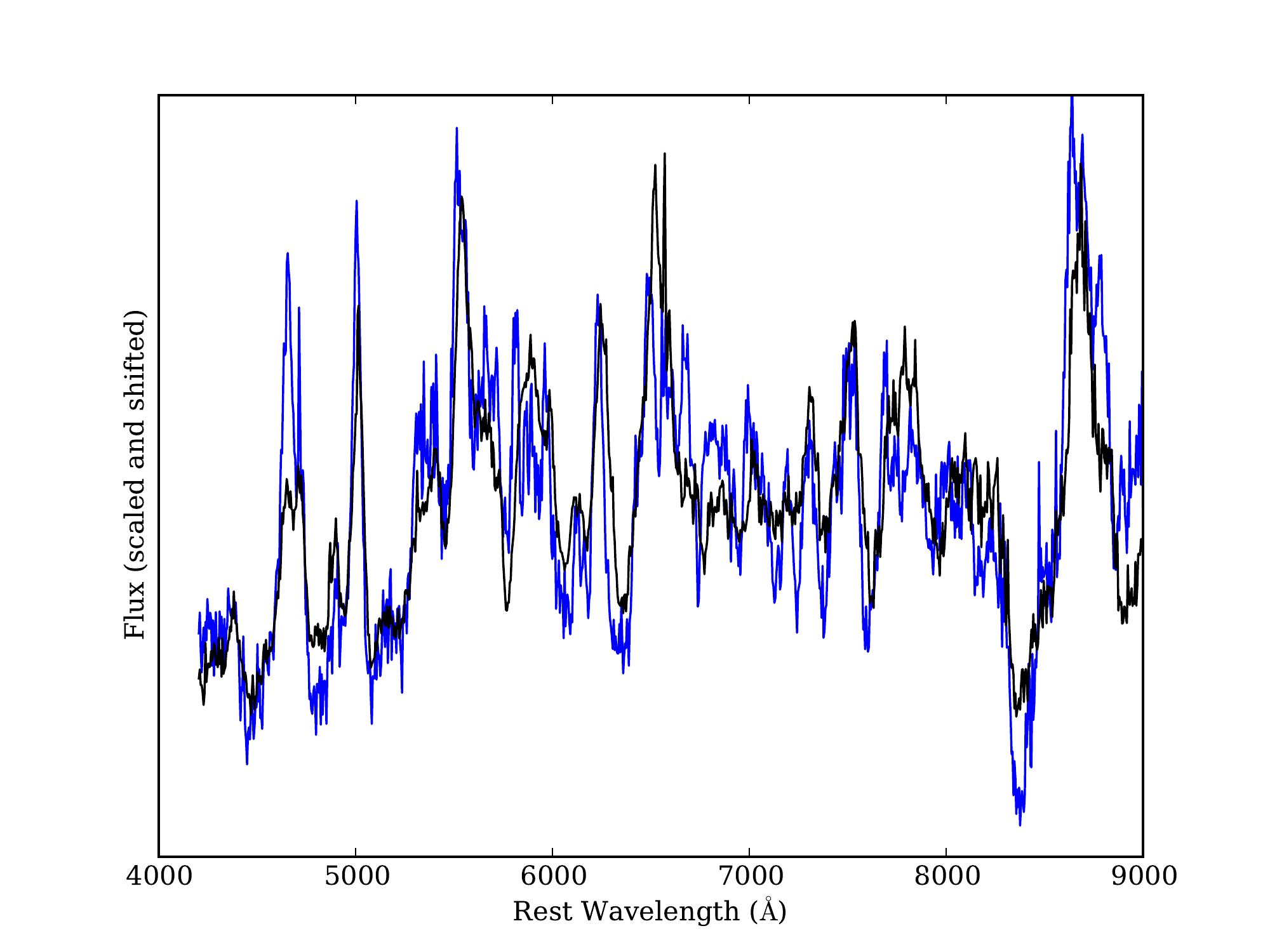}
  \caption{Comparison between PTF~09eoi (blue, $+21~\days$) and SN~2002cx (black, $+56~\days$). The former has had galaxy light subtracted via Superfit. \label{fig:09eoi_spec}}
\end{figure*}

\begin{figure*}
  \centering
  \includegraphics[height=0.5\textwidth]{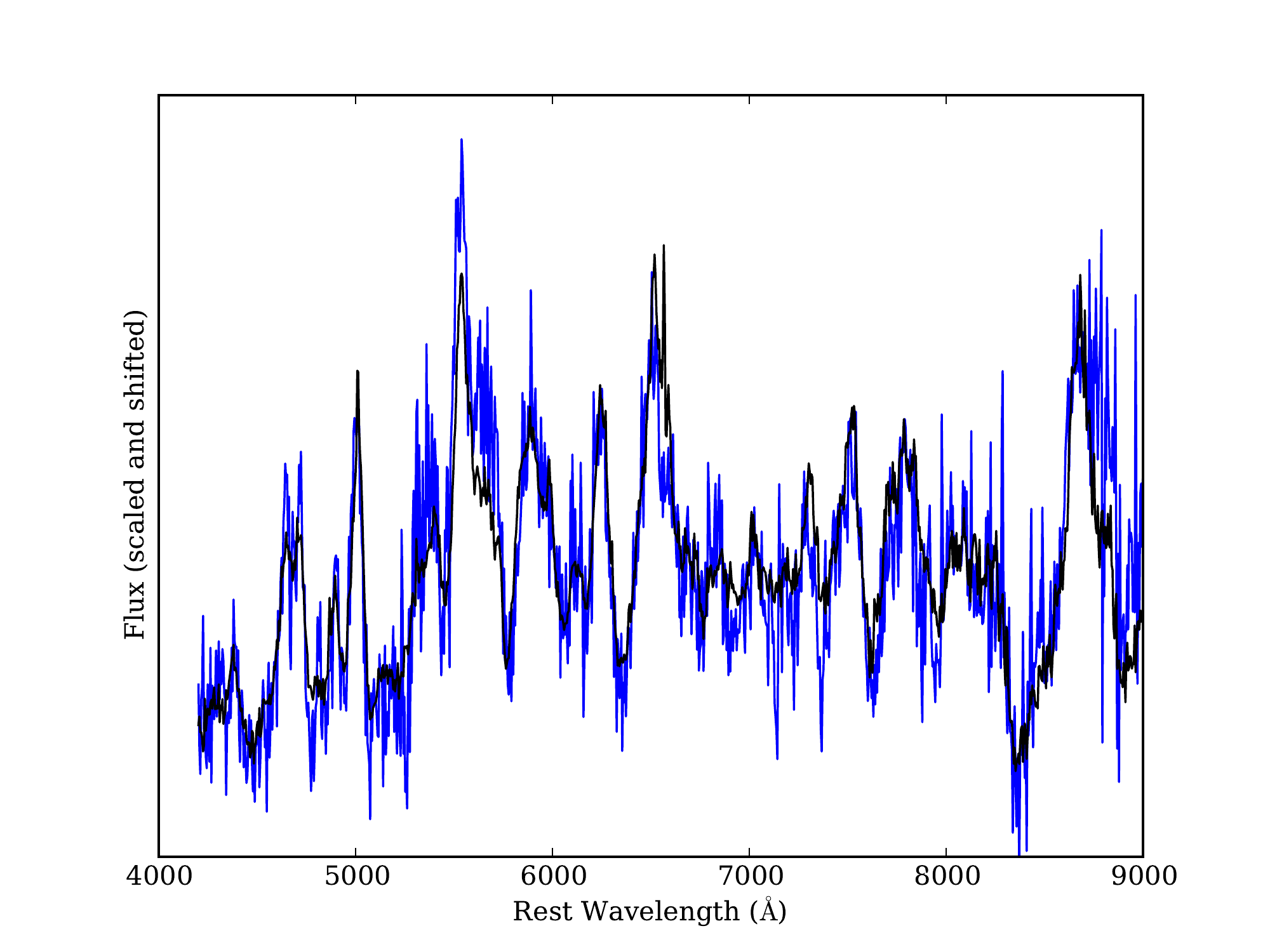}
  \caption{Comparison between PTF~10xk (blue, $+36~\days$) and SN~2002cx (black, $+56~\days$). The former has had galaxy light subtracted via Superfit, and the fiducial redshift of $0.066$ was used to align spectra. \label{fig:10xk_spec}}
\end{figure*}

\begin{figure*}
  \centering
  \includegraphics[height=0.5\textwidth]{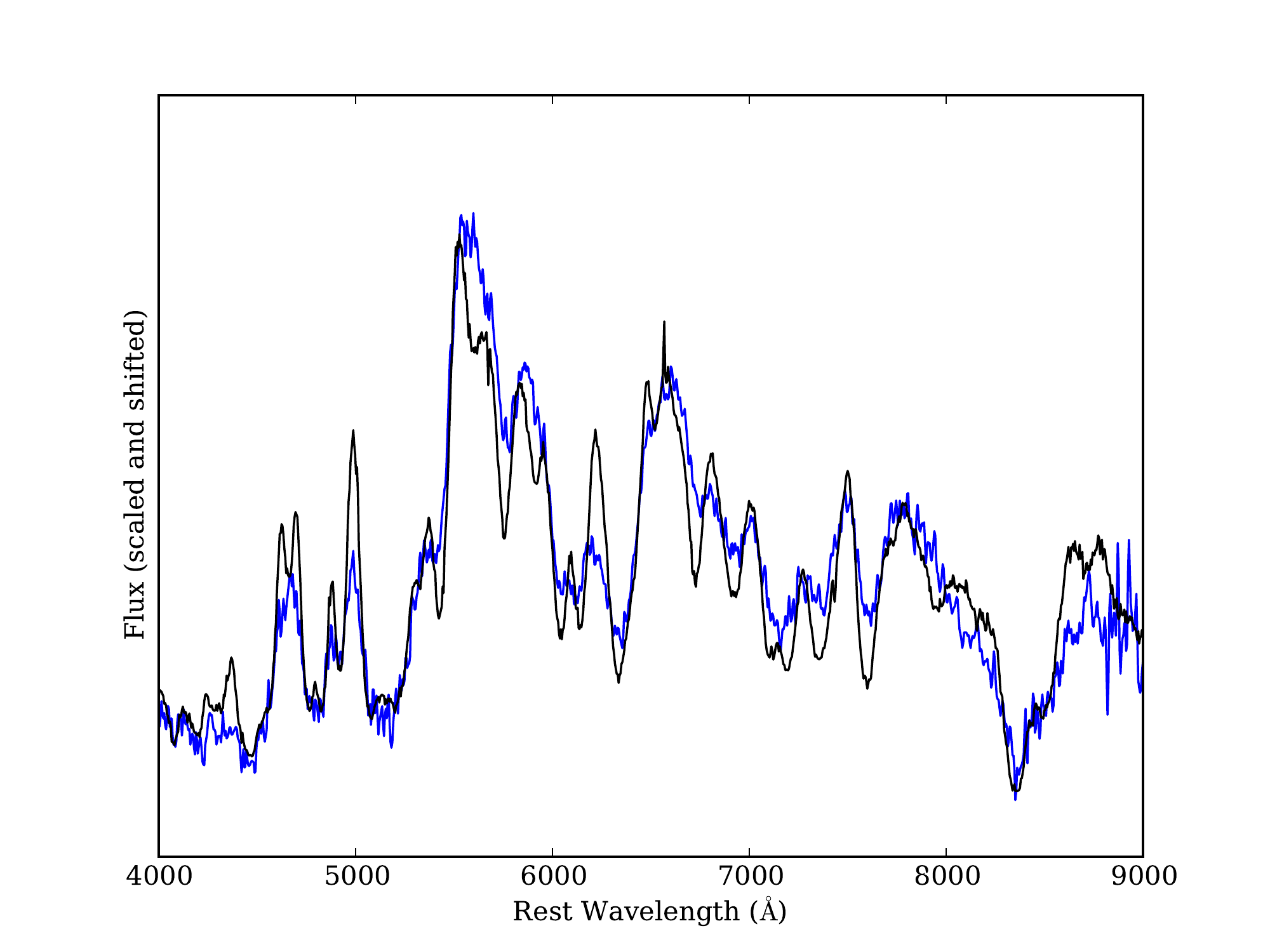}
  \caption{Comparison between PTF~11hyh (blue, $+25~\days$) and SN~2002cx (black, $+25~\days$). The former has had galaxy light subtracted via Superfit, and the fiducial redshift of $0.057$ was used to align spectra. \label{fig:11hyh_spec}}
\end{figure*}

\begin{figure*}
  \centering
  \includegraphics[height=0.5\textwidth]{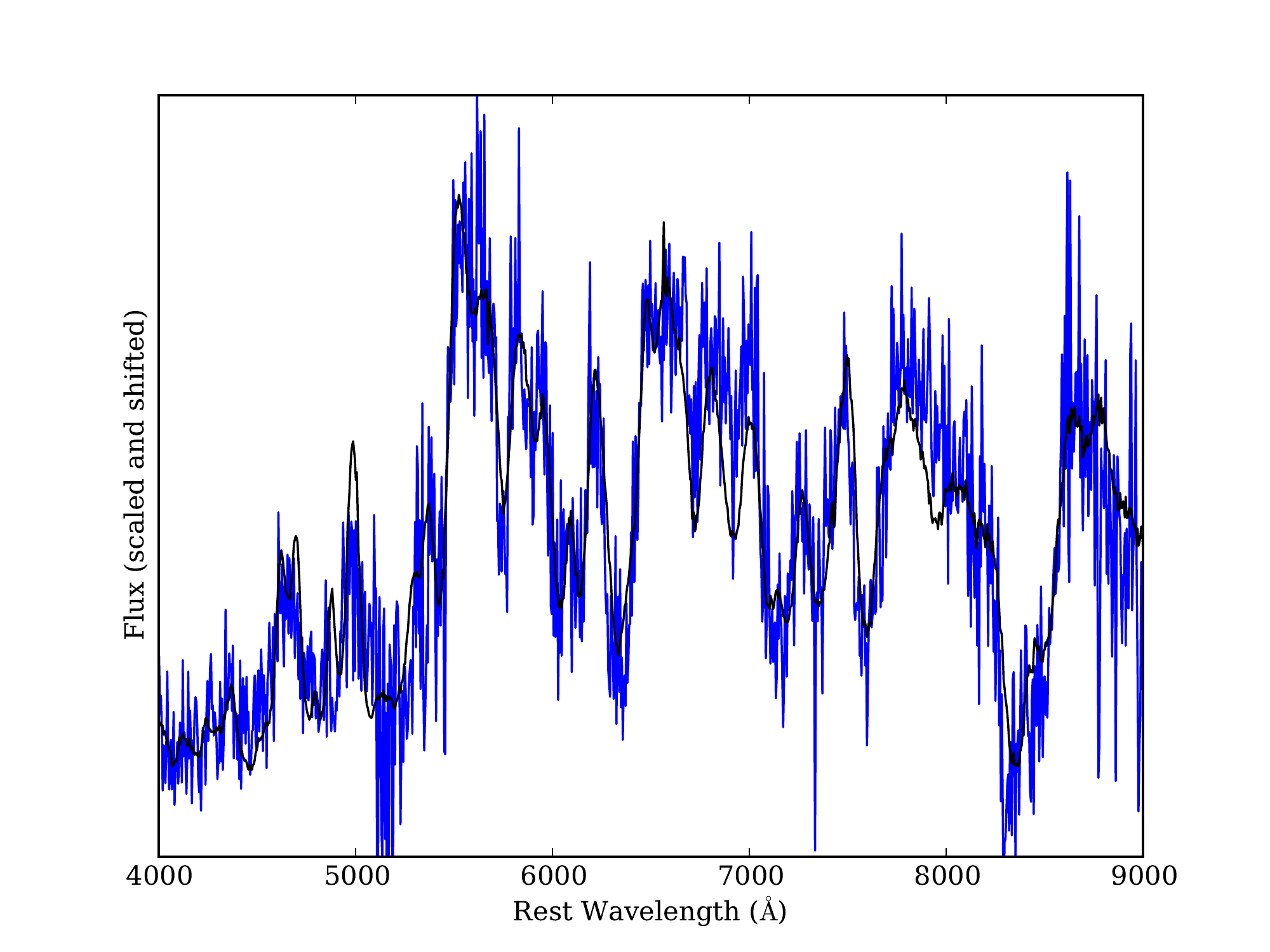}
  \caption{Comparison between iPTF~13an (blue, at least $+13~\days$) and SN~2002cx (black, $+25~\days$). The former has had galaxy light subtracted via Superfit. \label{fig:13an_spec}}
\end{figure*}

\begin{figure*}
  \centering
  \includegraphics[height=0.5\textwidth]{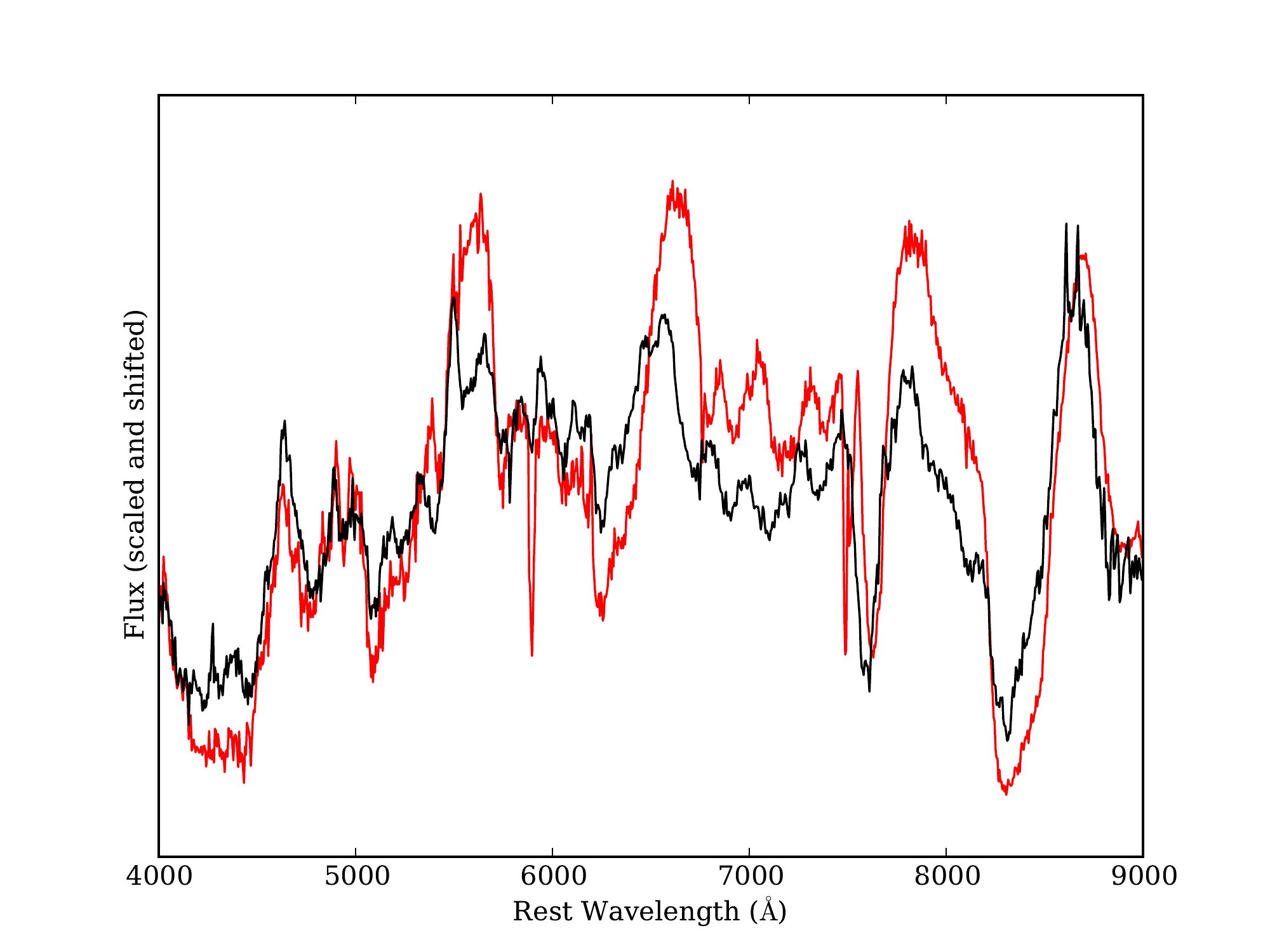}
  \caption{Comparison between PTF~10bvr (red, $+9~\days$) and SN~2002es (black, $+13~\days$). The former has had galaxy light subtracted via Superfit. \label{fig:10bvr_spec}}
\end{figure*}

\begin{figure*}
  \centering
  \includegraphics[height=0.5\textwidth]{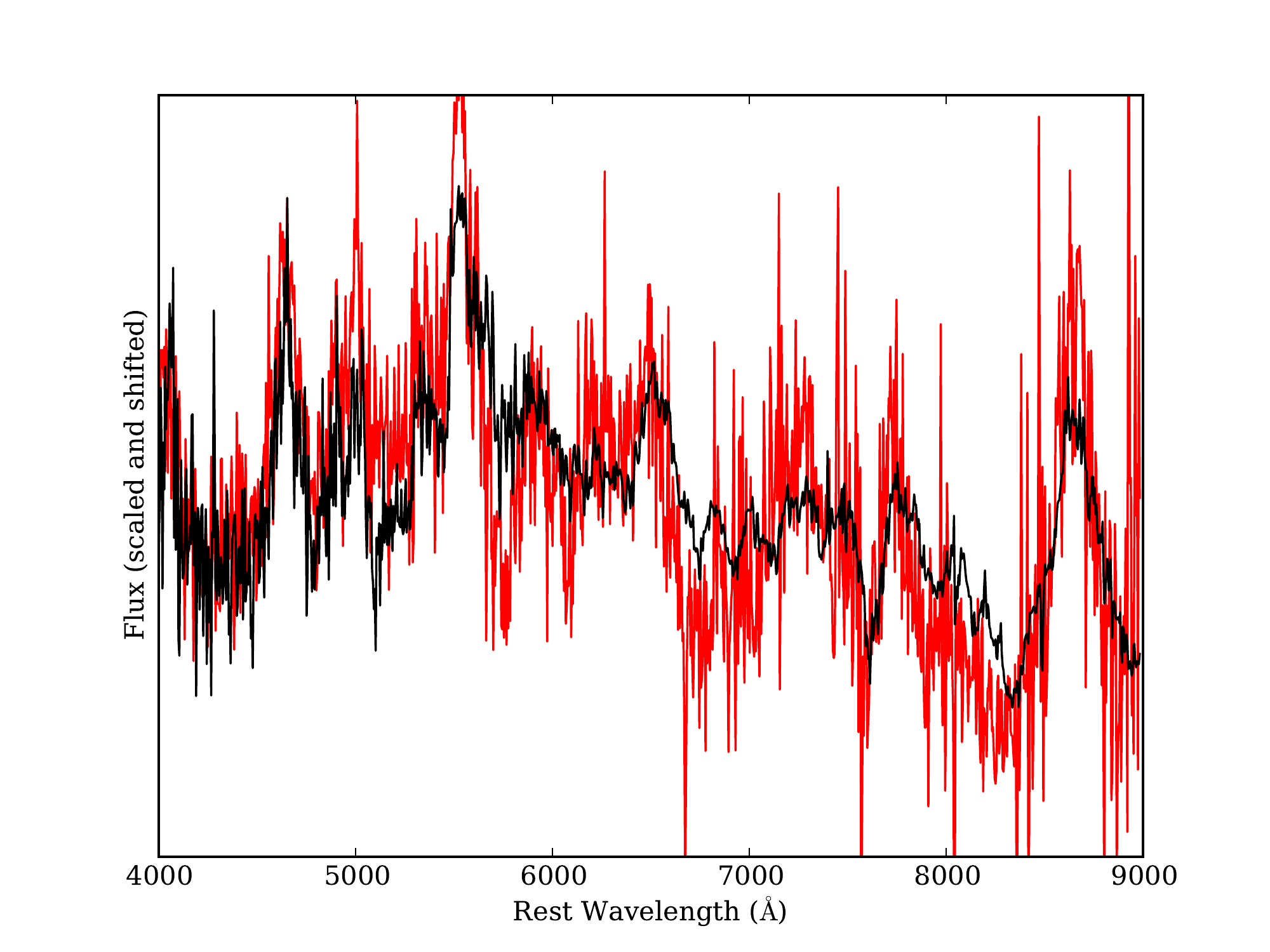}
  \caption{Comparison between PTF~10ujn (red, $+56~\days$) and SN~2002es (black, $+30~\days$). The former has had galaxy light subtracted via Superfit. \label{fig:10ujn_spec}}
\end{figure*}

\begin{figure*}
  \centering
  \includegraphics[height=0.5\textwidth]{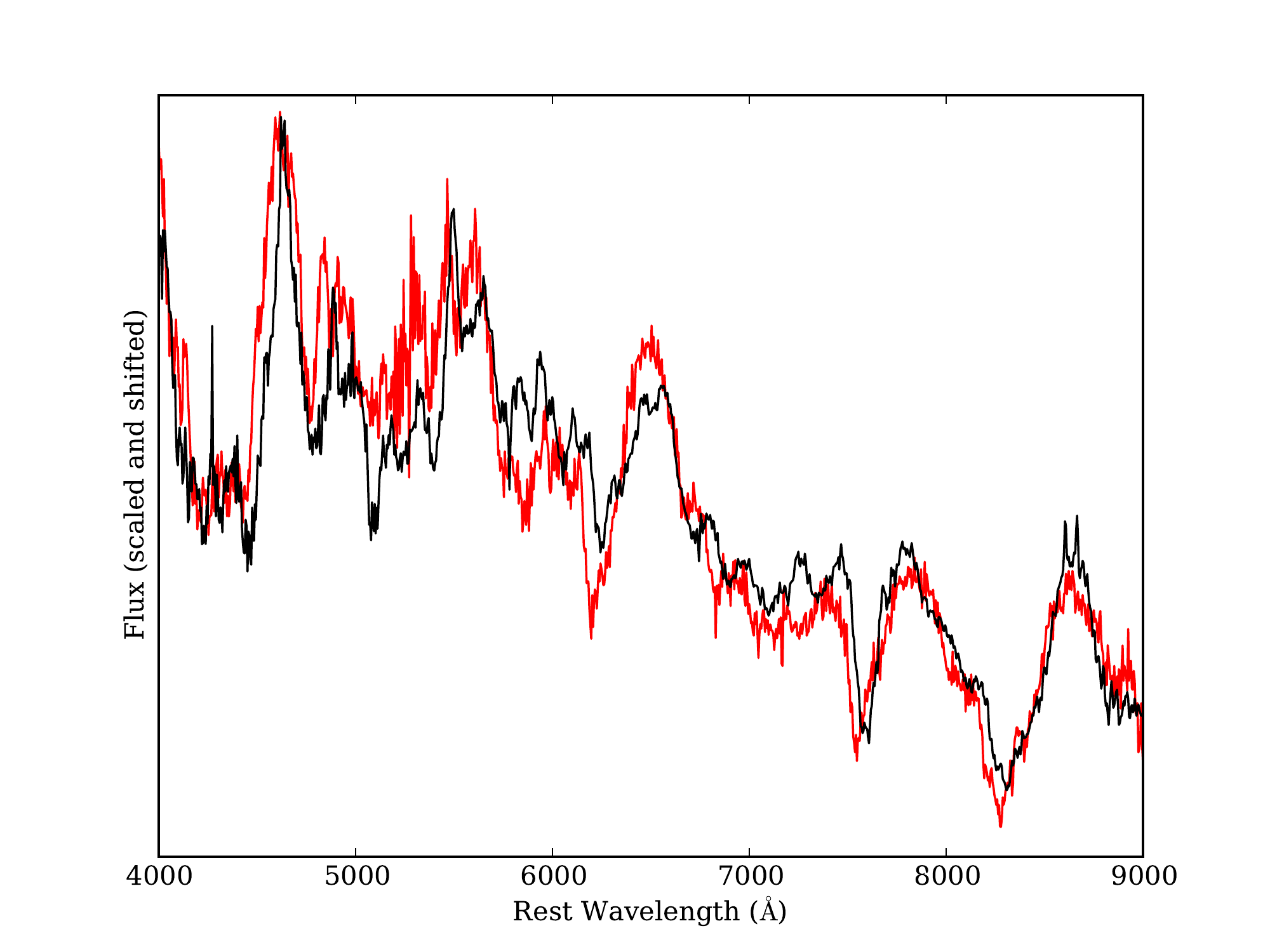}
  \caption{Comparison between PTF~10acdh (red, $+8~\days$) and SN~2002es (black, $+13~\days$). The former has had galaxy light subtracted via Superfit. \label{fig:10acdh_spec}}
\end{figure*}

\placefigure{fig:09ego_spec}
\placefigure{fig:09eiy_spec}
\placefigure{fig:09eoi_spec}
\placefigure{fig:10xk_spec}
\placefigure{fig:11hyh_spec}
\placefigure{fig:13an_spec}
\placefigure{fig:10bvr_spec}
\placefigure{fig:10ujn_spec}
\placefigure{fig:10acdh_spec}

\section{Spectra of Objects Determined to Not Match}
\label{sec:rejected_spectra}

Here we present overlays for the nine rejected objects discussed in \S\ref{sec:selection:ptf}. The first four objects---PTF~10vzj, 10xfh, 11cfm, and 11pzq---were automatically flagged by Superfit but did not pass further scrutiny. The other five are examples of slow-speed supernovae that are poor spectral matches to SN~2002cx, SN~2002es, or other related objects. Each overlay shows the best match we could get to an 02cx-like or SN~2002es template of interest.

\begin{figure*}
  \centering
  \includegraphics[height=0.5\textwidth]{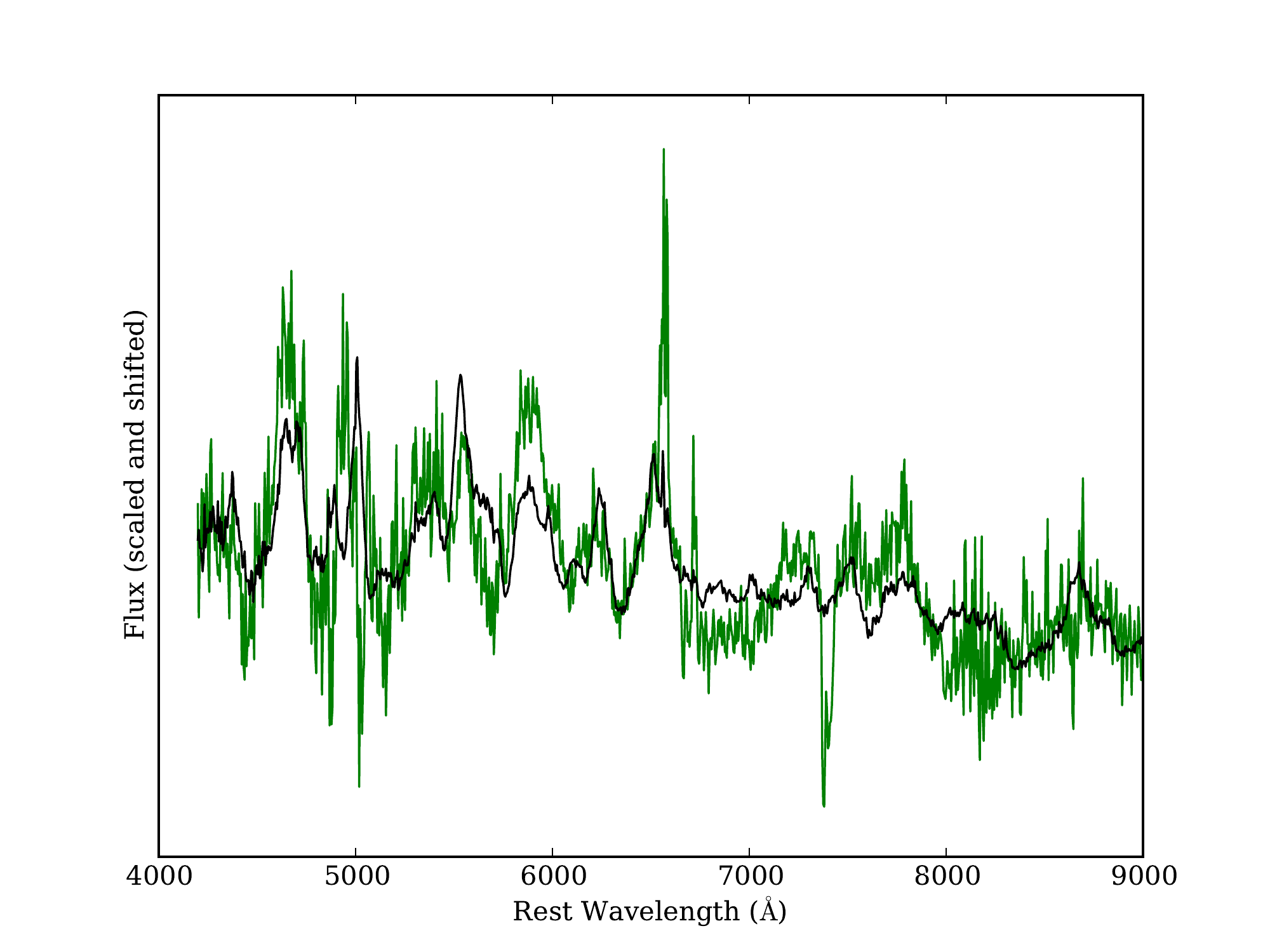}
  \caption{Comparison between PTF~10vzj (green, $+51~\days$) and SN~2002cx (black, $+56~\days$). The former has had galaxy light subtracted via Superfit. \label{fig:10vzj_spec}}
\end{figure*}

\begin{figure*}
  \centering
  \includegraphics[height=0.5\textwidth]{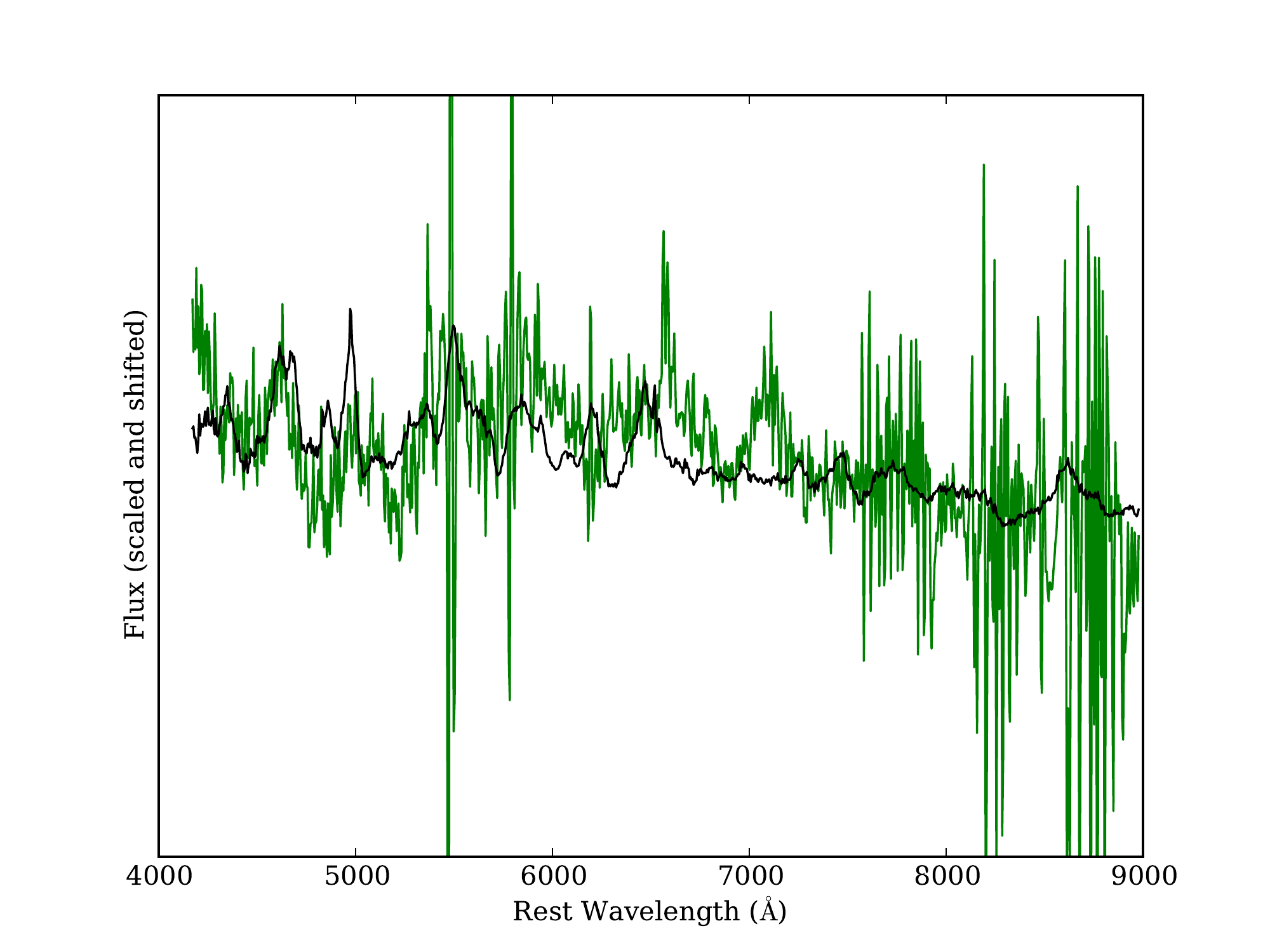}
  \caption{Comparison between PTF~10xfh (green, unknown phase, spectrum taken 2010~October~8) and SN~2002cx (black, $+56~\days$). The former has had galaxy light subtracted via Superfit. \label{fig:10xfh_spec}}
\end{figure*}

\begin{figure*}
  \centering
  \includegraphics[height=0.5\textwidth]{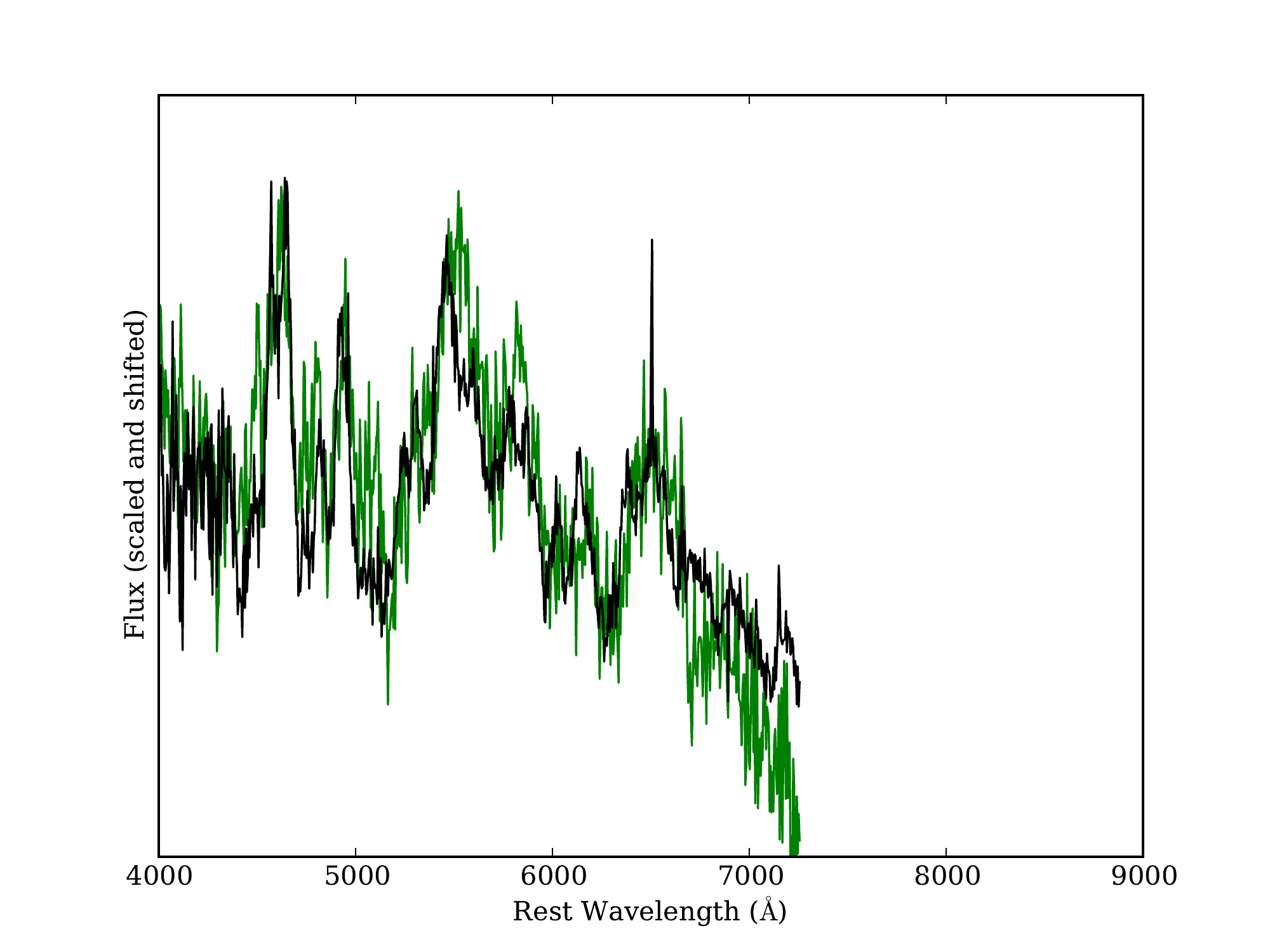}
  \caption{Comparison between PTF~11cfm (green, $+24~\days$) and SN~2002cx (black, $+15~\days$). The former has had galaxy light subtracted via Superfit. \label{fig:11cfm_spec}}
\end{figure*}

\begin{figure*}
  \centering
  \includegraphics[height=0.5\textwidth]{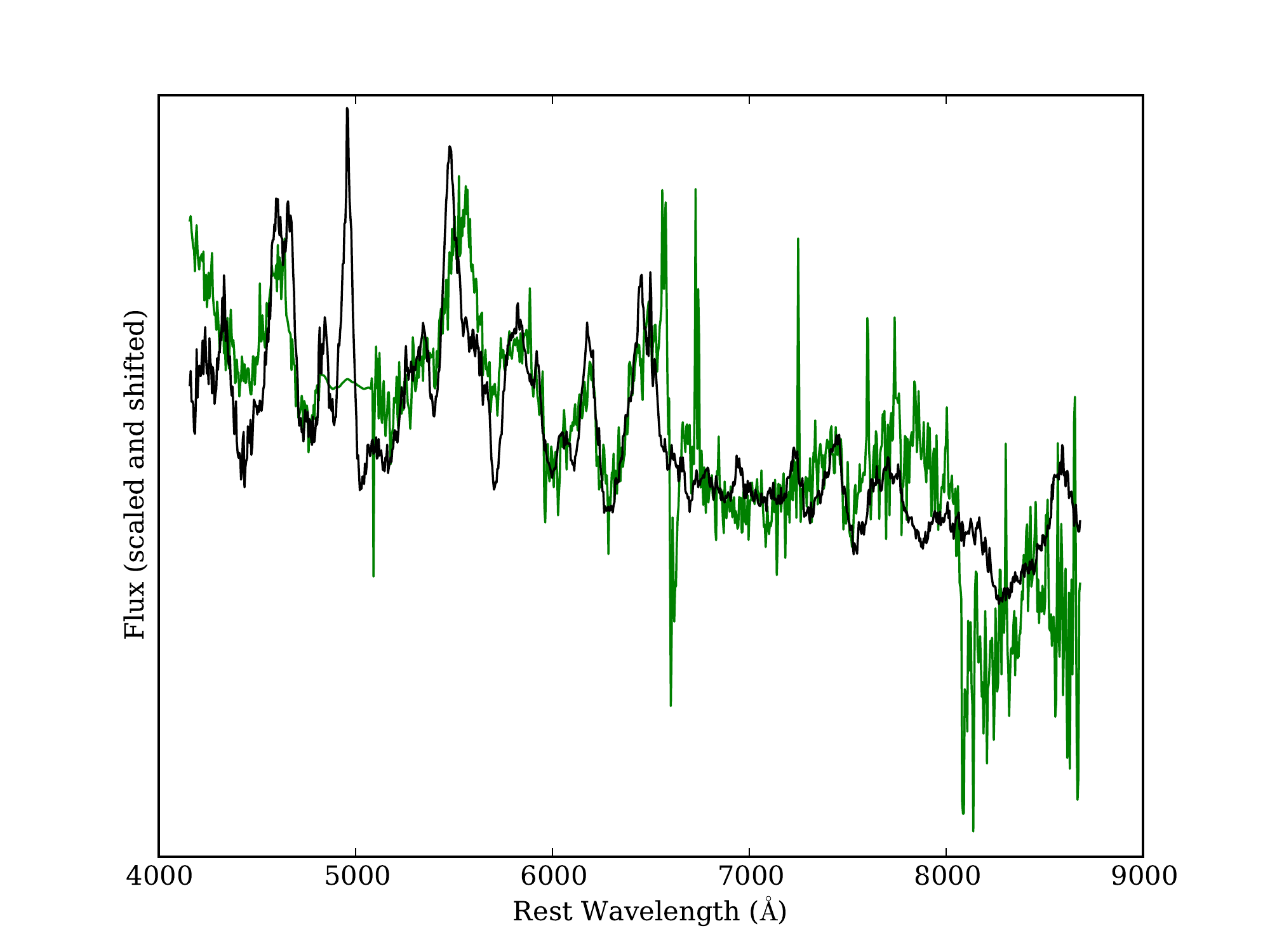}
  \caption{Comparison between PTF~11pzq (green, $+18~\days$) and SN~2002cx (black, $+56~\days$). The former has had galaxy light subtracted via Superfit. \label{fig:11pzq_spec}}
\end{figure*}

\begin{figure*}
  \centering
  \includegraphics[height=0.5\textwidth]{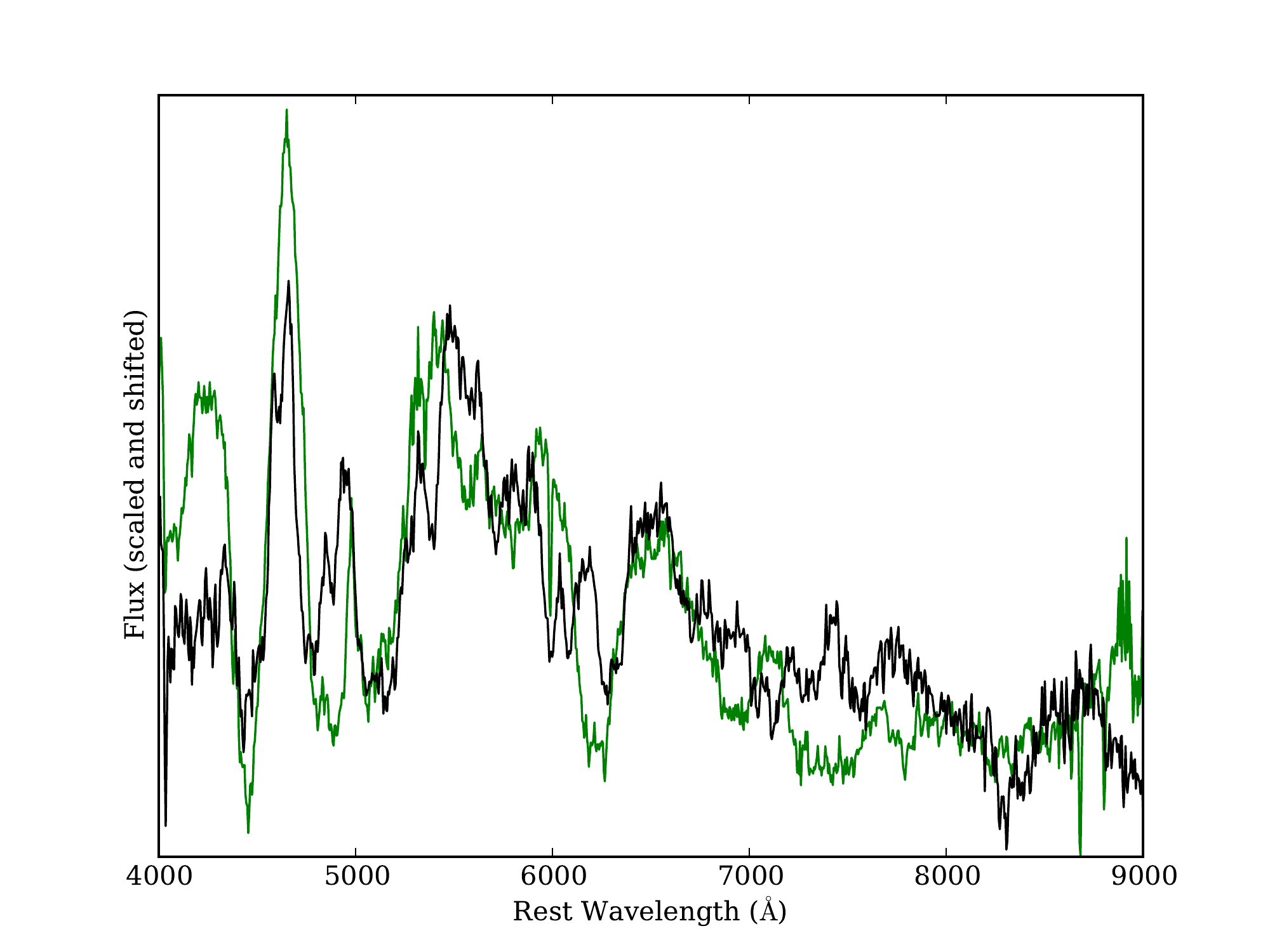}
  \caption{Comparison between PTF~09aly (green, $+6~\days$) and SN~2005hk (black, $+15~\days$). The former has had galaxy light subtracted via Superfit. \label{fig:09aly_spec}}
\end{figure*}

\begin{figure*}
  \centering
  \includegraphics[height=0.5\textwidth]{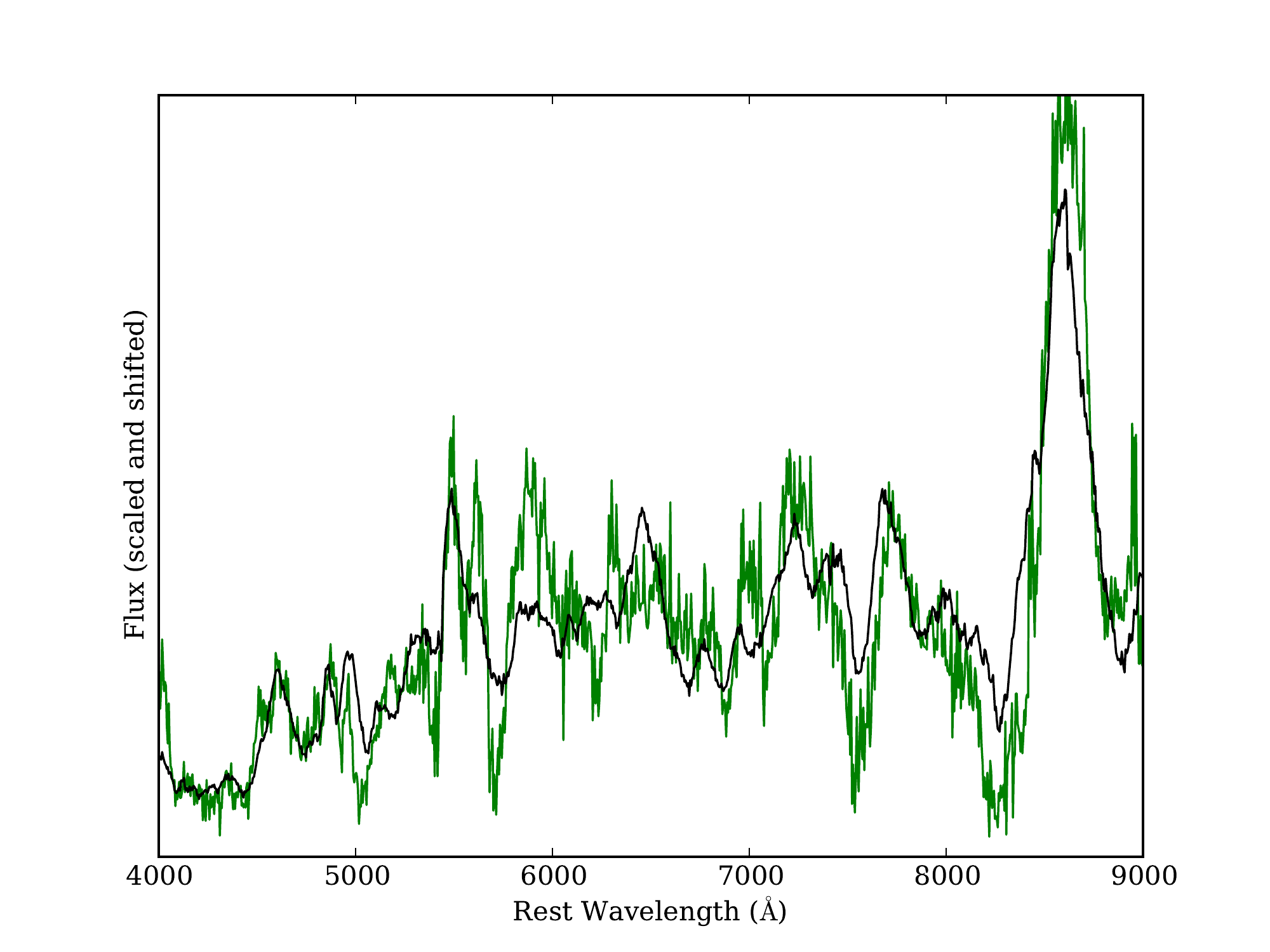}
  \caption{Comparison between PTF~09dav (green, $+12~\days$) and SN~2002es (black, $+37~\days$). The former has had galaxy light subtracted via Superfit. \label{fig:09dav_spec}}
\end{figure*}

\begin{figure*}
  \centering
  \includegraphics[height=0.5\textwidth]{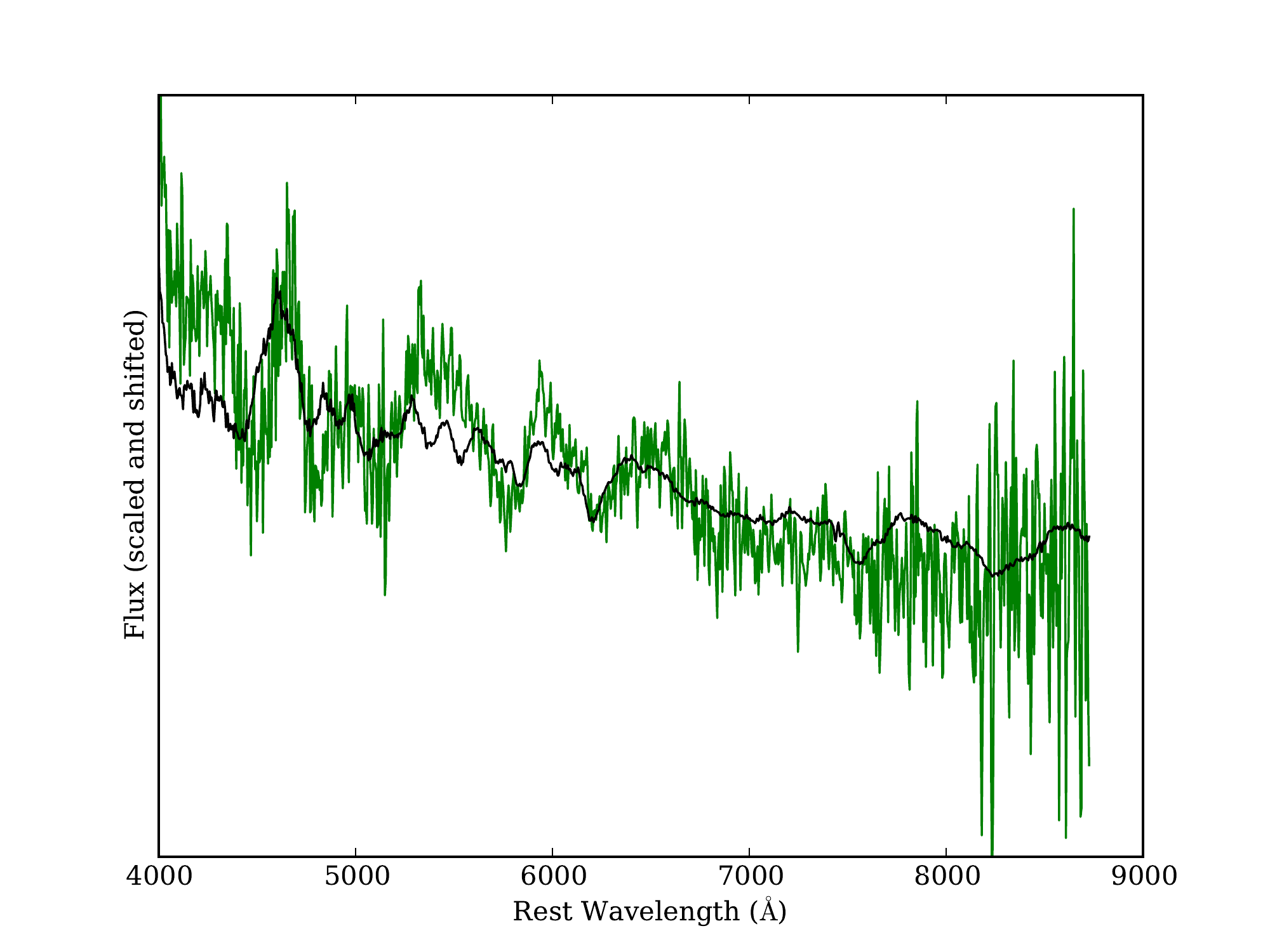}
  \caption{Comparison between PTF~10pko (green, $+8~\days$) and SN~2002es (black, $+3~\days$). The former has had galaxy light subtracted via Superfit. \label{fig:10pko_spec}}
\end{figure*}

\begin{figure*}
  \centering
  \includegraphics[height=0.5\textwidth]{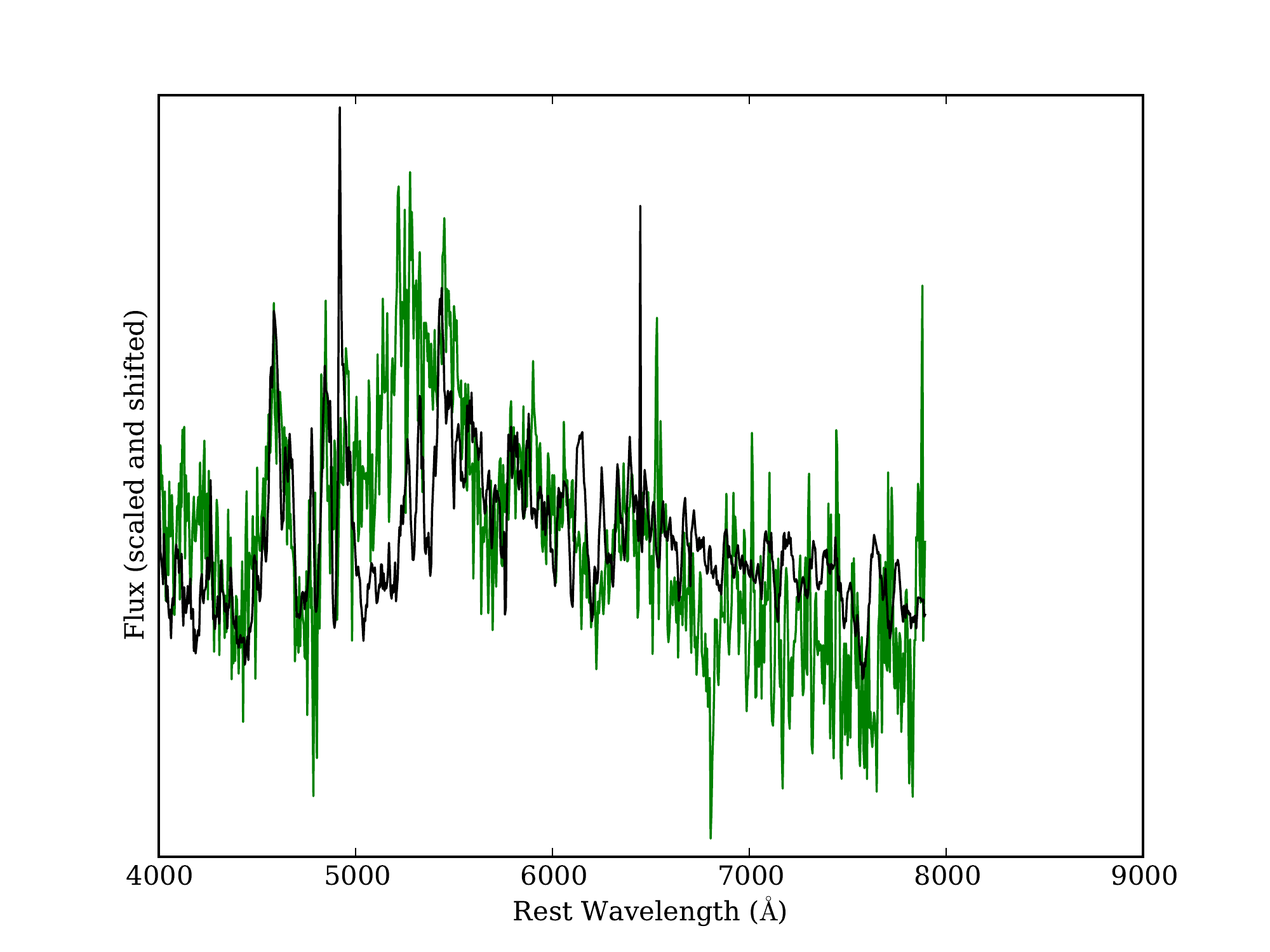}
  \caption{Comparison between PTF~10xfv (green, $+17~\days$) and SN~2008ha (black, $+11~\days$). The former has had galaxy light subtracted via Superfit. \label{fig:10xfv_spec}}
\end{figure*}

\begin{figure*}
  \centering
  \includegraphics[height=0.5\textwidth]{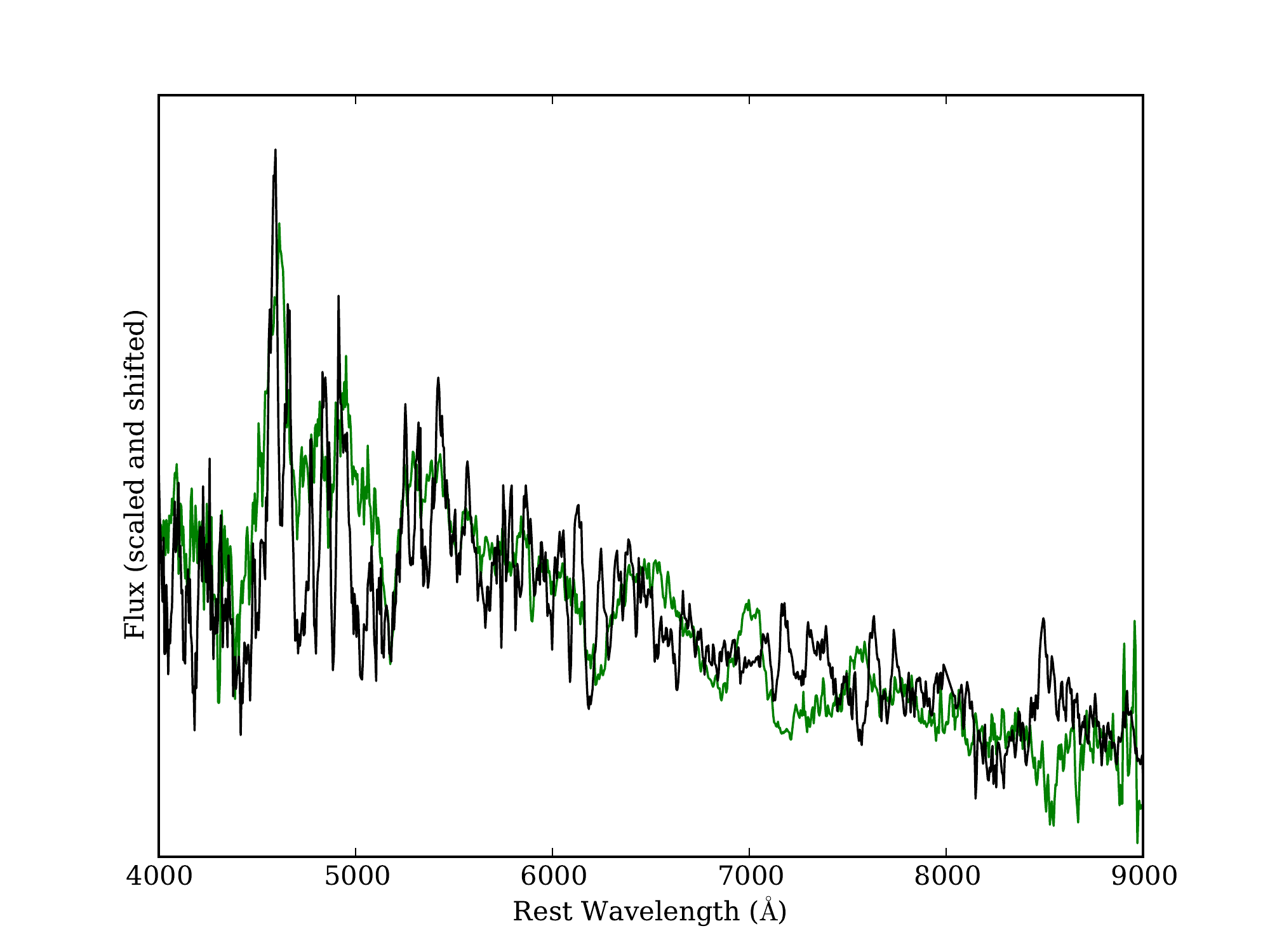}
  \caption{Comparison between PTF~11sd (green, $+8~\days$) and SN~2008ha (black, $+8~\days$). The former has had galaxy light subtracted via Superfit. \label{fig:11sd_spec}}
\end{figure*}

\placefigure{fig:10vzj_spec}
\placefigure{fig:10xfh_spec}
\placefigure{fig:11cfm_spec}
\placefigure{fig:11pzq_spec}
\placefigure{fig:09aly_spec}
\placefigure{fig:09dav_spec}
\placefigure{fig:10pko_spec}
\placefigure{fig:10xfv_spec}
\placefigure{fig:11sd_spec}

\clearpage
\begin{deluxetable}{cccccccc}
  \tabletypesize{\scriptsize}
  \tablecaption{Visual Vetting \label{tab:spectrum_properties}}
  \tablewidth{0pt}
  \tablehead{\colhead{Name} & \colhead{Phase (days)} & \colhead{\# Peaks ($6000\text{--}8000~\A$)\tablenotemark{a}} & \colhead{$4700~\A$ Resolved?} & \colhead{$6200~\A$ Feature?} & \colhead{\ion{Ti}{2} Trough?}}
  \startdata
    \sidehead{Templates}
      SN~2002cx  & $+25$                  & $9$  & yes   & yes   & no    \\
      SN~2002es  & $+13$                  & $12$ & no    & yes   & yes   \\
      SN~2011fe  & $+16$                  & $2$  & no    & no    & no    \\
    \sidehead{02cx matches}
      PTF~09ego  & $+9$                   & $6+$ & yes   & yes   & no    \\
      PTF~09eiy  & $+55$                  & $5+$ & no    & yes   & no    \\
      PTF~09eoi  & $+38$                  & $12$ & yes   & yes   & no    \\
      PTF~10xk   & $+31$                  & $12$ & yes   & yes   & no    \\
      PTF~11hyh  & $+23$                  & $7$  & no    & yes   & no    \\
      iPTF~13an  & $+13$\tablenotemark{b} & $11$ & yes   & yes   & no    \\
    \sidehead{02es matches}
      PTF~10bvr  & $+9$                   & $8$  & maybe & maybe & yes   \\
      PTF~10ujn  & $+119$                 & $9+$ & no    & maybe & maybe \\
      PTF~10acdh & $+42$                  & $7$  & no    & maybe & yes   \\
    \sidehead{False positives}
      PTF~09aly  & $+6$                   & $4$  & no    & no    & no    \\
      PTF~10pko  & $+8$                   & $14$ & yes   & no    & no    \\
      PTF~10vzj  & $+51$                  & $10$ & yes   & no    & no    \\
      PTF~10xfh  & \nodata                & $10$ & maybe & yes   & yes   \\
      PTF~10xfv  & $+17$                  & $13$ & maybe & no    & no    \\
      PTF~11sd   & $+8$                   & $6$  & no    & no    & no    \\
      PTF~11cfm  & $+24$                  & $8+$ & maybe & no    & no    \\
      PTF~11pzq  & $+18$                  & $9$  & no    & no    & no
  \enddata
  \tablenotetext{a}{A ``$+$'' indicates the spectrum does not fill the entire range and so a more complete spectrum would likely have more points. The counting is performed using the spectra as shown in Figures~\ref{fig:02cx_zoomed_spectra} and~\ref{fig:02es_zoomed_spectra}.}
  \tablenotetext{b}{Assumes first observation corresponds to peak brightness.}
\end{deluxetable}
\begin{deluxetable}{ccccccccc}
  \tabletypesize{\scriptsize}
  \tablecaption{PTF Sample Positions \label{tab:positions}}
  \tablewidth{0pt}
  \tablehead{\colhead{Name} & \colhead{RA (J2000)} & \colhead{DEC (J2000)}}
  \startdata
    PTF~09ego  & $\RA{17}{26}{25.16}$ & $+\DEC{62}{58}{22.1}$ \\
    PTF~09eiy  & $\RA{01}{54}{16.68}$ & $-\DEC{15}{05}{01.6}$ \\
    PTF~09eoi  & $\RA{23}{24}{12.87}$ & $+\DEC{12}{46}{42.6}$ \\
    PTF~10xk   & $\RA{01}{41}{02.86}$ & $+\DEC{30}{13}{38.7}$ \\
    PTF~10bvr  & $\RA{16}{31}{18.85}$ & $+\DEC{39}{09}{20.4}$ \\
    PTF~10ujn  & $\RA{07}{53}{13.72}$ & $+\DEC{72}{20}{13.5}$ \\
    PTF~10acdh & $\RA{09}{43}{07.58}$ & $+\DEC{09}{39}{31.3}$ \\
    PTF~11hyh  & $\RA{01}{45}{50.50}$ & $+\DEC{14}{35}{00.0}$ \\
    iPTF~13an  & $\RA{12}{14}{15.35}$ & $+\DEC{15}{32}{09.6}$
  \enddata
\end{deluxetable}

\clearpage

\clearpage
\begin{deluxetable}{cccccccc}
  \tabletypesize{\scriptsize}
  \tablecaption{Host Properties \label{tab:hosts}}
  \tablewidth{0pt}
  \tablehead{\colhead{Name} & \colhead{Host} & \colhead{$z$\tablenotemark{a}} & \colhead{$\mu$\tablenotemark{b}} & \colhead{$m(R)$ ($\magn$)\tablenotemark{c}} & \colhead{$M(R)$ ($\magn$)\tablenotemark{d}} & \colhead{\gi{} ($\magn$)\tablenotemark{e}} & \colhead{Type\tablenotemark{f}}}
  \startdata
    \sidehead{Literature sample}
      SN~1991bj   & IC~344                         & $0.018146$          & $34.26$ & $\phe14.18$               & $\phe{-}20.08$               & \nodata & Sb      \\
      SN~2002cx   & CGCG~044-035                   & $0.023963$          & $35.18$ & $\phe15.39$               & $\phe{-}19.79$               & $0.83$  & Sb      \\
      SN~2003gq   & NGC~7407                       & $0.021448$          & $34.72$ & \nodata                   & \nodata                      & \nodata & Sbc     \\
      SN~2004gw   & PGC~16812                      & $0.017018$          & $34.28$ & \nodata                   & \nodata                      & \nodata & Sbc     \\
      SN~2005hk   & UGC~272                        & $0.012993$          & $33.53$ & $\phe14.42$               & $\phe{-}19.11$               & $0.70$  & Sd      \\
      SN~2006hn   & UGC~6154                       & $0.017199$          & $34.34$ & \nodata                   & \nodata                      & \nodata & Sa      \\
      SN~2007qd   & SDSS~J020932.72-005959.7       & $0.043146$          & $36.35$ & $\phe16.27$               & $\phe{-}20.08$               & $0.59$  & Sc      \\
      SN~2008ge   & NGC~1527                       & $0.004043$          & $31.12$ & $\phe10.25$               & $\phe{-}21.02$               & \nodata & S0      \\
      SN~2008ha   & UGC~12682                      & $0.004647$          & $30.83$ & $\phe14.03$               & $\phe{-}17.47$               & $0.59$  & Irr     \\
      SN~2009ku   & APMUKS(BJ) B032747.73-281526.1 & $0.0792\phn\phn$    & $37.76$ & \nodata                   & \nodata                      & \nodata & Sc      \\[\tablerowsep]
      SN~2002es   & UGC~2708                       & $0.0182\phn\phn$    & $34.42$ & $\phe13.22$               & $\phe{-}22.20$               & $1.58$  & S0      \\
    \sidehead{SN~2002cx matches}
      PTF~09ego   & SDSS~J172625.23+625821.4       & $0.104\phn\phn\phn$ & $38.40$ & $\phe20.45$               & $\phe{-}17.95$               & $0.86$  & \nodata \\
      PTF~09eiy   & \nodata                        & $(0.06)\phn$        & $37.14$ & ${>}22.7\phn$             & ${>}{-}14.4\phn$             & \nodata & \nodata \\
      PTF~09eoi   & SDSS~J232412.96+124646.6       & $0.0415\phn\phn$    & $36.30$ & $\phe17.60$               & $\phe{-}18.70$               & $0.92$  & \nodata \\
      PTF~10xk    & \nodata                        & $(0.066)$           & $37.35$ & ${\gtrsim}23.5\phn$       & ${\gtrsim}{-}13.9\phn$       & \nodata & \nodata \\
      PTF~11hyh   & SDSS~J014550.57+143501.9       & $(0.057)$           & $37.02$ & $\phe21.81$               & $\phe{-}15.21$               & $0.90$  & \nodata \\
      iPTF~13an   & 2MASX~J12141590+1532096        & $0.080315$          & $37.70$ & $\phe16.58$               & $\phe{-}21.12$               & $1.11$  & \nodata \\
    \sidehead{SN~2002es matches}
      PTF~10bvr   & CGCG~224-067\tablenotemark{g}  & $0.015\phn\phn\phn$ & $34.05$ & $\phe14.11$               & $\phe{-}19.94$               & $1.26$  & E       \\
      PTF~10ujn   & \nodata                        & $0.113\phn\phn\phn$ & $38.59$ & ${\approx20}\phd\phn\phn$ & ${\approx}{-}19\phd\phn\phn$ & \nodata & \nodata \\
      PTF~10acdh  & 2MASX~J09430730+0939285        & $0.059471$          & $37.06$ & $\phe16.63$               & $\phe{-}20.43$               & $1.34$  & \nodata \\
    \sidehead{Other objects from literature}
      SN~1991T    & NGC~4527                       & $0.005791$          & $32.35$ & $\phe11.50$               & $\phe{-}19.16$               & $2.07$  & Sb      \\
      SN~1991bg   & M84                            & $0.003392$          & $31.17$ & $\phe\phn9.88$            & $\phe{-}21.29$               & $1.29$  & E       \\
      SN~2009dc   & UGC~10064                      & $0.021391$          & $34.83$ & $\phe13.63$               & $\phe{-}21.20$               & $1.33$  & S0
  \enddata
  \tablenotetext{a}{Heliocentric redshifts from the following sources: \citealt{Catinella2005} (SN~1991bj), \citealt{Falco1999} (SNe~2002cx and 2009dc), \citealt{RC3} (SN~2003gq), \citealt{Paturel2003} (SN~2004gw), \citealt{Barnes2001} (SN~2005hk), \citealt{Fisher1995} (SN~2006hn), SDSS (SN~2007qd, iPTF~13an, PTF~10acdh), \citealt{Ogando2008} (SN~2008ge), \citealt{Lu1993} (SN~2008ha), \citealt{Narayan2011} (SN~2009ku), \citealt{Ganeshalingam2012} (SN~2002es), \citealt{Strauss1992} (SN~1991T), \citealt{Cappellari2011} (SN~1991bg). Otherwise the values are from PTF spectra (multiple narrow lines for PTF~09ego and 09eoi, Na~D for PTF~10bvr, Na~D and Ca~IR for PTF~10ujn) or fits to SN~2002cx spectra (PTF~09eiy, 10xk, 11hyh).}
  \tablenotetext{b}{Distance moduli taken to be as follows: the median redshift-independent value listed on NED, excluding any determinations made using the supernova in question (SNe~1991bj, 2003gq, 2008ge, 2008ha, 1991T, 1991bg); calculated from the CMB-frame redshift as reported on NED (all other literature objects except SN~2009ku); calculated from the heliocentric redshift (all remaining objects). Where necessary we assume $H_0 = 70.4~\kms\Mpc^{-1}$, $\Omega_\mathrm{m} = 0.273$, and $\Omega_\Lambda = 0.728$.}
  \tablenotetext{c}{From \citealt{Jones2009} (SN~1991bj), \citealt{SPCE} (SN~2008ge), the PTF reference image (PTF~09eiy, 10xk, 10ujn), or SDSS (all others).}
  \tablenotetext{d}{Defined as $m(R) - \mu$.}
  \tablenotetext{e}{From SDSS.}
  \tablenotetext{f}{Where listed, morphological types are given in the following sources: \citealt{Ganeshalingam2012} (SN~2002es), \citealt{Narayan2011} (SN~2009ku), NED (PTF~10bvr), \citealt{Lira1998} (SN~1991T), \citealt{Filippenko1992} (SN~1991bg), \citealt{Silverman2011} (SN~2009dc), \citealt{Foley2009} (all others).}
  \tablenotetext{g}{This large elliptical galaxy is closest based on SDSS imaging and has a redshift of $0.03$. However, here we employ a redshift of $0.015$ based on narrow Na~D found in the PTF spectrum of the transient.}
\end{deluxetable}

\clearpage
\begin{deluxetable}{ccccc}
  \tabletypesize{\scriptsize}
  \tablecaption{Light Curve Properties in $R$-Band \label{tab:light_curves}}
  \tablewidth{0pt}
  \tablehead{\colhead{Name} & \colhead{Peak MJD\tablenotemark{a}} & \colhead{$\Mpeak(R)$ ($\magn$)\tablenotemark{b}} & \colhead{Rise Time (days)} & \colhead{$\Dm(R)$ ($\magn$)}}
  \startdata
    \sidehead{Literature sample\tablenotemark{c}}
      SN~1991bj  & ${\approx}48593.0$                  & \nodata                   & \nodata & \nodata                 \\
      SN~2002cx  & $\phe52421.2$                       & $\phe\phd{-}17.7$         & \nodata & $0.52$                  \\
      SN~2003gq  & $\phe52854.5$                       & $\phe\phd{-}17.2$         & \nodata & $0.71$                  \\
      SN~2004gw  & ${\approx}53362.0$                  & \nodata                   & \nodata & \nodata                 \\
      SN~2005hk  & $\phe53691.8$                       & $\phe\phd{-}18.0$         & $20$    & $0.67$                  \\
      SN~2006hn  & ${\approx}54003.0$                  & \nodata                   & \nodata & \nodata                 \\
      SN~2007qd  & $\phe54408.4$                       & $\phe\phd{-}15.6$         & \nodata & $0.28$                  \\
      SN~2008ge  & ${\approx}54725.0$\tablenotemark{e} & \nodata                   & \nodata & \nodata                 \\
      SN~2008ha  & $\phe54787.3$                       & $\phe\phd{-}14.2$         & \nodata & $0.97$                  \\
      SN~2009ku  & $\phe55097.2$                       & $\phe\phd{-}18.5$         & \nodata & $0.32$                  \\[\tablerowsep]
      SN~2002es  & $\phe52521.0$                       & $\phe\phd{-}18.0$         & $16$    & $0.57$                  \\
    \sidehead{SN~2002cx matches}
      PTF~09ego  & $\phe55088.3$                       & $\phe\phd{-}18.3$         & $26$    & $0.25$                  \\
      PTF~09eiy  & $\phe55084.4$                       & $<-18.0$                  & \nodata & $0.38$                  \\
      PTF~09eoi  & $\phe55076.3$                       & $\phe\phd{-}16.9$         & $24$    & \nodata                 \\
      PTF~10xk   & $\phe55203.3$                       & $\phe\phd{-}17.3$         & \nodata & $0.57$                  \\
      PTF~11hyh  & $\phe55755.0$                       & $\phe\phd{-}18.7$         & \nodata & $0.44$                  \\
      iPTF~13an  & ${\approx}56328.0$\tablenotemark{f} & $<-17.5$\tablenotemark{f} & \nodata & $0.61$\tablenotemark{f} \\
    \sidehead{SN~2002es matches}
      PTF~10bvr  & $\phe55263.5$                       & $<-16.1$                  & \nodata & \nodata                 \\
      PTF~10ujn  & $\phe55452.5$                       & $\phe\phd{-}17.9$         & $17$    & $0.63$                  \\
      PTF~10acdh & $\phe55553.0$                       & $\phe\phd{-}18.1$         & $14$    & $0.70$                  \\
    \sidehead{Other objects from literature\tablenotemark{d}}
      SN~1991T   & $\phe48377.1$                       & $\phe\phd{-}19.2$         & $16$    & $0.60$                  \\
      SN~1991bg  & $\phe48605.5$                       & $\phe\phd{-}17.5$         & \nodata & $1.36$                  \\
      SN~2009dc  & $\phe54946.9$\tablenotemark{g}      & $\phe\phd{-}19.5$         & $23$    & $0.15$
  \enddata
  \tablenotetext{a}{Modified Julian Date, $\mathrm{MJD} = \mathrm{JD} - 2{,}400{,}000$.}
  \tablenotetext{b}{Uses distance modulii as reported in Table~\ref{tab:hosts}.}
  \tablenotetext{c}{Derived from data from the following sources: \citealt{Foley2013} (SNe~1991bj, 2003gq, 2004gw, 2006hn, 2008ge), \citealt{Li2003} (SN~2002cx), \citealt{Sahu2008} (SN~2005hk), \citealt{McClelland2010} (SN~2007qd), \citealt{Foley2009} (SN~2008ha), \citealt{Narayan2011} (SN~2009ku), \citealt{Ganeshalingam2012} (SN~2002es).}
  \tablenotetext{d}{Derived from data from the following sources: \citealt{Lira1998} (SN~1991T), \citealt{Filippenko1992} (SN~1991bg), \citealt{Silverman2011} (SN~2009dc).}
  \tablenotetext{e}{Peak in $V$-band.}
  \tablenotetext{f}{Assuming earliest observation corresponds to peak brightness.}
  \tablenotetext{g}{Peak in $B$-band.}
\end{deluxetable}

\clearpage
\begin{deluxetable}{cccccccc}
  \tabletypesize{\scriptsize}
  \tablecaption{Observations of Spectra \label{tab:observations_of_spectra}}
  \tablewidth{0pt}
  \tablehead{\colhead{Name} & \colhead{Date (UT)} & \colhead{MJD} & \colhead{Phase\tablenotemark{a} (days)} & \colhead{Telescope} & \colhead{Instrument} & \colhead{Wavelength Range ($\A$)}}
  \startdata
    \multirow{2}{*}{PTF~09ego}                  & 2009-09-23 & 55097.9       & $\phn\phn{+}9$ & Keck~1 & LRIS   & $3800\text{--}9200$ \\
                                                & 2010-05-15 & 55332\phd\phn & ${+}221$       & Keck~1 & LRIS   & $3400\text{--}8800$ \\[\tablerowsep]
    \multirow{5}{*}{PTF~09eiy}                  & 2009-09-23 & 55097.9       & $\phn{+}14$    & Keck~1 & LRIS   & $3500\text{--}9700$ \\
                                                & 2009-10-12 & 55117\phd\phn & $\phn{+}33$    & WHT    & ISIS   & $3800\text{--}7600$ \\
                                                & 2009-11-11 & 55147.0       & $\phn{+}63$    & Keck~1 & LRIS   & $3700\text{--}8700$ \\
                                                & 2009-12-19 & 55184.8       & ${+}100$       & Keck~1 & LRIS   & $3800\text{--}8800$ \\
                                                & 2010-01-09 & 55205.8       & ${+}121$       & Keck~1 & LRIS   & $3900\text{--}8000$ \\[\tablerowsep]
    \multirow{5}{*}{PTF~09eoi}                  & 2009-09-23 & 55098.0       & $\phn{+}21$    & Keck~1 & LRIS   & $3700\text{--}9600$ \\
                                                & 2009-10-11 & 55116\phd\phn & $\phn{+}38$    & WHT    & ISIS   & $3800\text{--}7800$ \\
                                                & 2009-10-16 & 55120.7       & $\phn{+}43$    & Keck~1 & LRIS   & $3500\text{--}8400$ \\
                                                & 2009-11-11 & 55146.8       & $\phn{+}68$    & Keck~1 & LRIS   & $3700\text{--}8900$ \\
                                                & 2009-12-19 & 55184.7       & ${+}104$       & Keck~1 & LRIS   & $4000\text{--}8000$ \\[\tablerowsep]
    \multirow{2}{*}{PTF~10xk}                   & 2010-02-11 & 55238.8       & $\phn{+}36$    & Keck~1 & LRIS   & $3500\text{--}9100$ \\
                                                & 2010-03-07 & 55262.8       & $\phn{+}60$    & Keck~1 & LRIS   & $3500\text{--}7200$ \\[\tablerowsep]
    PTF~10bvr                                   & 2010-03-07 & 55263.1       & $\phn\phn{+}9$ & Keck~1 & LRIS   & $3500\text{--}9000$ \\[\tablerowsep]
    PTF~10ujn                                   & 2010-11-07 & 55508.1       & $\phn{+}56$    & Keck~1 & LRIS   & $3900\text{--}8300$ \\[\tablerowsep]
    \multirow{2}{*}{PTF~10acdh}                 & 2010-12-13 & 55544\phd\phn & $\phn\phn{-}8$ & P200   & DBSP   & $3800\text{--}8000$ \\
                                                & 2010-12-30 & 55561\phd\phn & $\phn\phn{+}8$ & Keck~1 & LRIS   & $3500\text{--}9500$ \\[\tablerowsep]
    PTF~11hyh                                   & 2011-08-06 & 55780.0       & $\phn{+}25$    & P200   & DBSP   & $3600\text{--}8800$ \\[\tablerowsep]
    \multirow{2}{*}{iPTF~13an\tablenotemark{b}} & 2013-02-18 & 56341.9       & $\phn{+}13$    & P200   & DBSP   & $3100\text{--}9300$ \\
                                                & 2013-06-06 & 56450\phd\phn & ${+}113$       & Keck~2 & DEIMOS & $4500\text{--}8900$ 
  \enddata
  \tablenotetext{a}{Time since $R$-band maximum; adjusted for host redshift except in the cases of PTF~09eiy, 10xk, and 11hyh.}
  \tablenotetext{b}{Phases given are lower bounds, as observations began after maximum.}
\end{deluxetable}
\begin{deluxetable}{cccccccc}
  \tabletypesize{\scriptsize}
  \tablewidth{0pt}
  \tablecolumns{8}
  \tablecaption{Relative Velocities Compared to Templates \label{tab:velocities}}
  \tablehead{\colhead{Object} & \colhead{Phase (days)\tablenotemark{a}} & \multicolumn{6}{c}{Velocity ($\kms$) Relative to} \\ \colhead{} & \colhead{} & \colhead{SN~2002cx $+12$} & \colhead{SN~2002cx $+25$} & \colhead{SN~2002cx $+56$} & \colhead{SN~2005hk $+54$} & \colhead{SN~2008ha $+11$} & \colhead{SN~2002es $+6$} \\ \colhead{} & \colhead{} & \colhead{($7000~\kms$)} & \colhead{($5000~\kms$)} & \colhead{($2000~\kms$)} & \colhead{} & \colhead{} & \colhead{($6000~\kms$)}}
  \startdata
    \sidehead{Templates}
      \multirow{3}{*}{SN~2002cx}                  & $\phe\phn{+}12$                  & $\phn\phn\phn\phn{-}4$ & $\phn{+}1070$          & $\phn{+}1811$           & $\phn{+}2270$          & $\phn{+}3041$          & $\phn{-}140$       \\
                                                  & $\phe\phn{+}25$                  & $\phn{-}1096$          & $\phn\phn\phn\phn{-}6$ & $\phn\phn{+}857$        & $\phn\phn{+}948$       & $\phn{+}1925$          & $-1901$            \\
                                                  & $\phe\phn{+}56$                  & $\phn{-}1813$          & $\phn\phn{-}844$       & $\phn\phn\phn\phn{-}5$  & $\phn\phn{+}140$       & $\phn\phn{+}873$       & $-2239$            \\[\tablerowsep]
      \multirow{2}{*}{SN~2005hk}                  & $\phe\phn\phn{+}4$               & $\phn\phn{+}895$       & $\phn{+}2405$          & $\phn{+}3100$           & $\phn{+}3003$          & $\phn{+}4844$          & $\phn{+}503$       \\
                                                  & $\phe\phn{+}54$                  & $\phn{-}2278$          & $\phn\phn{-}963$       & $\phn\phn{-}142$        & $\phn\phn\phn\phn{-}4$ & $\phn\phn{+}771$       & $-2888$            \\[\tablerowsep]
      \multirow{3}{*}{SN~2008ha}                  & $\phe\phn\phn{+}8$               & $\phn{-}3118$          & $\phn{-}1752$          & $\phn\phn{-}870$        & $\phn\phn{-}655$       & $\phn\phn{+}151$       & $-3467$            \\
                                                  & $\phe\phn{+}11$                  & $\phn{-}2908$          & $\phn{-}1843$          & $\phn\phn{-}878$        & $\phn\phn{-}761$       & $\phn\phn\phn\phn{-}2$ & $-3304$            \\
                                                  & $\phe\phn{+}22$                  & $\phn{-}3979$          & $\phn{-}2920$          & $\phn{-}1645$           & $\phn{-}1561$          & $\phn\phn{-}243$       & $-3974$            \\[\tablerowsep]
      \multirow{2}{*}{SN~2002es}                  & $\phe\phn\phn{+}6$               & $\phn\phn{+}183$       & $\phn{+}1904$          & $\phn{+}2252$           & $\phn{+}2873$          & $\phn{+}2362$          & $\phn\phn\phn{-}3$ \\
                                                  & $\phe\phn{+}67$                  & $\phn\phn{-}539$       & $\phn\phn{-}200$       & $\phn\phn{+}683$        & $\phn\phn{+}862$       & $\phn{+}2040$          & $\phn\phn{-}97$    \\[\tablerowsep]
    \sidehead{02cx-likes}
      \multirow{2}{*}{SN~1991bj}                  & ${\approx}\phn{+}29$             & $\phn{-}1050$          & $\phn\phn{+}158$       & $\phn{+}1044$           & $\phn{+}1277$          & $\phn{+}3635$          & $\phn{-}577$       \\
                                                  & ${\approx}\phn{+}39$             & $\phn{+}4025$          & $\phn{+}5120$          & $\phn{+}6236$           & $\phn{+}6365$          & $\phn{+}7907$          & $+2464$            \\[\tablerowsep]
      \multirow{3}{*}{SN~2003gq}                  & $\phe\phn\phn{-}6$               & $\phn{+}2972$          & $\phn{+}3928$          & $\phn{+}3594$           & $\phn{+}5260$          & $\phn{+}4376$          & $+1034$            \\
                                                  & $\phe\phn\phn{-}4$               & $\phn{+}1881$          & $\phn{+}4336$          & $\phn{+}4752$           & $\phn{+}5553$          & $\phn{+}3854$          & $\phn{+}195$       \\
                                                  & $\phe\phn{+}55$                  & $\phn{-}2386$          & $\phn{-}1549$          & $\phn{-}1083$           & $\phn\phn{-}683$       & \nodata                & $-2797$            \\[\tablerowsep]
      \multirow{2}{*}{SN~2004gw}                  & ${\approx}\phn{+}24$             & $\phn\phn\phn{-}85$    & $\phn\phn{+}948$       & $\phn{+}2024$           & $\phn{+}2103$          & $\phn{+}3783$          & $\phn{-}694$       \\
                                                  & ${\approx}\phn{+}52$             & $\phn{-}1279$          & $\phn\phn{-}221$       & $\phn\phn{+}764$        & $\phn\phn{+}926$       & $\phn{+}2416$          & $-1148$            \\[\tablerowsep]
      SN~2006hn                                   & ${\approx}\phn{+}30$             & $\phn{-}1284$          & $\phn\phn{-}144$       & $\phn\phn{+}750$        & $\phn{+}1115$          & \nodata                & $-2095$            \\[\tablerowsep]
      \multirow{2}{*}{SN~2007qd}                  & $\phe\phn{+}10$                  & $\phn{-}2303$          & $\phn\phn{-}986$       & $\phn\phn\phn{+}42$     & $\phn\phn{+}211$       & $\phn{+}625$           & $-2806$            \\
                                                  & $\phe\phn{+}14$                  & $\phn{-}3124$          & $\phn{-}1650$          & $\phn\phn{-}675$        & $\phn\phn{-}410$       & $\phn{+}134$           & $-3370$            \\[\tablerowsep]
      SN~2008ge                                   & $\phe\phn{+}41$                  & $\phn\phn{-}677$       & $\phn\phn{+}500$       & $\phn{+}1628$           & $\phn{+}1740$          & $\phn{+}2600$          & $-1756$            \\[\tablerowsep]
      \multirow{2}{*}{PTF~09ego}                  & $\phe\phn\phn{+}9$               & $\phn\phn{-}283$       & $\phn\phn{+}254$       & $\phn\phn{+}181$        & $\phn\phn{+}175$       & $\phn{+}1438$          & $-1025$            \\
                                                  & $\phe{+}221$                     & $\phn\phn{-}879$       & $\phn\phn{-}751$       & $\phn\phn{-}266$        & $\phn\phn{-}445$       & $\phn\phn\phn{-}60$    & $\phn\phn{+}53$    \\[\tablerowsep]
      \multirow{5}{*}{PTF~09eiy\tablenotemark{b}} & $\phe\phn{+}14$                  & $\phn{+}2642$          & $\phn{+}3637$          & $\phn{+}4403$           & $\phn{+}4885$          & $\phn{+}4740$          & $+1380$            \\
                                                  & $\phe\phn{+}33$                  & $\phn\phn{-}295$       & $\phn\phn{+}222$       & $\phn\phn{+}961$        & $\phn{+}1020$          & $\phn{+}1944$          & $-1594$            \\
                                                  & $\phe\phn{+}63$                  & $\phn{-}1709$          & $\phn{-}1276$          & $\phn\phn{-}388$        & $\phn\phn{-}260$       & $\phn\phn{+}360$       & $-2712$            \\
                                                  & $\phe{+}100$                     & $\phn{-}2970$          & $\phn{-}2200$          & $\phn{-}1174$           & $\phn\phn{-}964$       & $\phn\phn{-}500$       & $-3895$            \\
                                                  & $\phe{+}121$                     & $\phn{-}3110$          & $\phn{-}2581$          & $\phn{-}1501$           & $\phn{-}1403$          & $\phn\phn{-}495$       & $-3671$            \\[\tablerowsep]
      \multirow{5}{*}{PTF~09eoi}                  & $\phe\phn{+}21$                  & $\phn{-}1583$          & $\phn\phn{-}530$       & $\phn\phn{+}479$        & $\phn\phn{+}624$       & $\phn{+}1100$          & $-2375$            \\
                                                  & $\phe\phn{+}38$                  & $\phn{-}2176$          & $\phn{-}1468$          & $\phn\phn{-}486$        & $\phn\phn{-}503$       & $\phn\phn{+}456$       & $-3043$            \\
                                                  & $\phe\phn{+}43$                  & $\phn{-}2522$          & $\phn{-}1254$          & $\phn\phn{-}266$        & $\phn\phn{-}166$       & $\phn\phn{+}477$       & $-2824$            \\
                                                  & $\phe\phn{+}68$                  & $\phn{-}2782$          & $\phn{-}1222$          & $\phn\phn{-}392$        & $\phn\phn{-}234$       & $\phn\phn{+}156$       & $-3163$            \\
                                                  & $\phe{+}104$                     & $\phn{-}2284$          & $\phn\phn{+}518$       & $\phn\phn{-}273$        & $\phn\phn{-}117$       & $\phn\phn{+}276$       & $\phn{+}637$       \\[\tablerowsep]
      \multirow{2}{*}{PTF~10xk\tablenotemark{b}}  & $\phe\phn{+}36$                  & $\phn{-}1902$          & $\phn\phn{-}804$       & $\phn\phn{+}151$        & $\phn\phn{+}311$       & $\phn{+}1302$          & $-2312$            \\
                                                  & $\phe\phn{+}60$                  & $\phn{-}1955$          & $\phn{-}1031$          & $\phn\phn{-}191$        & $\phn\phn{+}188$       & $\phn\phn{+}790$       & $-1185$            \\[\tablerowsep]
      PTF~11hyh\tablenotemark{b}                  & $\phe\phn{+}25$                  & $\phn{-}1113$          & $\phn\phn\phn{-}62$    & $\phn\phn{+}554$        & $\phn\phn{+}719$       & $\phn{+}1544$          & $-1590$            \\[\tablerowsep]
      \multirow{2}{*}{iPTF~13an}                  & $\phe\phn{+}13$\tablenotemark{c} & $\phn\phn{-}683$       & $\phn\phn{+}567$       & $\phn{+}1590$           & $\phn{+}1685$          & $\phn{+}2139$          & $-1539$            \\
                                                  & $\phe{+}113$\tablenotemark{c}    & $\phn{-}2541$          & $\phn{-}1500$          & $\phn\phn{-}701$        & $\phn\phn{-}483$       & $\phn\phn{+}170$       & $-2625$            \\[\tablerowsep]
    \sidehead{02es-likes}
      PTF~10bvr                                   & $\phe\phn\phn{+}9$               & $\phn{-}1457$          & $\phn\phn{-}239$       & $\phn\phn{-}252$        & $\phn\phn{-}664$       & $\phn{+}2594$          & $-1952$            \\[\tablerowsep]
      PTF~10ujn                                   & $\phe\phn{+}50$                  & $\phn\phn{-}508$       & $\phn\phn{+}400$       & $\phn\phn{+}966$        & $\phn{+}1302$          & $\phn{+}2003$          & $\phn{-}822$       \\[\tablerowsep]
      \multirow{2}{*}{PTF~10acdh}                 & $\phe\phn\phn{-}8$               & $\phn{+}7782$          & $\phn{+}8777$          & $\phn{+}9894$           & $+10353$               & $+10503$               & $+2334$            \\
                                                  & $\phe\phn\phn{+}8$               & $\phn{+}2086$          & $\phn{+}2764$          & $\phn{+}3098$           & $\phn{+}3270$          & $\phn{+}4284$          & $\phn{+}865$       \\[\tablerowsep]
    \sidehead{Other objects from literature}
      \multirow{2}{*}{SN~1991T}                   & $\phe\phn\phn{+}6$               & $\phn{+}5500$          & $\phn{+}6730$          & $\phn{+}7797$           & $\phn{+}7988$          & $\phn{+}7809$          & $+4330$            \\
                                                  & $\phe\phn{+}40$                  & $\phn{+}3648$          & $\phn{+}4536$          & $\phn{+}5534$           & $\phn{+}5763$          & $\phn{+}7390$          & $+2569$            \\[\tablerowsep]
      \multirow{2}{*}{SN~1991bg}                  & $\phe\phn\phn{\phe}0$            & $+10947$               & $\phn{+}12124$         & $\phn{+}12936$          & $+12978$               & $+14192$               & $+2990$            \\
                                                  & $\phe\phn{+}53$                  & $\phn{+}2134$          & $\phn{+}3360$          & $\phn{+}3463$           & $\phn{+}3975$          & $\phn{+}5570$          & $+2441$            \\[\tablerowsep]
      \multirow{11}{*}{SN~2009dc}                 & $\phe\phn\phn{-}7$               & $\phn{+}4837$          & $\phn{-}2202$          & $\phn{-}1203$           & $\phn{-}1439$          & $\phn\phn{+}402$       & $+2389$            \\
                                                  & $\phe\phn{+}23$                  & $\phn\phn{+}611$       & $\phn{+}1390$          & $\phn{+}1152$           & $\phn{+}2594$          & $\phn{+}3958$          & $\phn{+}505$       \\
                                                  & $\phe\phn{+}35$                  & $\phn\phn{-}818$       & $\phn\phn{+}199$       & $\phn\phn{+}961$        & $\phn{+}1412$          & \nodata                & $\phn{-}827$       \\
                                                  & $\phe\phn{+}52$                  & $\phn{-}1114$          & $\phn\phn{-}211$       & $\phn\phn\phn{-}61$     & $\phn\phn{+}644$       & \nodata                & $\phn{-}923$       \\
                                                  & $\phe\phn{+}64$                  & $\phn\phn{-}646$       & $\phn\phn\phn{-}15$    & $\phn\phn{+}147$        & $\phn\phn{+}832$       & $\phn{+}3083$          & $\phn{-}250$       \\
                                                  & $\phe\phn{+}79$                  & $\phn\phn{-}123$       & $\phn\phn{+}420$       & $\phn\phn{+}762$        & $\phn{+}1378$          & $\phn{+}3009$          & $\phn{-}217$       \\
                                                  & $\phe\phn{+}80$                  & $\phn\phn{-}368$       & $\phn\phn{+}241$       & $\phn\phn{+}640$        & $\phn{+}1256$          & $\phn{+}2829$          & $\phn{-}158$       \\
                                                  & $\phe\phn{+}87$                  & $\phn\phn\phn{-}23$    & $\phn\phn{+}479$       & $\phn\phn{+}923$        & $\phn{+}1597$          & $\phn{+}2897$          & $\phn{-}358$       \\
                                                  & $\phe\phn{+}92$                  & $\phn\phn\phn\phn{+}8$ & $\phn\phn{+}580$       & $\phn\phn{+}930$        & $\phn{+}1622$          & $\phn{+}2877$          & $\phn{-}149$       \\
                                                  & $\phe{+}109$                     & $\phn\phn{+}147$       & $\phn\phn{+}635$       & $\phn{+}1090$           & $\phn{+}1829$          & $\phn{+}2330$          & $-1636$            \\
                                                  & $\phe{+}281$                     & $\phn{-}2318$          & $\phn{-}1317$          & $\phe\phn\phn\phn\phn0$ & $\phn\phn{+}549$       & $\phn{+}1302$          & $-2024$            \\[\tablerowsep]
      \multirow{3}{*}{SN~2011fe}                  & $\phe\phn\phn{+}7$               & $\phn{+}4869$          & $\phn{+}6858$          & $\phn{+}9519$           & $\phn{+}8943$          & $\phn{+}5290$          & $+3910$            \\
                                                  & $\phe\phn{+}16$                  & $\phn{+}2915$          & $\phn{+}3660$          & $\phn{+}4140$           & $\phn{+}5018$          & $\phn{+}5316$          & $+2520$            \\
                                                  & $\phe\phn{+}78$                  & $\phn{+}3227$          & $\phn{+}1932$          & $\phn{+}2303$           & $\phn{+}2782$          & $\phn{+}4198$          & $\phn{+}956$
  \enddata
  \tablenotetext{a}{Time since $R$-band maximum; adjusted for host redshift except in the cases of PTF~09eiy, 10xk, and 11hyh.}
  \tablenotetext{b}{Shift depends on chosen value for redshift.}
  \tablenotetext{c}{Assuming first observation corresponds to peak luminosity.}
\end{deluxetable}

\clearpage

\clearpage
\begin{deluxetable}{ccc}
  \tabletypesize{\scriptsize}
  \tablewidth{0pt}
  \tablecaption{Properties of SN~2002cx and SN~2002es Families \label{tab:comparison}}
  \tablehead{\colhead{Property} & \colhead{02cx-like} & \colhead{02es-like}}
  \startdata
    Host type                                 & generally late      & early               \\
    Host $g\text{--}i$ color ($\magn$)        & $0.59\text{--}1.47$ & $1.26\text{--}1.58$ \\
    Host luminosity ($R$-band, $\magn$)       & $-14$ to $-21$      & $-19$ to $-22$      \\
    Peak luminosity ($R$-band, $\magn$)       & $-13$ to $-19$      & $-18$               \\
    Rise time (days)                          & $20\text{--}30$     & $14\text{--}17$     \\
    Decline rate ($\Dm$, $\magn$)             & $0.2\text{--}1.0$   & $0.6\text{--}0.7$   \\
    \ion{Ti}{2} trough?                       & no                  & yes                 \\
    Ejecta velocity ($\kms$)\tablenotemark{a} & $4000\text{--}9000$ & $4000\text{--}7000$
  \enddata
  \tablenotetext{a}{At approximately $10~\days$ post-maximum.}
\end{deluxetable}


\clearpage

\clearpage

\end{document}